\documentclass{article}
\pdfoutput=1
\newcommand{\bigO}{{\mathcal{O}}} 
\newcommand{\bo}{\boldsymbol}
\newcommand{\tr}{\textrm} 
 
\newcommand{\wt}{\widetilde}
\newcommand{\ttt}{\texttt}

\usepackage[utf8]{inputenc}
\usepackage{amsmath}
\usepackage{amssymb}
\usepackage{polski}
\usepackage[english,polish,main=english]{babel}
\usepackage[makeroom]{cancel}
\usepackage{graphicx}
\usepackage{subcaption}
\usepackage{tikz}
\usepackage{tikz-3dplot}
\usepackage{tikz-layers}
\usepackage{pgfplots}
\usepackage{biblatex}
\usepackage{float}
\usepackage{ upgreek }
\usepackage{ esint }
\usepackage[mode=buildnew]{standalone}
\usepackage{tabularx}
\usepackage{subfiles}
\usepackage{enumerate}
\usepackage{xcolor}
\usepackage{ifthen}
\usepackage{makecell}
\usetikzlibrary{math}
\usetikzlibrary{through}

\pgfplotsset{compat=1.17}
\usetikzlibrary{arrows.meta}
\usetikzlibrary{intersections,calc,pgfplots.fillbetween}
\usetikzlibrary{calc,3d,intersections, positioning,intersections,shapes}
\definecolor{myBlue}{rgb}{0.75,0.9,1}

\usepackage{subfiles}

\title{Formation of Thermal Vortex Rings}
\author{Paweł Jędrejko\\ {\small Interdisciplinary Centre for Mathematical and} \\ [-0.15cm]{\small Computational Modelling, University of Warsaw}\\[0.3 cm]{Jun-Ichi Yano,} \\ {\small CNRM UMR3589 (CNRS), M{\'e}t{\'e}o-France,} \\ [-0.15cm]{\small 31057 Toulouse Cedex, France}\\[0.3 cm]{Marta Wacławczyk,}\\{\small Institute of Geophysics, University of Warsaw}}
\date{}
\begin{document}

\maketitle
\begin{abstract}
    An evolution of a spherical region, subjected to uniform buoyancy force, is investigated. Incompressibility and axial symmetry are assumed, together with a buoyancy discontinuity at the boundary. 
    The boundary turns into a vortex sheet and the system evolves into a ring. Contrary to the case of mechanically generated rings, buoyancy-driven rings are unstable. This is due to the generation of negative vorticity at the bottom.
    Furthermore, a sequence of Kelvin-Helmholtz instabilities arises along the buoyancy anomaly boundary. This sequence transfers the energy toward large scales with $\sim\kappa^{-3}$ distribution. \\
    The vortex blob method has been used to simulate the system numerically. An optimization algorithm, used previously in two dimensions, has been extended to the axisymmetric case.
    It reduces computational complexity from $N^2$ to $N\log N$, where $N$ is the number of nodes.
    Additionally, a new algorithm has been developed as a remedy for the exponential growth of the number of nodes required. It exploits a tendency of the vortex sheet to form many parallel stripes, by merging them together.
\end{abstract}
\section{Introduction}
    
Thermal vortex rings play an important role in the formation of cumulus clouds. Rising thermal vortex rings correspond to the initial phase of atmospheric convection which is considered adiabatic \cite{PAC}. In this work, we adopt this assumption and model them as regions of a fluid subjected to uniform buoyancy force. We neglect stratification and focus on the high Reynolds number regime.
The problem, which details are described in the next section was already considered 40 years ago.
Since that time, a lot has changed in computational physics and available computational resources. We approach the problem from the very same perspective of vortex dynamics in lagrangian formulation, although enriching it with modern capacities. Its advantage over standard, eulerian mesh methods is an insight into flow coherencies.
Moreover, high Reynolds numbers might be troublesome for Eulerian methods.
We present a few developments on an algorithmic basis and take advantage of modern hardware architecture. We will also refer to mechanically generated vortex rings. 
The main body of this article consists of four parts. In the next one - the second section, we present a mathematical statement of the problem and derivation of the governing equations. The third section describes the basics of the numerical method that we used and details of the surgery procedure. The fourth one is dedicated to optimization of the numerics - a fast velocity induction algorithm is described as well as fast surgery. A few test cases support their correctness. We present results and postprocessing in the fifth section. The article finishes with a conclusion and suggestions for further work.

\section{Statement of the problem}

\subsection{Geometry and governing equations}
We will consider an evolution of an initially spherical region, of an incompressible fluid with increased temperature. We will let it be constant in both regions, with a discontinuity at the interface. The system is assumed to have axial symmetry. Far enough we expect the impact of the hot region to vanish so we let the velocity approach zero at infinity:
\begin{equation}
    \lim_{r \to \infty}\boldsymbol{u} = \boldsymbol{0}
    \label{bc}
\end{equation}
\begin{figure}[h]
    \centering

\definecolor{myBlue}{rgb}{0.0,0.7,1}
\definecolor{myDarkBlue}{rgb}{0.3,    0.650,    0.8500}

\pgfdeclarelayer{ft2} 
\pgfdeclarelayer{ft}
\pgfdeclarelayer{bg} 
\pgfsetlayers{behind,main,ft,above} 

\newcommand{\vaOne}{60} 
\newcommand{\vaTwo}{115}

\tdplotsetmaincoords{\vaOne}{\vaTwo}
\pgfplotsset{compat=newest}

\begin{tikzpicture}[tdplot_main_coords]

\def\R{2.5}
\def\Ry{0.3}

\coordinate (O) at (0,0,0);  
\shade[ball color =myBlue , opacity = 0.4] (O) circle (\R cm);
\path[draw, name path=contour] (0 cm,0 cm) circle (\R cm);
  
    \draw[->] (0,0,{\R}) -- (0,0,{\R*1.3}) node[above] {$z$};
    \draw[dashed] (0,0,0) -- (0,0,\R);
\begin{scope}[canvas is yz plane at x = 0]
    \draw[->, semithick] ({\R},0) -- ({\R*1.3},0) node[right] {$\rho$};
    \draw[dashed] (0,0) -- ({\R},0);
\end{scope}

\begin{scope}[canvas is xy plane at z=0] 
    \path[draw,thin,name path=equator2] (\R,0) arc [start angle=0,end angle=90,radius=\R cm];
        \path[draw,thin,dashed,name path=equator2] (\R,0) arc [start angle=0,end angle=360,radius=\R cm];
\end{scope}  


\foreach \Th in {0} {
        \def\rv{cos(\Th)*\R};
        \def\hv{sin(\Th)*\R};
        
        \begin{scope} 
        \path[canvas is xy plane at z = \hv, name path=pilot] (0,0) circle (\rv);
        \path [name intersections={of=pilot and contour,by=K}];
        \clip($(K)+(1 cm,0 cm)$) rectangle (-3cm,-3cm);
    
        \path[draw,thin,canvas is xy plane at z = \hv] (0,{\rv}) arc [start angle=90,end angle=360,radius=\rv]; 
        \end{scope}
        } 
     
\node[] at ({-0.46*\R cm}, {-0.1*\R cm})  (a2)    {$T$};
\node[] at ({-1.3*\R cm}, {-0.1*\R cm})  (a2)    {$T_0$};

\end{tikzpicture}

    \caption{Initial condition}
    \label{fig_intro}
\end{figure}
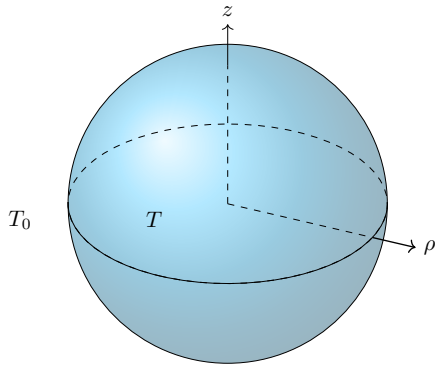
We start with a system of three, well-known, equations, representing the conservation of mass, momentum, and energy (without dissipative heating) respectively
\begin{equation}
    \frac{\partial \uprho}{\partial t} + \nabla \cdot (\uprho \boldsymbol{u}) = 0
    \label{mass}
\end{equation}
\begin{equation}
    \uprho\frac{D\boldsymbol{u}}{Dt} = -\nabla p + \mu\nabla ^2 \boldsymbol{u} - \uprho g \hat{k}
    \label{momentum}
\end{equation}
\begin{equation}
    \frac{DT}{Dt} = \frac{\lambda}{\uprho c_p} \nabla^2 T
    \label{energy}
\end{equation}
we also assumed that fluid properties are constant with exception of density. We will further assume that it depends on $T$ exclusively and it can be expanded in a Taylor series around some reference value $T_0$:
\begin{equation}
   \uprho(T) = \uprho(T_0) + \frac{d \uprho}{d T}\Big|_{T_0}(T-T_0) + \bigO \Big( (T-T_0)^2 \Big)
    \label{TTaylor}
\end{equation}
If we now consider a piece of fluid, from the mass conservation (de facto defining the piece) and product rule, we know that:
\begin{flalign}
            & \frac{d (\uprho V)}{d T} = 0 \\
            & \frac{d \uprho}{d T}V = -\frac{d V}{d T}\rho
        \end{flalign}
and by definition of thermal expansion coefficient:
\begin{equation}
    \frac{\partial \uprho}{\partial T} = -\beta \uprho
\end{equation}
Substituting this to (\ref{TTaylor}) and neglecting higher order terms we obtain:
\begin{equation}
    \uprho(T)\approx \uprho_0  -\rho_0\beta (T-T_0) 
\end{equation}
We will assume that changes of density are relevant only in the source term of momentum equation (\ref{momentum}), what is called the Boussinesq approximation. That turns (\ref{mass}) into:
\begin{equation}
    \nabla \cdot \boldsymbol{u} = 0
    \label{continuity}
\end{equation}
and (\ref{momentum}) into:
\begin{equation}
    \frac{D\boldsymbol{u}}{Dt} = -\frac{1}{\uprho_0}\nabla p + \nu\nabla ^2 \boldsymbol{u} -  g \hat{k}  -\beta (T-T_0)  g \hat{k}
\end{equation}
Introducing buoyancy as:
\begin{equation}
    b = -\beta(T-T_0)g
    \label{buoyancy}
\end{equation}
and taking $g\hat{k}$ under the gradient:
\begin{equation}
    \frac{D\boldsymbol{u}}{Dt} = -\frac{1}{\uprho_0}\nabla (p+\uprho_0gz) + \nu\nabla ^2 \boldsymbol{u}  +b \hat{k}
    \label{momentumB}
\end{equation}
We can also use (\ref{buoyancy}) to substitute for $T=-b/(\beta g)+T_0$ in (\ref{energy}). Because $T$ appears only under the derivative and $\beta$, $T_0$, $g$ are constants, $b$ will just replace $T$. \\ \\
Our last fundamental simplification will be neglecting the diffusive processes. This will turn buoyancy into a passive scalar, maintaining the discontinuous character of its distribution. The governing equations of the system are presented below:
\begin{equation}
    \nabla \cdot \bo{u} = 0
\end{equation}
\begin{equation}
    \frac{D\bo{u}}{Dt} = -\frac{1}{\uprho_0}\nabla(p+\uprho_0 gz) + b\hat{k}
\end{equation}
\begin{equation}
    \frac{Db}{Dt} = 0
\end{equation}

\subsection{Vorticity based formulation}
To reduce the number of unknowns we will turn to the vorticity equation. We will introduce vorticity defined as a curl of velocity:
\begin{equation}
    \bo{\omega} = \nabla \times \bo{u}
\end{equation}
where $C$ is a closed curve, bounding an area $A$. According to this definition, vorticity could be considered as a circulation density per unit area. \\ \\
Looking for a reverse relation we apply Helmholtz decomposition to the velocity field. Splitting it to divergence-free part ($\bo{u_{\omega}}$) and curl-free part ($\bo{u_q}$) we obtain:
\begin{equation}
    \boldsymbol{u} = \bo{u_\omega} + \bo{u_q}
    \label{helmholtz}
\end{equation}
We will also write down a few useful vector identities:
\begin{subequations}
    \begin{equation}
    \nabla \cdot (\nabla \times \boldsymbol{\psi}) = 0    
    \end{equation}
    \begin{equation}
     \nabla \times (\nabla \phi) = 0   
     \label{curlOfGrad}
    \end{equation}
    \begin{equation}
     \nabla \times (\nabla \times \boldsymbol{\psi}) =  \nabla (\nabla \cdot \boldsymbol{\psi}) -\nabla \cdot (\nabla \boldsymbol{\psi})
     \label{curlOfCurl}
     \end{equation}
     \begin{equation}
        \nabla\times(\phi\bo{\psi}) = \phi(\nabla\times\bo{\psi}) + \nabla\phi\times\psi = \phi(\nabla\times\bo{\psi}) - (\bo{\psi}\times\nabla)\phi
         \label{curlOfProd}
     \end{equation}
\end{subequations}
Using the first two ones, we write:
\begin{subequations}
    \begin{equation}
        \bo{u_\omega} = \nabla \times \bo{\psi}
        \label{helmholtzVort}
    \end{equation}
    \begin{equation}
        \bo{u_q} = \nabla \phi
    \end{equation}
    \label{helmholtzComps}
\end{subequations}
We will take for granted that all divergence-free fields can be represented as curls of some other field called vector potential.
\subsubsection{The vorticity-induced component of velocity}
Substituting (\ref{helmholtzComps}) to (\ref{helmholtz}), then taking the curl we get:
\begin{equation}
        \nabla \times \boldsymbol{u} = \nabla (\nabla \cdot \boldsymbol{\psi}) -\nabla \cdot (\nabla \boldsymbol{\psi})
        \label{praBS}
\end{equation}
In (\ref{helmholtzVort}), $\boldsymbol{\psi}$ is determined up to a potential field, due to identity (\ref{curlOfGrad}). Therefore we have a freedom to add such $\nabla f$ that makes it divergence-free:
\begin{equation}
    \nabla \cdot \boldsymbol{\psi} = \nabla \cdot (\boldsymbol{\psi'} + \nabla f) = 0
\end{equation}
This can be done by choosing some $\boldsymbol{\psi}'$ and solving Poisson equation for the potential of "correction" $f$. For convenience we will denote the source term with $q = \nabla \cdot \boldsymbol{\psi'}$
\begin{equation}
    \nabla^2 f = q
    \label{poissonf}
\end{equation}
The equation is linear so we can superimpose velocities ($v_i = \nabla f_i$) induced by separate infinitesimal sources. Let us say that at point $\boldsymbol{r}$ we got a mass source of strength $q_i dV$, closed in a ball-shaped control volume of radius $\rho$. Assuming that there is no preferable direction ($v_i = v_i(\rho)$) we can determine the velocity at point $\boldsymbol{r_0}$, belonging to the boundary of the control volume, by the integral form of mass conservation:
\begin{equation}
    q_i dV = 4\pi \rho^2 \frac{df_i}{d\rho}
\end{equation}
\begin{equation}
    v_i(\boldsymbol{r_0}) = \frac{df_i}{d\rho} = \frac{q_i dV}{4\pi \rho^2}
\end{equation}
which satisfies the boundary condition of vanishing at infinity.
By radial integration, we can obtain a potential (let constant be equal 0):
\begin{equation}
    f_i(\boldsymbol{r_0}) = -\frac{q_i dV}{4\pi \rho} = -\frac{q_i dV}{4\pi |\boldsymbol{r_0}-\boldsymbol{r}|}
\end{equation}
By superposition of all such point sources, we obtain a solution of (\ref{poissonf}):
\begin{equation}
    f(\boldsymbol{r_0}) = \frac{-1}{4\pi}\int \frac{q(\boldsymbol{r})}{ |\boldsymbol{r_0}-\boldsymbol{r}|} dV
    \label{poissonSol}
\end{equation}
Now, having divergence-free $\boldsymbol{\psi}$ we can go back to equation (\ref{praBS}) and simplify it to a system of three Poisson equations:
\begin{equation}
    \omega_i = \nabla^2 \psi_i
\end{equation}
(where $i$ can be $x$, $y$ or $z$). Solving them in exactly the same way as before, we get:
\begin{equation}
    \psi_i(\boldsymbol{r_0}) = -\frac{1}{4\pi}\int \frac{\omega_i(\boldsymbol{r})}{ |\boldsymbol{r_0}-\boldsymbol{r}|} dV
    \label{poissonSol2}
\end{equation}
then:
\begin{equation}
    \frac{\partial \psi_i}{\partial x_{0j}} = \frac{1}{4\pi}\int \frac{\omega_i (x_{0j}-x_j)}{|\boldsymbol{r_0}-\boldsymbol{r}|^3} dV
\end{equation}
Finally we compute the vorticity-induced velocity:
\begin{equation}
    \boldsymbol{u_{\omega}}(\boldsymbol{r_0}) = \nabla_0 \times \boldsymbol{\psi} = \frac{1}{4\pi}\int \frac{\boldsymbol{\omega}\times (\boldsymbol{r_0}-\boldsymbol{r})}{ |\boldsymbol{r_0}-\boldsymbol{r}|^3} dV
    \label{BS1}
\end{equation}


\subsubsection{The remaining component of velocity}
Taking the divergence of (\ref{helmholtz}) and using continuity (\ref{continuity}) on the left hand side we obtain:
\begin{equation}
    0 = \nabla^2 \phi
\end{equation}
so $\phi$ is a harmonic function. That means that it can take maximal and minimal values only at the boundary (maximum modulus principle). That is very intuitve if we interpret it as a steady state of diffusion.
Boundary condition (\ref{bc}) implies that $\nabla \phi$ is zero, so the flow is uniquely determined by the vorticity distribution.


\subsubsection{The vorticity equation}
\begin{figure}
    \centering
    \definecolor{myBlue}{rgb}{0.75,0.9,1}
\definecolor{myDarkBlue}{rgb}{0.3,    0.650,    0.8500}

\pgfdeclarelayer{ft2} 
\pgfdeclarelayer{ft}
\pgfdeclarelayer{bg} 
\pgfsetlayers{behind,main,ft,above} 

\newcommand{\vaOne}{60} 
\newcommand{\vaTwo}{115}

\tdplotsetmaincoords{\vaOne}{\vaTwo}
\pgfplotsset{compat=newest}

\begin{tikzpicture}[tdplot_main_coords]

\def\R{2.5}
\def\Ry{0.3}

\coordinate (O) at (0,0,0);  
\shade[ball color =lightgray , opacity = 0.4] (O) circle (\R cm);
\path[draw, name path=contour] (0 cm,0 cm) circle (\R cm);
  

\path[name path=rightSec, thick,canvas is zy plane at x=0] (\R,0) arc [start angle=0,end angle=180,radius=\R];
    
\path[name path=leftSec, thick,canvas is xz plane at y=0] (0,\R) arc [start angle=90,end angle=-90,radius=\R];

\path [name intersections={of=contour and rightSec,by=rsc}];
\path [name intersections={of=leftSec and contour,by=lsc}];

\coordinate (Q) at (0,0,-\R*0.2);
\coordinate (rsc2) at ($(rsc)+(Q)$);
\coordinate (lsc2) at ($(lsc)+(Q)$);
\fill[white] (0,0,-\R) -- (rsc) -- (rsc2) -- (lsc2)  -- (lsc) -- cycle; 

\begin{scope}[on above layer]
\path[draw,name path=rightSec, thin,canvas is zy plane at x=0] (\R,0) arc [start angle=0,end angle=180,radius=\R];
    
\path[draw,name path=leftSec, thin,canvas is xz plane at y=0] (0,\R) arc [start angle=90,end angle=-90,radius=\R];

\tikzfillbetween[of=leftSec and rightSec, on layer=ft]{myBlue, opacity=1};


\begin{scope}[canvas is xy plane at z=0]
    \coordinate (L) at ({cos(\vaTwo-180)*\R cm}, {sin(\vaTwo-180)*\R cm});
    \path[draw,thin,name path=equator1] (L) arc [start angle={\vaTwo-180},end angle=0,radius=\R cm];
    \path[draw,thin,name path=equator2,dashed] (\R,0) arc [start angle=0,end angle=90,radius=\R cm];
    \path[draw,thin,name path=equator3] (0,\R) arc [start angle=90,end angle={\vaTwo},radius=\R cm];
\end{scope}

\begin{scope}[canvas is yz plane at x = 0]
    \draw[->, semithick] (O) -- ({\R*cos(45)},{\R*sin(45)}) node[left] {$\textbf{r}(s,t)$};
    \draw[->, semithick] (O) -- ({\R*1.3},0) node[right] {$\rho$};
    \path[draw,->,semithick] (0,-\R) arc [start angle=-90,end angle=-70,radius=\R] node[midway,below] {$s$};
\end{scope}
\draw[->] (0,0,{-\R*1.2}) -- (0,0,{\R*1.3}) node[above] {$z$};

\begin{scope}
        \foreach \Th in {-40,-30,...,50} {
        \def\rv{cos(\Th)*\R};
        \def\hv{sin(\Th)*\R};
        
        \begin{scope} 
        \path[canvas is xy plane at z = \hv, name path=pilot] (0,0) circle (\rv);
        \path [name intersections={of=pilot and contour,by=K}];
        \clip($(K)+(1 cm,0 cm)$) rectangle (-3cm,-3cm);
    
        \path[draw,thin,->,gray,canvas is xy plane at z = \hv] (0,{\rv}) arc [start angle=90,end angle=360,radius=\rv]; 
        \end{scope}
        } 
       \foreach \Th in {60,70} {
        \def\rv{cos(\Th)*\R};
        \def\hv{sin(\Th)*\R};
        \path[draw,thin,->,gray,canvas is xy plane at z = \hv] (0,{\rv}) arc [start angle=90,end angle=360,radius=\rv]; 
        }
\end{scope}

\draw [gray] ({-0.78*\R cm}, {0.78*\R cm}) node {$\boldsymbol{\omega}$};
\draw [->] ({0.4*\R},0,0) arc [start angle=0,end angle=90,radius={0.4*\R}] node[midway,below] {$\phi$};
\draw [->,myDarkBlue] ({\R*0.45},0,{\R*0.25}) -- ({\R*0.45},0,{\R*0.6}) node[midway, right] {$\boldsymbol{b}$};

\end{scope}

\end{tikzpicture}
    \caption{Region of non-zero buoyancy (blue), covered with parametrized vortex sheet (grey). Shown in half-section.}
    \label{fig_param}
\end{figure}
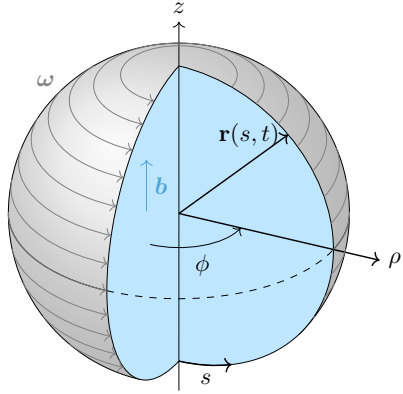

We will now take the curl of (\ref{momentumB}) to take an advantage of the fact that curl of gradient is zero.
\begin{equation}
    \nabla \times \Bigg(\frac{\partial \bo{u}}{\partial t} + (\bo{u} \cdot \nabla) \bo{u}\Bigg)= \nabla \times \Bigg(-\frac{1}{\rho_0}\nabla (p + \rho_0gz) + \nu \nabla^2 \bo{u} + b\hat{k} \Bigg)
    \label{curlOfNS}
\end{equation}
Using the identity (\ref{curlOfProd}):
\begin{equation}
    \nabla \times \Big( (\bo{u}\cdot\nabla)\bo{u}\Big) = (\bo{u}\cdot\nabla)(\nabla\times\bo{u}) - \Big(\nabla(\bo{u}\cdot\nabla)\Big)\times\bo{u} 
\end{equation}
and using Schwarz theorem (switching derivatives), we get the vorticity equation:
\begin{equation}
    \frac{\partial \bo{\omega}}{\partial t} + (\bo{u} \cdot \nabla) \bo{\omega} = (\bo{\omega} \cdot \nabla) \bo{u} +  \nu \nabla^2 \bo{\omega} + \nabla\times (b\hat{k})
    \label{vorticity}
\end{equation}
The hitch is that the last term is not differentiable at the interface due to assumed discontinuity in the buoyancy distribution. Nevertheless, we will assume that the derivative exists but is singular. Of course, in all other regions it is zero. The source term will generate a vortex sheet at the interface, with the same type of singularity, what suggests a walk-around. Curl of buoyancy points in the azimuthal direction, so does the vorticity. No vorticity will appear in non-interface regions, due to Kelvin's theorem, assuming that viscosity is negligible. \\ \\
Despite some complications, at the current state we can take a step forward and reformulate the problem. We will describe the vortex sheet at the interface as a parametric surface. Due to axial symmetry we will skip the dependence on azimuth $\phi$. 
\begin{equation}
    \bo{r}(s,t) = \begin{bmatrix}
                       \rho(s, t)\\
                        z(s, t)
                  \end{bmatrix}
\end{equation}
\begin{equation*}
    s \in [-\tfrac{1}{2}\pi, \tfrac{1}{2}\pi]
\end{equation*}
Where $s$ is a parameter in a fixed range, coincident to initial length of the $\phi = \textrm{const}$ section of the surface.
The boundary conditions, given by the axial symmetry, are:
    \begin{equation*}
        \rho(-\tfrac{1}{2}\pi,t) = 0
    \end{equation*}
    \begin{equation}
        \rho(\tfrac{1}{2}\pi,t) = 0
        \label{bcSurf}
    \end{equation}
    \begin{equation*}
        \frac{\partial z}{\partial s}(-\tfrac{1}{2}\pi,t) = 0
    \end{equation*}
    \begin{equation*}
        \frac{\partial z}{\partial s}(\tfrac{1}{2}\pi,t) = 0
    \end{equation*}
and the initial condition:
\begin{equation*}
    \rho(s, 0) = R\cos(s)
\end{equation*}
\begin{equation}
    z(s, 0) = R\sin(s)
\end{equation}
Figure (\ref{fig_param}) presents both the parametrization and the initial condition.
\subsubsection{Nondimensionalization}
Basic quantities characterizing the system are:
\begin{itemize}
    \item $b$   [L T$^2$] buoyancy
    \item $\nu$   [L$^2$ T$^{-1}$] kinematic viscosity
    \item $R$   [L] initial radius
\end{itemize}
(in brackets we denote the physical dimension)\\
then we can express (buoyancy-linked) time, velocity and vorticity scales as:
\begin{itemize}
    \item T = $\sqrt{\frac{R}{b}}$   [T] \\
    \item U = $\frac{R}{T} = \sqrt{bR}$   [L T$^{-1}$]\\
    \item $\Omega = \frac{U}{R} = \sqrt{\frac{b}{R}}$   [T$^{-1}$]
\end{itemize}
Let us now introduce non-dimensional quantities (denoted by tilde): $\bo{u} = U\wt{\bo{u}}$ etc. Substituting this to equation (\ref{vorticity}) and dividing both sides by the vorticity scale at the very beginning, we obtain:
\begin{equation}
    \frac{\partial \wt{\bo{\omega}}}{\partial \wt{t}}\frac{1}{\tr{T}} + (\wt{\bo{u}} \cdot \wt{\nabla}) \wt{\bo{\omega}} \frac{\tr{U}}{R}= (\wt{\bo{\omega}} \cdot \wt{\nabla)} \wt{\bo{u}}\frac{\tr{U}}{R} +  \frac{\nu}{R^2} \wt{\nabla}^2 \wt{\bo{\omega}} + \wt{\nabla}\times (1\hat{k})\frac{b}{R{\Omega}}
\end{equation}
 multiplying by the time scale, introducing:
 \begin{equation}
     \varepsilon = \frac{\nu T}{R^2} = \frac{\nu}{RU}
 \end{equation}
 and skipping tyldas for convenience, we get:
 \begin{equation}
         \frac{\partial \bo{\omega}}{\partial t} + (\bo{u} \cdot \nabla) \bo{\omega} = (\bo{\omega} \cdot \nabla) \bo{u} +  \varepsilon \nabla^2 \bo{\omega} + \nabla\times (1\hat{k})
         \label{vortNonDim}
 \end{equation}
 $\varepsilon$ might be interpreted as an inverse of the Reynolds number. By taking the limit $\varepsilon \to 0$ or equivalently $\tr{Re}\to \infty$ we end up with:
  \begin{equation}
         \frac{\partial \bo{\omega}}{\partial t} + (\bo{u} \cdot \nabla) \bo{\omega} = (\bo{\omega} \cdot \nabla) \bo{u}  + \nabla\times (b\hat{k})
 \end{equation}
 Although non-dimensional body force $b=1$, we will still write it explicitly, just keeping in mind its value of one. 
\subsubsection{The curl of buoyancy and the circulation density}
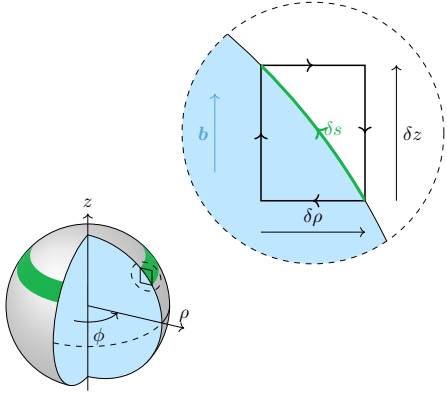
\begin{figure}
    \centering
    \definecolor{myBlue}{rgb}{0.75,0.9,1}
\definecolor{myGreen}{rgb}{0.1,0.7,0.3}
\definecolor{myDarkBlue}{rgb}{0.3,    0.650,    0.8500}

\pgfdeclarelayer{ft2} 
\pgfdeclarelayer{ft}
\pgfdeclarelayer{bg} 
\pgfsetlayers{behind,main,ft,above} 

\newcommand{\vaOne}{60} 
\newcommand{\vaTwo}{115}

\tdplotsetmaincoords{\vaOne}{\vaTwo}

\tdplotsetmaincoords{\vaOne}{\vaTwo}
\pgfplotsset{compat=newest}

\begin{tikzpicture}[tdplot_main_coords]

\def\R{1.5}
\def\Sc{8}
\coordinate (O) at (0,0,0);  

\shade[ball color =lightgray , opacity = 0.4] (O) circle (\R cm);
\path[draw, name path=contour] (0 cm,0 cm) circle (\R cm);
  

\coordinate (A0) at (0,0,{\R*sin(30)});

\begin{scope}
        \def\Th{30}
        \def\ThT{45}
        \def\rv{cos(\Th)*\R};
        \def\hv{sin(\Th)*\R};
        \def\rvT{cos(\ThT)*\R};
        \def\hvT{sin(\ThT)*\R};
        \path[canvas is xy plane at z = \hv, name path=uGclp] (0,{\rv}) arc [start angle=90,end angle=360,radius=\rv]; 
        \path[canvas is xy plane at z = \hvT, name path=uGTclp] (0,{\rvT}) arc [start angle=90,end angle=360,radius=\rvT]; 

        \path [name intersections={of=uGTclp and contour,by=K}];

        \def\THlow{60};
        \def\THup{45};
        
        \foreach \Ph in {-90,-85,...,140} {
            \def\PhT{\Ph+5.5};
            
            \coordinate (AA) at ({sin(\THlow)*cos(\Ph)*\R},{sin(\THlow)*sin(\Ph)*\R},{cos(\THlow)*\R} );
            \coordinate (BB) at ({sin(\THup)*cos(\Ph)*\R},{sin(\THup)*sin(\Ph)*\R},{cos(\THup)*\R} );
            \coordinate (CC) at ({sin(\THup)*cos(\PhT)*\R},{sin(\THup)*sin(\PhT)*\R},{cos(\THup)*\R} );
            \coordinate (DD) at ({sin(\THlow)*cos(\PhT)*\R},{sin(\THlow)*sin(\PhT)*\R},{cos(\THlow)*\R} );

            \fill[myGreen] (AA) -- (BB) -- (CC) -- (DD) -- cycle;
            \begin{scope}[canvas is xy plane at z =0]
            \end{scope}
        }

\end{scope}

\path[name path=rightSec, thick,canvas is zy plane at x=0] (\R,0) arc [start angle=0,end angle=180,radius=\R];
    
\path[name path=leftSec, thick,canvas is xz plane at y=0] (0,\R) arc [start angle=90,end angle=-90,radius=\R];

\path [name intersections={of=contour and rightSec,by=rsc}];
\path [name intersections={of=leftSec and contour,by=lsc}];

\coordinate (Q) at (0,0,-\R*0.2);
\coordinate (rsc2) at ($(rsc)+(Q)$);
\coordinate (lsc2) at ($(lsc)+(Q)$);
\fill[white] (0,0,-\R) -- (rsc) -- (rsc2) -- (lsc2)  -- (lsc) -- cycle; 

\begin{scope}[on above layer]
\path[draw,name path=rightSec, thin,canvas is zy plane at x=0] (\R,0) arc [start angle=0,end angle=180,radius=\R];
    
\path[draw,name path=leftSec, thin,canvas is xz plane at y=0] (0,\R) arc [start angle=90,end angle=-90,radius=\R];

\tikzfillbetween[of=leftSec and rightSec, on layer=ft]{myBlue, opacity=1};


\begin{scope}[canvas is xy plane at z=0]
    \coordinate (L) at ({cos(\vaTwo-180)*\R cm}, {sin(\vaTwo-180)*\R cm});
    \path[draw,thin,name path=equator2,dashed] (\R,0) arc [start angle=0,end angle=90,radius=\R cm];
    \path[draw,thin,name path=equator3] (0,\R) arc [start angle=90,end angle={\vaTwo},radius=\R cm];
    

\end{scope}  
\draw[->] (0,0,{-\R*1.2}) -- (0,0,{\R*1.3}) node[above] {$z$};
\draw[->] (0,0,0) -- (0,{\R*1.3},0) node[above] {$\rho$};
\draw [->] ({0.4*\R},0,0) arc [start angle=0,end angle=90,radius={0.4*\R}] node[midway,below] {$\phi$};
\begin{scope}[canvas is yz plane at x=0]
    \coordinate (A) at ({\R*cos(30)},{\R*sin(30)});
    \coordinate (C) at ({\R*cos(45)},{\R*sin(45)});
    \coordinate (B) at ({\R*cos(30)},{\R*sin(45)});
    \coordinate (D) at ({\R*cos(45)},{\R*sin(30)});
    \coordinate (M) at ($0.5*(C)+0.5*(A)$);
    \path[draw,->] (A) -- (B) -- (C) -- (D) -- cycle;
    \path [draw,dashed, name path=dashCirc2] (M) circle ({\R*0.2 cm});
\end{scope}

\begin{scope}[rotate around z=25, rotate around y=-30,canvas is yz plane at x=0]




\coordinate (A) at ({\R*cos(30)},{\R*sin(30)});
\coordinate (C) at ({\R*cos(45)},{\R*sin(45)});
\coordinate (B) at ({\R*cos(30)},{\R*sin(45)});
\coordinate (D) at ({\R*cos(45)},{\R*sin(30)});

\coordinate (M) at ($0.5*(C)+0.5*(A)$);
\coordinate (OO) at ($-4.5*(M)$); 
\coordinate (M1) at ($\Sc*(M)+(OO)$);
\path [name path=dashCirc2] (M) circle ({\R*0.2 cm});

    \coordinate (A2) at ($(M1)+\Sc*(A)-\Sc*(M)$);
    \coordinate (B2) at ($(M1)+\Sc*(B)-\Sc*(M)$);
    \coordinate (C2) at ($(M1)+\Sc*(C)-\Sc*(M)$);
    \coordinate (D2) at ($(M1)+\Sc*(D)-\Sc*(M)$);
    \coordinate (M2) at ($0.5*(A2)+0.5*(C2)$);

\begin{scope}
    \clip(M1) circle ({\Sc*\R*0.2 cm});
    \fill [myBlue] (OO) circle ({\R*\Sc cm});
    \draw (OO) circle ({\R*\Sc cm});
    \path[draw,->, thick] (A2) -- (B2) -- (C2) -- (D2) -- cycle;
\end{scope}
\path [draw,dashed,name path =dashCirc] ($(M1)$) circle ({\Sc*\R*0.2 cm});

\coordinate (M0) at ($(M)+{1/\Sc}*(M) - {1/\Sc}*(M1)$);
\coordinate (Mc) at ($0.5*(M0)+0.5*(M1)$);
\coordinate (Mcc) at ($0.5*(M0)+0.5*(M)$);

\tikzmath{coordinate \Q;
\Q = (Mc)-(M1); 
\dist = sqrt((\Qx)^2+(\Qy)^2); 
\Q = (Mcc)-(M0);
\distt = sqrt((\Qx)^2+(\Qy)^2);
}
\path[name path=centCirc] (Mc) circle (\dist pt);
\path [name intersections={of=centCirc and dashCirc}];
  \coordinate (TG1)  at (intersection-1);
  \coordinate (TG1a)  at (intersection-2);

\path[name path=centCirc2] (Mcc) circle (\distt pt);
\path [name intersections={of=dashCirc2 and centCirc2}];
  \coordinate (TG2)  at (intersection-1);
  \coordinate (TG2a)  at (intersection-2);

    \coordinate (tg3) at ($({0.2*\Sc*\R*cos(135)},{0.2*\Sc*\R*sin(135)})+(M2)$);

    \coordinate (tg4) at ($({0.2*\Sc*\R*cos(-90)},{0.2*\Sc*\R*sin(-90)})+(M2)$);


\def\cL{\R*\Sc*0.06}; 
\coordinate (bNode) at ($(M2)+(-2.5*\cL,-\cL)$);
\draw [->,myDarkBlue] (bNode) -- ($(bNode)+(0,\cL*2)$) node[midway, left] {$\boldsymbol{b}$};

\draw[->] ($(D2)-(0,\cL*0.8)$) -- ($(A2)-(0,\cL*0.8)$) node [midway, above] {$\delta \rho$};

\draw[->] ($(A2)+(\cL*0.8,0)$) -- ($(B2)+(\cL*0.8,0)$) node [midway, right] {$\delta z$};

\begin{scope}
    \clip (A2) -- (B2) -- (C2) -- (D2) -- cycle;
    \path [draw,ultra thick,name path=circ3,myGreen] (OO) circle ({\R*\Sc cm});
    \path [name path = diag3](D2)--(B2);
    \path [name intersections={of=circ3 and diag3, by=Arr}];
    \draw[->,very thick,myGreen] (Arr) -- ($(Arr)+(-0.001,0.001)$) node[above,right] {$\delta s$};
\end{scope}

\draw[->,thick] ($0.5*(A2)+0.5*(B2)$) --+ ($(0,-0.001)$);
\draw[->,thick] ($0.5*(B2)+0.5*(C2)$) --+ ($(0.001,0)$);
\draw[->,thick] ($0.5*(C2)+0.5*(D2)$) --+ ($(0,0.001)$);
\draw[->,thick] ($0.5*(D2)+0.5*(A2)$) --+ ($(-0.001,0)$);

\end{scope}

\end{scope}
\end{tikzpicture}

    \caption{A small piece of the vortex sheet and enclosing curve}
    \label{fig_circ_b}
\end{figure}

To walk-around the problem of singularities in the vorticity equation (\ref{vorticity}), we will take a step back to equation (\ref{curlOfNS}). Let us recall the definition of a curl operator.
\begin{equation}
    \Gamma^u = \oint_C \bo{u} \cdot d\bo{r}
\end{equation}
we will call the circulation of field $\bo{u}$ around curve $C$. Then the $i$th component of the curl is defined as:
\begin{equation}
   (\nabla\times\bo{u})_i  = \lim_{A\to 0}\frac{\Gamma^u}{A}
\end{equation}
where $C$ is assumed to be in a plane normal to $i$ direction (given by unit vector $\hat{e}_i$) and $A$ is an area bounded by $C$. Therefore (using Einstein's summation convention):
\begin{equation}
   \nabla\times\bo{u}  = \lim_{A_i\to 0}\frac{\Gamma^u_i}{A_i}\hat{e_i}
\end{equation}
Now, we will apply identity (\ref{curlOfProd}) to equation (\ref{curlOfNS}), as before, but this time using the definition of the curl.
\begin{equation}
    \frac{d}{dt}\Big( \frac{\Gamma^u_i}{A_i}\bo{e}_i \Big)  = \Big(\frac{\Gamma^u_i}{A_i}\hat{e}_i\;\cdot\nabla\Big)\bo{u} + \frac{\Gamma^b_i}{A_i}
\end{equation}
Expanding the vortex stretching term in cylindrical coordinates (keeping the axial symmetry in mind) and the time derivative with a product rule:
\begin{equation}
    \frac{1}{A_\phi}\frac{d\Gamma^u_\phi}{dt} + \Gamma^u_\phi\Big(-\frac{1}{A_\phi^2}\frac{dA_\phi}{dt} \Big)  = \frac{\Gamma^u_\phi}{A_\phi}\frac{u_\rho}{\rho} + \frac{\Gamma^b_\phi}{A_\phi}
    \label{halfg}
\end{equation}
The total time derivative of $A_\phi$ can be deduced from the incompressibility:
\begin{equation}
    \frac{dV}{dt} = \frac{d(A_\phi \rho \delta\phi)}{dt}= 0
\end{equation}
\begin{equation}
    \frac{dA_\phi}{dt}\rho\delta\phi + A_\phi \frac{d\rho}{dt}\delta\phi = 0
\end{equation}
so, by the definition of $u_\rho$:
\begin{equation}
    \frac{dA_\phi}{dt} = -A_\phi \frac{u_\rho}{\rho} 
\end{equation}
now substituting it to (\ref{halfg}):
\begin{equation}
    \frac{1}{A_\phi}\frac{d\Gamma^u_\phi}{dt} + \Gamma^u_\phi\Big(\frac{1}{A_\phi}\frac{u_\rho}{\rho} \Big)  = \frac{\Gamma^u_\phi}{A_\phi}\frac{u_\rho}{\rho} + \frac{\Gamma^b_\phi}{A_\phi}
\end{equation}
we see that this term cancels out with the vortex stretching. What is even more important, we can get rid of $A_\phi$ and remove the singularity (we will omit index $\phi$):
\begin{equation}
    \frac{d\Gamma^u}{dt}  =\Gamma^b
    \label{Gammaubeq}
\end{equation}
The last challenge left is to compute these two circulations. We will refer to fig. (\ref{fig_circ_b}). To compute the $\Gamma^b$ we will enclose a small piece of the interface (that is given by $\delta s$) in an oriented rectangle. Note that the orientation is clockwise due to the direction of $\phi$. One can easily notice, that only the left side will contribute to circulation.
we can express its length as:
\begin{equation}
    \delta z = \frac{\partial z}{\partial s} \delta s
\end{equation}
then:
\begin{equation}
    \Gamma^b = b\,\delta z = b\frac{\partial z}{\partial s} \delta s
    \label{Gammab}
\end{equation}
Now, it would be convenient to express $\Gamma^u$ also in terms of $\delta s$. We will do this by introducing circulation density per parameter length $\gamma$:
\begin{equation}
    \Gamma^u = \gamma \delta s
    \label{Gammau}
\end{equation}
It characterizes a uniform circulation around a ring of thickness $\delta s$.
Substituting (\ref{Gammab}) and (\ref{Gammau}) to (\ref{Gammaubeq}), taking an advantage of the fact that $\delta s$ by design does not depend on time and canceling it out, we obtain:
\begin{equation}
    \frac{d\gamma}{dt}  = b\frac{\partial z}{\partial s}
    \label{gammaEq}
\end{equation}
The corresponding Biot-Savart formula can be obtained by substituting:
\begin{equation}
    \bo{\omega}dV = \hat{\phi}\Gamma^u dl = \hat{\phi}\gamma ds \rho d\phi
\end{equation}
($dl$ is an infinitesimal length in the azimuthal direction) \\
to (\ref{BS1}), what gives:
\begin{equation}
    \boldsymbol{u}(\boldsymbol{r_0}) = \frac{1}{4\pi}\int_0^{2\pi}\int_{-\pi/2}^{\pi/2} \frac{ \gamma\hat{\phi}\times (\boldsymbol{r_0}-\boldsymbol{r})}{ |\boldsymbol{r_0}-\boldsymbol{r}|^3} \rho ds\, d\phi
    \label{BS2}
\end{equation}

    \subsection{Evolution of chosen integral quantities}
\label{integralsEvolution}
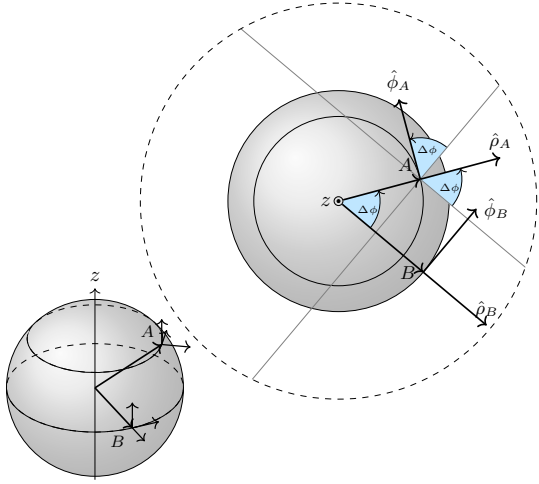
\begin{figure}[h]
    \centering
    \definecolor{myBlue}{rgb}{0.75,0.9,1}
\definecolor{myGreen}{rgb}{0.1,0.7,0.3}
\definecolor{myGray}{rgb}{0.8711,0.8711,0.8711}
\definecolor{myDarkBlue}{rgb}{0.3,    0.650,    0.8500}
\definecolor{myDarkBlue2}{rgb}{0.2,    0.450,    0.6500}
\pgfdeclarelayer{ft2} 
\pgfdeclarelayer{ft}
\pgfdeclarelayer{bg} 
\pgfsetlayers{behind,main,ft,above} 

\newcommand{\vaOne}{60} 
\newcommand{\vaTwo}{115}

\tdplotsetmaincoords{\vaOne}{\vaTwo}

\tdplotsetmaincoords{\vaOne}{\vaTwo}
\pgfplotsset{compat=newest}

\begin{tikzpicture}[tdplot_main_coords]

\def\R{1.5}
\def\Sc{8}
\coordinate (O) at (0,0,0);  

\shade[ball color =lightgray , opacity = 0.4] (O) circle (\R cm);
\path[draw, name path=contour] (0 cm,0 cm) circle (\R cm);

\begin{scope}[on above layer]

\draw[->] (0,0,{-\R*1.2}) -- (0,0,{\R*1.3}) node[above] {$z$};
\def\thA{50};
\def\phA{105};
\coordinate (A) at ({\R*sin(\thA)*cos(\phA)},{\R*sin(\thA)*sin(\phA)},{\R*cos(\thA)});

\def\thB{90};
\def\phB{50};
\coordinate (B) at ({\R*sin(\thB)*cos(\phB)},{\R*sin(\thB)*sin(\phB)},{\R*cos(\thB)});

\draw [->,thick] (O) -- (A) node[above left] {\footnotesize $A$};
\draw [->,thick] (O) -- (B) node[below left] {\footnotesize $B$};

\path[draw, dashed, ultra thin, name path=Acirc] (0,0,{\R*cos(\thA)}) circle ({\R*sin(\thA)});
\path[draw, dashed, ultra thin, name path=Bcirc] (0,0,{\R*cos(\thB)}) circle ({\R*sin(\thB)});

\path [name intersections={of=contour and Acirc}];
  \coordinate (AC1)  at (intersection-1);
   \coordinate (AC2)  at (intersection-4);

\path [name intersections={of=contour and Bcirc}];
  \coordinate (BC1)  at (intersection-1);
  \coordinate (BC2)  at (intersection-2);

\begin{scope}
    \clip (AC1) -- (AC2) -- ($1.25*(BC2)$) -- ($1.25*(BC1)$) -- cycle;
    \path[draw, ultra thin] (0,0,{\R*cos(\thA)}) circle ({\R*sin(\thA)});
\end{scope}

\begin{scope}
    \clip (BC1) -- (BC2) --+ ($(0,0,-\R)$) -- ($(BC1)+(0,0,-\R)$) -- cycle;
    \path[draw, ultra thin] (0,0,{\R*cos(\thB)}) circle ({\R*sin(\thB)});
\end{scope}

\def\uvs{0.5};
\coordinate (Ar) at ($(A)+({\uvs*cos(\phA)},{\uvs*sin(\phA)},0)$);
\coordinate (Aph) at ($(A)+({-\uvs*sin(\phA)},{\uvs*cos(\phA)},0)$);
\coordinate (Az) at ($(A)+(0,0,\uvs)$);
\draw [->,semithick] (A) -- (Ar);
\draw [->,semithick] (A) -- (Aph);
\draw [->,semithick] (A) -- (Az);

\coordinate (Br) at ($(B)+({\uvs*cos(\phB)},{\uvs*sin(\phB)},0)$);
\coordinate (Bph) at ($(B)+({-\uvs*sin(\phB)},{\uvs*cos(\phB)},0)$);
\coordinate (Bz) at ($(B)+(0,0,\uvs)$);
\draw [->,semithick] (B) -- (Br);
\draw [->,semithick] (B) -- (Bph);
\draw [->,semithick] (B) -- (Bz);
\begin{scope}[rotate around z=25, rotate around y=-30,canvas is yz plane at x=0]

\coordinate (M) at ($0.5*({\R*cos(45)},{\R*sin(45)})+0.5*({\R*cos(30)},{\R*sin(30)})$);
\coordinate (OO) at ($-4.5*(M)$); 
\coordinate (M1) at ($\Sc*(M)+(OO)$);
\path [name path=dashCirc2] (M) circle ({\R*0.2 cm});


\begin{scope}
    \def\scu{1.25}; 
    \def\scuv{\scu*0.75}; 
    \clip(M1) circle ({\Sc*\R*0.28 cm});
    
    \shade[ball color =lightgray , opacity = 0.4] (M1) circle ({\scu*\R*sin(\thB)});
    
    \draw(M1) circle ({\scu*\R*sin(\thB)});
    \draw(M1) circle ({\scu*\R*sin(\thA)});
    \coordinate (a) at ($(M1)+({\scu*\R*sin(\thA)*cos(\phA-90)},{\scu*\R*sin(\thA)*sin(\phA-90)})$);
    \coordinate (b) at ($(M1)+({\scu*\R*sin(\thB)*cos(\phB-90)},{\scu*\R*sin(\thB)*sin(\phB-90)})$);

    \coordinate (ar) at ($({\scuv*\R*cos(\phA-90)},{\scuv*\R*sin(\phA-90)})$);
    \coordinate (aph) at ($({-\scuv*\R*sin(\phA-90)},{\scuv*\R*cos(\phA-90)})$);

    \coordinate (br) at ($({\scuv*\R*cos(\phB-90)},{\scuv*\R*sin(\phB-90)})$);
    \coordinate (bph) at ($({-\scuv*\R*sin(\phB-90)},{\scuv*\R*cos(\phB-90)})$);
    
 \draw[fill,myBlue] (a)--($(a)+0.5*(br)$) arc (\phB-90:\phA-90:\scuv*0.5*\R)--cycle;
 \draw[->] ($(a)+0.5*(br)$) arc (\phB-90:\phA-90:\scuv*0.5*\R) node[midway] {\tiny $\Delta \phi\quad\;\;$};

 \draw[fill,myBlue] (a)--($(a)+0.5*(bph)$) arc (\phB:\phA:\scuv*0.5*\R)--cycle;
 \draw[->] ($(a)+0.5*(bph)$) arc (\phB:\phA:\scuv*0.5*\R) node[below, midway] {\tiny $\Delta \phi\;$};

 \draw[fill,myBlue] (M1)--($(M1)+0.5*(br)$) arc (\phB-90:\phA-90:\scuv*0.5*\R)--cycle;
 \draw[->] ($(M1)+0.5*(br)$) arc (\phB-90:\phA-90:\scuv*0.5*\R) node[midway] {\tiny $\Delta \phi\quad\;\;$};

    \draw[ultra thin, gray] ($(a)-10*(br)$) -- ($(a)+10*(br)$);
    \draw[ultra thin, gray] ($(a)-10*(bph)$) -- ($(a)+10*(bph)$);

    \draw[->,thick] (b) --+ (br) node[above] {\small $\hat{\rho}_B$};
    \draw[->,thick] (b) --+ (bph)node[right] {\small $\hat{\phi}_B$};
    \draw[->,thick] (a) --+ (ar) node[above] {\small $\hat{\rho}_A$};
    \draw[->,thick] (a) --+ (aph) node[above] {\small$\hat{\phi}_A$};
    \draw[->,thick] (M1) -- (a) node[above left] {\small $A$};
    \draw[->,thick] (M1) -- (b) node[left] {\small $B$};

\draw[fill, myGray] (M1) circle (0.075cm);
\draw (M1) circle (0.075cm);
\draw[fill] (M1) circle (0.025cm) node[left] {$z$};   
 
\end{scope}

\path [draw,dashed,name path =dashCirc] ($(M1)$) circle ({\Sc*\R*0.28 cm});

\end{scope}

\end{scope}
\end{tikzpicture}
    \caption{Unit vectors of points $A$ and $B$ in axonometric view and seen from above}
    \label{unitVecs}
\end{figure}

We can deduce the evolution of certain integral quantities characterizing the region of non-zero buoyancy (we will call it interior). This can be used to monitor the simulation quality and possibly improve the time-stepping scheme. Volume integrals are rather inconvenient due to the difficulties of dynamic volume discretization. For that reason, we will transform proper expressions to surface integrals using Gauss and Stokes theorems. We will start by recalling a few useful formulas in cylindrical coordinates. \\ \\
The derivatives of unit vectors and their cross products:
\begin{subequations}
    \begin{equation}
        \frac{d \hat{\rho}}{d \phi} = \hat{\phi}
    \end{equation}
    \begin{equation}
        \frac{d \hat{\phi}}{d \phi} = -\hat{\rho}
    \end{equation}
    \begin{equation}
        \hat{z}\times\hat{\rho}=\hat{\phi}
    \end{equation}
    \begin{equation}
        \hat{\phi}\times\hat{z}=\hat{\rho}
    \end{equation}
    \begin{equation}
        \hat{\rho}\times\hat{\phi} = \hat{z}
        \label{rhoCrossPhi}
    \end{equation}
    \label{crosses}
\end{subequations}
It would also be useful to decompose unit vectors given
at point $A$ to unit vectors from point $B$ (fig. \ref{unitVecs}).
\begin{subequations}
    \begin{equation}
        \hat{\rho}_A = \hat{\rho}_B \cos(\Delta\phi) + \hat{\phi}_B \sin(\Delta\phi) 
        \label{rhoDec}
    \end{equation}
    \begin{equation}
        \hat{\phi}_A = \hat{\rho}_B \sin(\Delta\phi) + \hat{\phi}_B \cos(\Delta\phi)
        \label{phiDec}
    \end{equation}
\end{subequations}
where $\Delta \phi = \phi_A - \phi_B$.\\
Divergence of some vector field $\bo{q}$ in axisymmetric case is:
\begin{equation}
    \nabla\cdot\bo{q} = \frac{1}{\rho}\frac{\partial(\rho \,q_\rho)}{\partial\rho} + \frac{\partial q_z}{\partial z}
\end{equation}
Let the parametrized, axisymmetric surface be given by:
\begin{equation}
    \bo{r}(s) = \rho(s)\hat{\rho}(\phi) + z(s)\hat{z}
\end{equation}
then its infinitesimal piece is given by:
\begin{equation}
    d\bo{S} = \frac{\partial \bo{r}}{\partial\phi}d\phi  \times \frac{\partial \bo{r}}{\partial s}ds =
    (\rho\hat{\phi}d\phi) \times \Big(\frac{\partial \rho}{\partial s}\hat{\rho} + \frac{\partial z}{\partial s}\hat{z}\Big)ds =
    \Big( \frac{\partial z}{\partial s}\hat{\rho}-\frac{\partial \rho}{\partial s}\hat{z}\Big)\rho d\phi ds
\end{equation}
In some cases, integration over the azimuth is trivial and it will be convenient to introduce also:
\begin{subequations}
    \begin{equation}
        d\bo{n} =  \Big( \frac{\partial z}{\partial s}\hat{\rho}-\frac{\partial \rho}{\partial s}\hat{z}\Big) ds
        \label{dNormal}
    \end{equation}
    \begin{equation}
        d\bo{\tau} =  \Big( \frac{\partial \rho}{\partial s}\hat{\rho}+\frac{\partial z}{\partial s}\hat{z}\Big) ds
        \label{dTangent}
    \end{equation}
\end{subequations}
understood, respectively, as vector normal and vector tangent to the $\phi=\tr{const}$ section of the surface.
\subsubsection{Mass}
We assumed constant density, so mass conservation in the interior is expressed by the constancy of the enclosed volume:
\begin{equation}
    \frac{dV}{dt} = \frac{d}{dt}\int_V 1 dV = 0
\end{equation}
we can interpret the integrand as a divergence of some vector field of unit divergence, then use the Gauss theorem. In the end, we also perform trivial integration over azimuth. Below are two exemplary choices:
\begin{subequations}
    \begin{equation}
        V = \int_V \nabla\cdot \Big(\frac{\rho}{2}\hat{\rho}\Big) dV = \oiint \frac{\rho}{2}\hat{\rho} \cdot d\bo{S} = \pi\int_{-\pi/2}^{\pi/2} \rho^2\frac{\partial z}{\partial s} ds
        \label{mass1}
    \end{equation}
    \begin{equation}
         V = \int_V \nabla\cdot z\hat{z}dV = \oiint z\hat{z} \cdot d\bo{S} = -2\pi\int_{-\pi/2}^{\pi/2} z\rho\frac{\partial \rho}{\partial s} ds
    \end{equation}
\end{subequations}
\subsubsection{Total circulation in meridional plane}
After trivial integration of (\ref{gammaEq}), we obtain the increase of the total circulation
\begin{equation}
    \frac{d}{dt} \int_{-\pi/2}^{\pi/2} \gamma ds = b(z_2-z_1)
    \label{totalCircEvolution}
\end{equation}
where $z_1$ and $z_2$ denote the coordinates at s equal $-\pi/2$ and $\pi/2$ respectively and as a difference - the thickness of buoyant region at the $z$ axis.
\subsubsection{Moment of vorticity}
The first moment is a conserved quantity for cases with no buoyancy. Their further importance is highlighted in section \ref{optimization}.
\begin{equation}
   \frac{d}{dt}\int \bo{r}\times\bo{\omega} dV =
   \frac{d}{dt}\int (\rho\hat{\rho}+z\hat{z})\times\bo{\omega} dV
      \label{zMoment}
\end{equation}
\begin{equation*}
  = \frac{d}{dt}\int_{-\pi/2}^{\pi/2}\int_{0}^{2\pi} (\rho\hat{\rho}+z\hat{z})\times\hat{\phi}\gamma \rho d\phi ds
\end{equation*}
\begin{equation*}
  = \frac{d}{dt}\int_{-\pi/2}^{\pi/2}\int_{0}^{2\pi} (\rho\hat{z}-z\hat{\rho})\gamma \rho d\phi ds
\end{equation*}
the $\hat{\rho}$ component will be zero due to axial symmetry, so we will focus on the other one. By the product rule, we get:
\begin{equation*}
    = \hat{z}\int_{-\pi/2}^{\pi/2} \int_{-\pi/2}^{\pi/2} \Bigg( 2\rho\frac{d\rho}{dt}\gamma + \frac{d\gamma}{dt} \rho^2\Bigg) ds d\phi
\end{equation*}
where $d\rho/dt$ is just $v_\rho$ determined by formula (\ref{BS2}). This means that the first term on the right-hand side is built of all the interactions between pairs of points, which sums up to zero by the following argument:\\
Consider two points belonging to the vortex sheet - $A$ and $B$. Contribution from the interaction where $A$ is the probing point and $B$ is the source is:
\begin{equation}
         i_{AB} = \Bigg(\frac{\hat{\rho}_A}{2\pi}\cdot \frac{ \gamma_B\hat{\phi}_B\times (\boldsymbol{r_A}-\boldsymbol{r_B})}{ |\boldsymbol{r_A}-\boldsymbol{r_B}|^3} \rho_B ds_B\, d\phi_B \Bigg) \gamma_A \rho_A ds_A ds_A
\end{equation}
It is enough to consider only the $\hat{z}$ component of the difference in the numerator because the rest will give $\hat{z}$ after the cross product. Using the formula (\ref{rhoDec}) we can take the dot product and obtain:
\begin{equation}
         i_{AB} = \Bigg(\gamma_B \frac{ (z_A-z_B)\cos(\phi_A-\phi_B)}{ 2\pi|\boldsymbol{r_A}-\boldsymbol{r_B}|^3} \rho_B ds_B\, d\phi_B \Bigg) \gamma_A \rho_A ds_A ds_A
\end{equation}
We see that, if we swap $A$ and $B$, the formula only changes sign, so $i_{AB} + i_{BA} = 0$
Going back to equation (\ref{zMoment}), substituting from (\ref{gammaEq}) and computing the trivial integral over azimuth:
\begin{equation}
   \frac{d}{dt}\int \bo{r}\times\bo{\omega} dV = 
   2\pi \hat{z} \int_{-\pi/2}^{\pi/2} b\frac{\partial z}{\partial s} \rho^2 ds
\end{equation}
recalling formula (\ref{mass1}):
\begin{equation}
   \frac{d}{dt}\int \bo{r}\times\bo{\omega} dV = 
   2bV\hat{z}
\end{equation}
\subsubsection{Generalized momenta}
To represent the volume integral of velocity as a divergence, we will use the Stokes stream function:
\begin{subequations}
    \begin{equation}
        u_\rho = -\frac{1}{\rho}\frac{\partial \psi}{\partial z}
    \end{equation}
    \begin{equation}
        u_z = \frac{1}{\rho}\frac{\partial\psi}{\partial\rho}
    \end{equation}
\end{subequations}
Starting with the momentum conjugated with the radial coordinate, and proceeding with Gauss divergence theorem:
\begin{equation}
    \int u_\rho dV = - \int \frac{1}{\rho}\frac{\partial\psi}{\partial z} dV = -\int \nabla\cdot\Big( \frac{\psi}{\rho}\hat{z}\Big) dV = -\int \Big( \frac{\psi}{\rho}\hat{z}\Big) \cdot d\bo{S}
\end{equation}
this can be expanded, integrated in $\phi$ with ease and integrated by parts in $s$:
\begin{equation}
    = 2\pi\int_{-\pi/2}^{\pi/2} \psi\frac{\partial \rho}{\partial s} ds  = 2\pi (\psi\rho) \Big|_{-\pi/2}^{\pi/2} - 2\pi\int_{-\pi/2}^{\pi/2}\rho\frac{\partial \psi}{\partial s}ds
\end{equation}
the first term is zero by the boundary conditions (\ref{bcSurf}) for $\rho$ and the second can be further expanded with the chain rule:
\begin{equation}
    = -2\pi\int_{-\pi/2}^{\pi/2}\rho\Big(\frac{\partial\psi}{\partial\rho}\frac{\partial\rho}{\partial s} + \frac{\partial\psi}{\partial z}\frac{\partial z}{\partial s} \Big) ds
    = 2\pi\int_{-\pi/2}^{\pi/2}\rho^2 \Big(-u_z\frac{\partial\rho}{\partial s} + u_\rho\frac{\partial z}{\partial s} \Big) ds
\end{equation}
so we can conclude:
\begin{equation}
    \int u_\rho dV = 2\pi \int \rho^2 \bo{u}\cdot d\bo{n} = \int \rho\bo{u} \cdot d\bo{S}
\end{equation}
Proceeding analogically with $z$ component:
\begin{equation}
    \int u_z dV = \int \frac{1}{\rho} \frac{\partial}{\partial \rho}\Big( \frac{\psi}{\rho}\rho\Big) dV = \int \Big(\frac{\psi}{\rho}\hat{\rho}\Big) \cdot d\bo{S}
\end{equation}
expanding the dot product, integrating in $\phi$, then integrating by parts in $s$:
\begin{equation}
    = 2\pi \int_{-\pi/2}^{\pi/2} \psi \frac{\partial z}{\partial s}ds = 2\pi (\psi z )\Big|_{-\pi/2}^{\pi/2}-2\pi\int_{-\pi/2}^{\pi/2} z \frac{\partial\psi }{\partial s} ds
\end{equation}
the boundary term is zero because both boundaries of the curve lay on the axial streamline $\psi=0$.
\begin{equation}
    =-2\pi \int_{-\pi/2}^{\pi/2} z\Big(\frac{\partial \psi}{\partial \rho}\frac{\partial \rho}{\partial s} + \frac{\partial \psi}{\partial z}\frac{\partial z}{\partial s} \Big)ds = 2\pi \int_{-\pi/2}^{\pi/2} z (-u_z \frac{\partial\rho}{\partial s} + u_\rho \frac{\partial z}{\partial s})\rho ds
\end{equation}
and we can conclude:
\begin{equation}
    \int u_z dV = 2\pi \int z\rho \bo{u}\cdot d\bo{n} = \int z\bo{u} \cdot d\bo{S}
\end{equation}

    \subsection{Axial symmetry of induced velocity}
       
\label{sectionAxialSym}
Using the Biot-Savart formula (\ref{BS2}), we will now take an advantage of the axial symmetry and integrate the induced velocity over the azimuth. Starting by expanding the numerator:
\begin{equation}
    \gamma\hat{\phi}\times(\bo{r_0}-\bo{r}) 
    = \gamma\hat{\phi}\times(\rho_0\hat{\rho}_0+z_0\hat{z} - \rho\hat{\rho}-z\hat{z} )
\end{equation}
and substituting for $\hat{\rho}$ with the help of formula (\ref{rhoDec}):
\begin{equation}
    = \gamma\hat{\phi}\times\Big(
    \hat{\rho}(\rho_0\cos(\Delta\phi) -\rho) + \hat{\phi}\rho_0\sin(\Delta\phi)
    +\hat{z}(z_0-z) \Big)
\end{equation}
where $\Delta\phi=\phi_0-\phi$ \\
Now, taking the cross product according to formulas (\ref{crosses}):
\begin{equation}
    = \gamma\Big(
    \hat{z}(\rho-\rho_0\cos(\Delta\phi)) 
    +\hat{\rho}(z_0-z) \Big)
\end{equation}
and using (\ref{rhoDec}) again; but this time, keeping in mind that $\hat{\phi}_0$ component has to integrate to zero by axial symmetry. Therefore, we will just skip this term and write:
\begin{equation}
   \gamma\hat{\phi}\times(\bo{r_0}-\bo{r})  = \gamma\Big(
    \hat{z}(\rho-\rho_0\cos(\Delta\phi)) 
    +\hat{\rho}_0\cos(\Delta\phi)(z_0-z) \Big)
    \label{numerator}
\end{equation}
Proceeding to the denominator of (\ref{BS2}), we expand it and take the dot product. Using (\ref{rhoDec}) we can notice that $\hat{\rho}_0\cdot\hat{\rho} = \cos(\Delta\phi)$.
\begin{equation}
    |\bo{r_0}-\bo{r}|^3 =( \hat{\rho}_0\rho_0 + \hat{z}z_0 - \hat{\rho}\rho - \hat{z}z)^3
\end{equation}
\begin{equation}
    =\Big( \rho_0^2 + \rho^2 -2\rho\rho_0 \cos(\Delta\phi) + (z_0-z)^2\Big)^{3/2}
    \label{denominator}
\end{equation}
Now we can plug both results (\ref{numerator}) and (\ref{denominator}) to (\ref{BS2}) obtaining:
\begin{equation}
    \bo{u}(\bo{r_0}) = \frac{1}{4\pi}\int_{-\pi/2}^{\pi/2}\int_0^{2\pi}
    \gamma\frac{ 
    \hat{z}(\rho-\rho_0\cos(\Delta\phi)) 
    +\hat{\rho}_0\cos(\Delta\phi)(z_0-z) }{\Big( \rho_0^2 + \rho^2 -2\rho\rho_0 \cos(\Delta\phi) + (z_0-z)^2\Big)^{3/2}}\rho d\phi ds
    \label{BS3}
\end{equation}
Proceeding in an analogical way as in \cite{lamb} (art. 161), we introduce:
\begin{subequations}
    \begin{equation}
        {R_1}^2 = (\rho_0-\rho)^2+(z_0-z)^2
    \end{equation}
    \begin{equation}
    {R_2}^2 = (\rho_0+\rho)^2+(z_0-z)^2
    \end{equation}
    \label{R1R2}
\end{subequations}
which can be interpreted as the least ($R_1$) and the greatest ($R_2$) distances from the circular cross-section ($z$ const) to the point $\bo{r_0}$. If we now add equations (\ref{R1R2}) and divide by 2, we obtain:
\begin{equation}
    \frac{{R_1}^2+{R_2}^2}{2} = \rho_0^2 + \rho^2 +(z_0-z)^2    
\end{equation}
by analogous subtraction:
\begin{equation}
    \frac{{R_2}^2-{R_1}^2}{2} = 2\rho_0\rho  
\end{equation}
We can now use these formulas to transform the denominator of (\ref{BS3}). For convenience we will skip the root, considering just the dot product:
\begin{equation}
    |\bo{r_0}-\bo{r}|^2 = \frac{1}{2}\Big(
    {R_1}^2+{R_2}^2- ({R_2}^2-{R_1}^2)\cos(\Delta\phi)
    \Big)
\end{equation}
Now, using the double-angle formula:
\begin{equation*}
     = \frac{1}{2}\Big(
    {R_1}^2+{R_2}^2- ({R_2}^2-{R_1}^2)\big(\,\cos^2(\tfrac{\Delta\phi}{2})-\sin^2(\tfrac{\Delta\phi}{2})\,\big)
    \Big)
\end{equation*}
some further algebra:
\begin{equation}
    ={R_1}^2\,\cos^2(\tfrac{\Delta\phi}{2})+ {R_2}^2\,\sin^2(\tfrac{\Delta\phi}{2})
\end{equation}
\begin{equation}
    ={R_2}^2\,\bigg(\,\big(\frac{R_1}{R_2}\big)^2\cos^2(\tfrac{\Delta\phi}{2})+ \sin^2(\tfrac{\Delta\phi}{2})\,\bigg)
\end{equation}
trigonometric identity again:
\begin{equation}
    ={R_2}^2\,\bigg(1-\,k^2\,\cos^2(\tfrac{\Delta\phi}{2})\,\bigg)
\end{equation}
where
\begin{equation}
    k^2 = 1-\bigg(\frac{R_1}{R_2}\bigg)^2
    \label{k}
\end{equation}
Let us also substitute:
\begin{equation}
    \theta = \tfrac{\pi}{2}-\tfrac{\Delta\phi}{2}
\end{equation}
Then, the denominator will take the form:
\begin{equation}
    |\bo{r_0}-\bo{r}|^2 = {R_2}^2\bigg(1-\,k^2\,\sin^2\theta\,\bigg)
    \label{denom}
\end{equation}
and for the numerator of (\ref{BS3}) we use double-angle formula:
\begin{equation}
    \cos(\Delta\phi) = \cos^2(\tfrac{\Delta\phi}{2}) - \sin^2(\tfrac{\Delta\phi}{2})
    \label{phi2theta}
\end{equation}
\begin{equation*}
    = 1 - 2\sin^2(\tfrac{\pi}{2}-\theta)
    =1 - 2\cos^2\theta
\end{equation*}
to obtain:
\begin{equation}
    \hat{\phi}\times(\bo{r_0}-\bo{r}) = \hat{z}\Big(\rho-\rho_0 +2\rho_0 \cos^2\theta)\Big) 
    +\hat{\rho}_0(1 - 2\cos^2\theta)(z_0-z)
    \label{num}
\end{equation}
We can now represent the radial component of (\ref{BS3}) as follows:
\begin{equation*}
    u_\rho(s_0)=\frac{1}{4\pi}\int_{-\pi/2}^{\pi/2}\;g\,\rho\,(z_0-z)\int_{\pi/2}^{3\pi/2}  
    \frac{(1-2\cos^2\theta)\,(2\,d\theta)}{
    {R_2}^3\bigg(1-\,k^2\,\sin^2\theta\,\bigg)^{3/2}} \, ds
\end{equation*}
\begin{equation}
    =\frac{1}{4\pi}\int_{-\pi/2}^{\pi/2}\;\frac{g\,\rho\,(z_0-z)}{{R_2}^3}\int_{0}^{\pi}  
    \frac{2\,(1-2\cos^2\theta)\,d\theta}{
    \bigg(1-\,k^2\,\sin^2\theta\,\bigg)^{3/2}} \, ds
\end{equation}
For convenience, we will split integral over $\theta$ into two parts, defining the following:
\begin{equation}
    I_1 = \int_0^{\pi} \frac{2\,d\theta}{\bigg( 1-k^2\,\sin^2\theta\bigg)^{3/2}} 
\end{equation}
\begin{equation}
    I_2 = \int_0^{\pi} \frac{4\cos^2\theta\,d\theta}{\bigg( 1-k^2\,\sin^2\theta\bigg)^{3/2}} 
\end{equation}
and then:
\begin{equation}
u_\rho(s_0)=\frac{1}{4\pi}\int_{-\pi/2}^{\pi/2}\frac{\gamma\,r\,(z_0-z)}{{R_2}^3}(I_1 - I_2) \, ds
\label{ur}
\end{equation}
proceeding in a similar fashion with the vertical component of (\ref{BS3}), we obtain:
\begin{equation}
    u_z(s_0)=\frac{1}{4\pi}\int_{-\pi/2}^{\pi/2}\;\frac{\gamma\,\rho\,\rho_0}{{R_2}^3}\bigg(\frac{\rho-\rho_0}{\rho_0}I_1+I_2\bigg) \, ds
    \label{uz}
\end{equation}
Integrals $I_1$ and $I_2$ can be computed analytically resulting in:
\begin{equation}
    I_1 = \frac{4}{1-k^2}E(k^2)
\end{equation}
\begin{equation}
    I_2 = \frac{8}{k^2}\bigg( K(k^2) - E(k^2)  \bigg)
\end{equation}
where $K()$ and $E()$ are the complete elliptic integrals of the first and the second kind respectively. An equivalent set of equations was derived in \cite{krasnyRing}.

    \subsection{Induced velocity is finite}
        
Formulas (\ref{ur}) and (\ref{uz}) have a singularity, when $s\to s_0$ i.e. when we approach the case of autoinduction of velocity. In this case $k \to 1$, what gives a singularity in $I_2$ due to $K$ and in $I_1$ due to $(1-k^2)^{-1}$.
Nevertheless, the resulting velocity intuitively should be finite. To show that, we will consider a small piece of the vortex sheet. Without losing generality, we can assume that it has $s_0 = 0$. Because $\gamma$ is always zero at the $z$ axis, we will also assume that $\rho>0$.
We will start by expanding $\rho(s)$ and $z(s)$ in a Taylor series around $s_0$ and substituting this into $k^2$:
\begin{equation}
    k^2 = 1 - \frac{\big(-\rho_0'+\bigO( s)\big)^2 + \big(-z_0'+\bigO(s)\big)^2}{\big(2\rho_0 + \rho_0' s + \bigO(s ^2)\big)^2 + \big(-z_0' s + \bigO( s^2)\big)^2}  s^2
\end{equation}
thus:
\begin{equation}
    k \approx 1 + \bigO(s^2)
    \label{bigO_k}
\end{equation}
Then we can use an asymptotic expansion from \cite{K_expansion} for $K()$ when $k\to1^-$:
\begin{equation}
    K(k) \approx \ln\Big(\frac{4}{\sqrt{1-k^2}}\Big)
\end{equation}
\begin{equation}
    K(k) \approx \ln\Big(\frac{4}{|s|}\Big)
\end{equation}
and this integrates to a finite value:
\begin{equation}
   \int _0^{\Delta s}K(k) ds \approx \int_0^{\Delta s} \ln\Big(\frac{4}{|s|}\Big) ds = \Delta s (1 - \ln\Big({\frac{\Delta s}{4}}\Big))
\end{equation}
All terms including $K()$, by partial integration, give also a finite value. Proceeding to the second source of the problem: we can note that $I_1$ is always multiplied by $(\rho_0-\rho)$ or $(z_0-z)$ which are both $\bigO(s)$. Using (\ref{bigO_k}), we can write:
\begin{equation}
    I_1 = \frac{4}{\bigO(s^2)} E(k^2)
\end{equation}
so for example:
\begin{equation}
    I_1(z_0-z) \approx \frac{-4z_0'}{s} E(k^2)
\end{equation}
and this integrates (in a sense of Cauchy principal value) to a finite quantity. Therefore, the overall induced velocity stays finite.

    \subsection{Velocity smoothing}
        
\label{section_delta}
\begin{figure}[H]
        \centering
        \begin{subfigure}[b]{0.49\textwidth}
            \centering
            \includegraphics[trim={3.5cm 9.0cm 4cm 9.5cm},clip,width=1.0\textwidth]{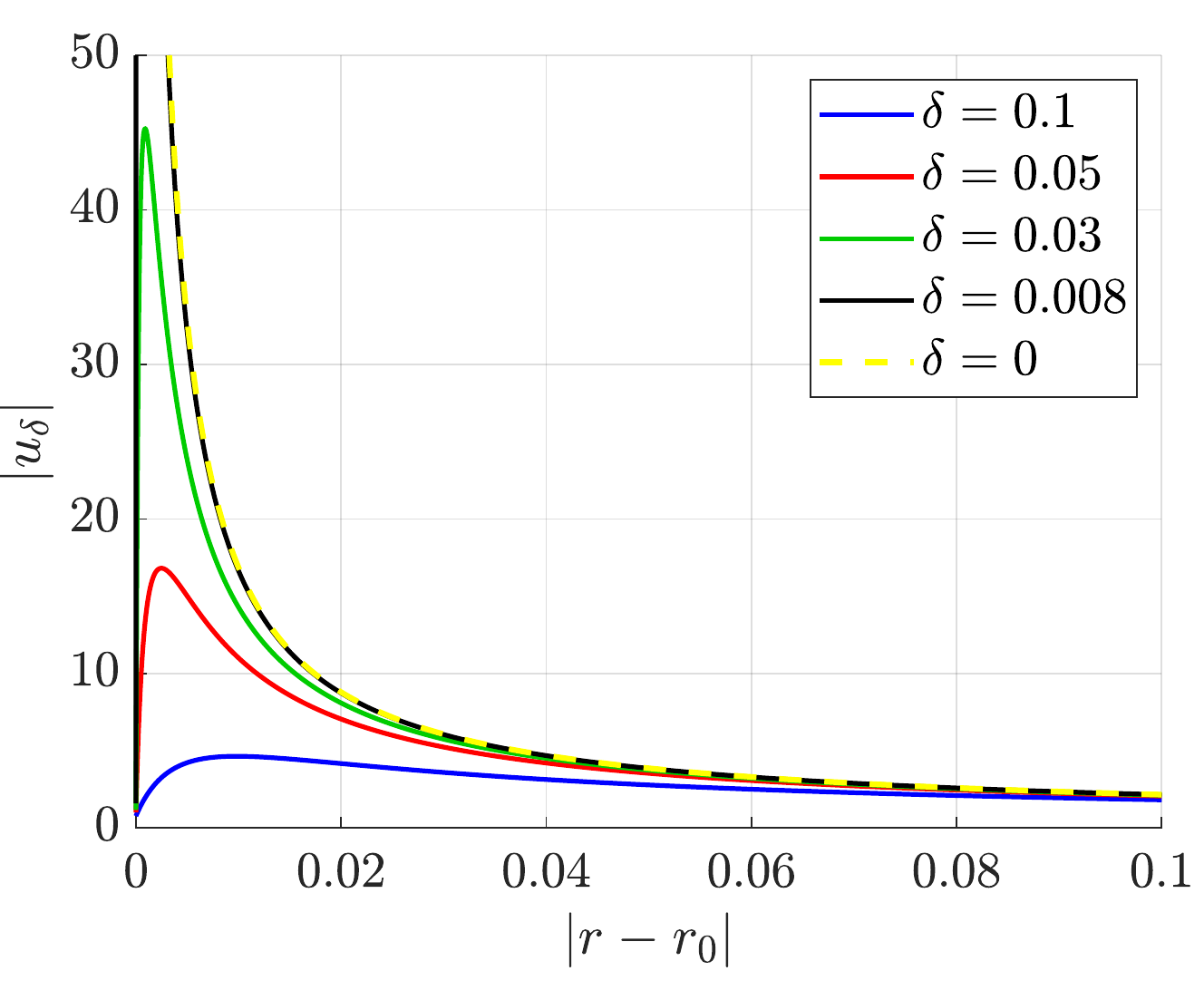}
            \caption{Induced velocity}
        \end{subfigure}
        \hfill 
        \begin{subfigure}[b]{0.49\textwidth}
            \centering
            \includegraphics[trim={3.5cm 9.0cm 4cm 9.5cm},clip,width=1.0\textwidth]{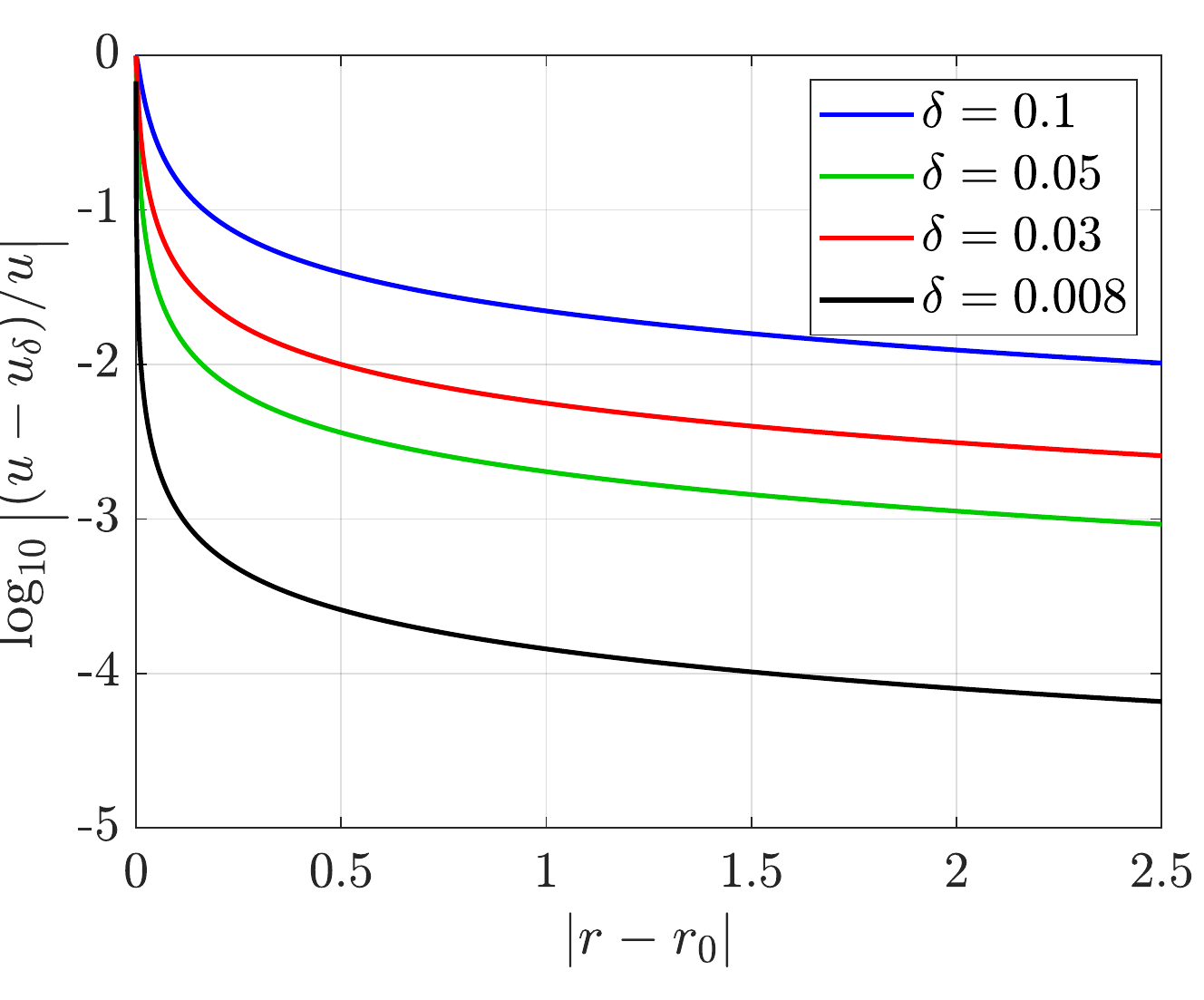}
            \caption{Relative error from $\delta=0$ case}
        \end{subfigure}
    \caption{Influence of smoothing parameter on the velocity field}
    \label{smoothingEffect}
\end{figure}

Although the induced velocity is finite, formulas (\ref{uz}), (\ref{ur}) are still troublesome. The problem of singularity in the integrand might be solved by various methods and an example is presented in \cite{Vooren}, where a function with the same singularity is subtracted from the integrand.
No matter of solution details, some areas of the sheet quickly get noisy and the resulting structure is growing cancer-like. The irregular movement of the nodes was studied in \cite{krasnyErrors} and \cite{Moore81} and is associated with unresolved Kelvin-Helmholtz instability. Modeling a physical vortex sheet, of finite thickness, with an idealized two-dimensional surface, introduces instability in all small wavenumbers. Its development can be launched by e.g. finite precision arithmetics. A well-established solution to this problem can be found in \cite{krasny_desingularization}, \cite{krasny_RollUp_Separation}, \cite{krasnyRing}. The core idea is to modify the Biot-Savart formula by adding a smoothing parameter $\delta$ to the denominator, i.e.:
\begin{equation}
    \bo{u}(\bo{r_0}) = \frac{1}{4\pi}\int_0^{2\pi}\int_{-\pi/2}^{\pi/2} \frac{ \gamma\hat{\phi}\times (\bo{r_0}-\bo{r})}{ \big(|\bo{r_0}-\bo{r}|^2 +\delta^2\big)^{3/2}} \rho ds\, d\phi
    \label{BS2delta}
\end{equation}
The exact value of $\delta$ determines the amount of damping that is applied to high wavenumbers and thus limits the range of scales present in a flow. The figure (\ref{smoothingEffect}) presents the smoothed velocity field.
\cite{deltaAsLO} shows that such a modification approaches the Lamb-Oseen solution of the viscous vortex decay, where 
\begin{equation}
    \delta(t) = \sqrt{5.03\nu t}
\end{equation}
although we used $\delta$ of fixed size, as in most of the articles mentioned. An alternative, Gaussian smoothing, leveraging Lamb-Oseen vortex directly is presented in \cite{gaussianDelta}. The comparison of the two methods might be found in \cite{deltaComparision}. Lagrangian, vorticity-based formulations of the flow, that include some kind of smoothing in the Biot-Savart formula are known in the literature as vortex blob method.

    \subsection{Summary of the governing equations}
        
Below, we present the governing equations that take an advantage of the axial symmetry (section \ref{sectionAxialSym}) and include smoothing (section \ref{section_delta}), applied in exactly the same way as in \cite{krasnyRing}.
\begin{equation}
    \frac{d\gamma}{dt} = b\frac{\partial z}{\partial s}
\end{equation}
\begin{equation}
    \frac{d\rho}{dt} = u^\delta_\rho
\end{equation}
\begin{equation}
    \frac{dz}{dt} = u^\delta_z
\end{equation}
\begin{equation}
    u_\rho^\delta(s_0)=\frac{1}{4\pi}\int_{-\pi/2}^{\pi/2}\;\frac{\gamma\,\rho\,(z_0-z)}{{R_2^\delta}^3}\,(I_1^\delta-I_2^\delta) \, ds
    \label{ur_delta}
\end{equation}
\begin{equation}
    u_z^\delta(s_0)=\frac{1}{4\pi}\int_{-\pi/2}^{\pi/2}\;\frac{\gamma\,\rho}{{R_2^\delta}^3}\bigg((\rho-\rho_0)\,I_1^\delta + \rho_0\,I_2^\delta \bigg)ds
    \label{uz_delta}
\end{equation}
where:
\begin{equation}
    I_1^\delta = \frac{4}{1-{k_\delta}^2}E({k_\delta}^2)
    \label{I1}
\end{equation}
\begin{equation}
    I_2^\delta = \frac{8}{{k_\delta}^2}\bigg( K({k_\delta}^2) - E({k_\delta}^2)  \bigg)
    \label{I2}
\end{equation}
\begin{equation}
    {k_\delta}^2 = 1-\bigg(\frac{R_1^\delta}{R_2^\delta}\bigg)^2 
\end{equation} 
\begin{align}
    {R_1^\delta}^2 = (\rho_0-\rho)^2+(z_0-z)^2 +\delta^2\\
    {R_2^\delta}^2 = (\rho_0+\rho)^2+(z_0-z)^2 +\delta^2
\end{align}
$K()$ and $E()$ denote complete elliptic integrals of the 1st and 2nd kind respectively. $\rho_0$ is a shortcut for $\rho(s_0)$ etc.

\section{Numerical algorithm}
    
    \subsection{General structure of the code}
    We demand from the discretization method the two following features: the ability to add new nodes in regions of intense stretching and the ability to merge nodes that are close enough. 
    For this purpose, the vortex sheet is represented as a set of nodes and a set of segments connecting the nodes. 
    Such a complication will be more understandable in subsection \ref{secSurgery}. 
    Allowing the merging of nearby nodes, we will change the sheet's topology. A single node could be connected with more than two other nodes so the sheet could no longer be considered a continuous, oriented curve, but rather an oriented graph. \\ \\
    Each node consists of:
    \begin{itemize}
        \item \ttt{iW} - an integration weight, such that: $\sum_{i=1}^{N} f_i \ttt{iW}_i \approx \int_{-\pi/2}^{\pi/2} f ds$, that is scheme-dependent
        \item \ttt{seg[]} - a list of segments linked to the node
        \item $\rho$, $z$ - position, alternatively represented by vector $\bo{r}$
        \item $\gamma$ - circulation density
    \end{itemize}
    and possibly some other values, necessary for memory management, etc. We will denote by \ttt{Nn} the total number of nodes and use the index $i$ while mentioning a particular one.\\ \\
    Each segment consists of:
    \begin{itemize}
        \item \ttt{start}, \ttt{end} - indices of the starting and the ending nodes of that segment
        \item \ttt{ds} - parameter length
        \item \ttt{bdry} - if the segment constitutes a boundary of a buoyant region
    \end{itemize}
    and possibly some other values. The total number of segments will be denoted by \ttt{Ns} and each particular one, will be distinguished by the index $j$. We will also denote the values at the starting node by subscript $s$ and at the ending node by $e$.\\
    We resigned from keeping the values of the parameter ($s$) in nodes because it would be difficult to track when nodes are merged. The same purpose can be reached by keeping $ds$ of every segment.
    \subsection{Computation of values at nodes}
    Quantities are assumed to vary linearly over the segments. Accessing nodes from the previous or next segment for higher-order schemes would be difficult because, due to merging, the order of segments is not properly defined (there could be multiple previous or next segments). Nevertheless, it could be done by turning segments into higher-order finite elements, with additional nodes inside. Although such improvement is left for further work and by now, the trapezoidal rule is used, which is second order in \texttt{ds}.
    \subsubsection{Integral weights}
    Computation of the trapezoidal integral weights is done by splitting \ttt{ds} from every segment in half, between its starting point and ending point. From the "node's perspective", we sum up all the \texttt{ds} from segments listed in \ttt{seg[]} and divide by 2:
    \begin{equation}
        \ttt{iW}_i = \tfrac{1}{2}\sum_{j\in\ttt{seg}} \ttt{ds}_j
        \label{iW}
    \end{equation}
    \subsubsection{Time derivative of the circulation density}
    The generation of $\gamma$ at every node is computed as follows: equation (\ref{gammaEq}) is integrated over the parameter length of every segment giving the total increase of circulation around it. Then, the circulation generation is split among the starting and ending node of the segment. At each node, after it is gathered from the adjacent segments, it is divided by \ttt{iW} and stored as \ttt{g}. This results in the following formula:
    \begin{equation}
        \frac{d \gamma_i}{dt} = \frac{b}{\ttt{iW}_i}\sum_{j\in\ttt{seg}} \tfrac{1}{2}(z_e-z_s)_j 
    \end{equation}
    which can also be seen as a linear staggered grid.
    \subsubsection{Velocities and coordinates}
    Velocities were computed in a procedure described in section \ref{section_dynnikova}. Time integration was performed with the 4th-order Runge-Kutta scheme.
    \subsection{Refinement}
\begin{figure}[h]
     \centering
     \begin{subfigure}[b]{0.5\textwidth}
         \centering
         \definecolor{myBlue}{rgb}{0.75,0.9,1}
\definecolor{myDarkBlue}{rgb}{0.3,    0.650,    0.8500}

\pgfdeclarelayer{ft2} 
\pgfdeclarelayer{ft}
\pgfdeclarelayer{bg} 
\pgfsetlayers{behind,main,ft,above} 

\newcommand{\vaOne}{60} 
\newcommand{\vaTwo}{115}

\tdplotsetmaincoords{\vaOne}{\vaTwo}
\pgfplotsset{compat=newest}

\begin{tikzpicture}[tdplot_main_coords]

\def\R{0.05cm}

\coordinate (e) at (3cm,1.5cm);  
\coordinate (s) at (0.75cm,1cm);  
\coordinate (e0) at (4cm,1.5cm);  
\coordinate (s0) at (-0.3cm,0.2cm);
\coordinate (n) at ($0.5*(s)+0.5*(e)$);
\coordinate (o2) at (4cm,0.2cm);

\path[fill, myBlue] (s0)--(s) -- (e)--(e0)--(o2)--cycle;
\path[draw,gray,semithick] (s0)--(s) -- (e)--(e0);
\draw[->,gray,semithick] (s0)--+($0.5*(s)-0.5*(s0)$);
\draw[->,gray,semithick] (e)--+($0.5*(e0)-0.5*(e)$);
\draw[->,gray,semithick] (s)--+($0.5*(e)-0.5*(s)$);

\fill[black] (s) circle (\R) node[above] {$s$};
\fill[black] (e) circle (\R) node[above] {$e$};
\fill[black] (e0) circle (\R);
\fill[black] (s0) circle (\R);
\path[gray] (n) node [below] {$1$};
\end{tikzpicture}
         \caption{}
     \end{subfigure}
     \hfill
     \begin{subfigure}[b]{0.5\textwidth}
         \centering
         \definecolor{myBlue}{rgb}{0.75,0.9,1}
\definecolor{myDarkBlue}{rgb}{0.3,    0.650,    0.8500}

\pgfdeclarelayer{ft2} 
\pgfdeclarelayer{ft}
\pgfdeclarelayer{bg} 
\pgfsetlayers{behind,main,ft,above} 

\newcommand{\vaOne}{60} 
\newcommand{\vaTwo}{115}

\tdplotsetmaincoords{\vaOne}{\vaTwo}
\pgfplotsset{compat=newest}

\begin{tikzpicture}[tdplot_main_coords]

\def\R{0.05cm}

\coordinate (e) at (3cm,1.5cm);  
\coordinate (s) at (0.75cm,1cm);  
\coordinate (e0) at (4cm,1.5cm);  
\coordinate (s0) at (-0.3cm,0.2cm);
\coordinate (n) at ($0.5*(s)+0.5*(e)$);
\coordinate (o2) at (4cm,0.2cm);

\path[fill, myBlue] (s0)--(s) -- (e)--(e0)--(o2)--cycle;
\path[draw,gray,semithick] (s0)--(s) -- (e)--(e0);
\draw[->,gray,semithick] (s0)--+($0.5*(s)-0.5*(s0)$);
\draw[->,gray,semithick] (e)--+($0.5*(e0)-0.5*(e)$);
\draw[->,gray,semithick] (s)--+($0.5*(n)-0.5*(s)$) node [below] {$2$};
\draw[->,gray,semithick] (n)--+($0.5*(e)-0.5*(n)$) node [below] {$3$};

\fill[black] (s) circle (\R) node[above] {$s$};
\fill[black] (e) circle (\R) node[above] {$e$};
\fill[black] (n) circle (\R) node[above] {$n$};
\fill[black] (e0) circle (\R);
\fill[black] (s0) circle (\R);
\end{tikzpicture}
         \caption{}
     \end{subfigure}
        \caption{Before and after refinement of a segment}
        \label{splitting}
\end{figure}
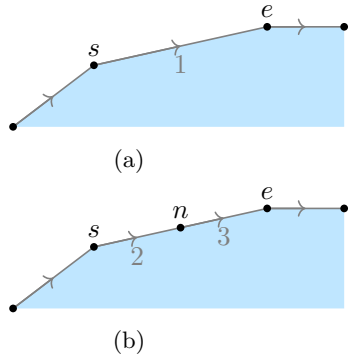

    The fundamental constant characterizing general discretization is 
    \begin{equation}
        \ttt{ds0} = \frac{\pi}{\ttt{Nn}_0-1}    
    \end{equation}
    which is the initial length of the segments, both, in the parameter sense and in the nondimensional physical sense. A segment, which exceeds prescribed maximal elongation $\kappa_E$:
    \begin{equation}
        (\bo{r}_e-\bo{r}_s)^2>\ttt{ds0}^2 \kappa_E^2 
    \end{equation}
    is split in half. Let us denote it by $1$ and let it start in node $s$ and end in node $e$, as depicted on fig. (\ref{splitting}). 
    A new node $n$ is inserted just between the starting and the ending nodes, averaging their values:
    \begin{equation}
        \bo{r}_n = \tfrac{1}{2}(\bo{r}_s + \bo{r}_e)
    \end{equation}
    \begin{equation*}
        \gamma_n = \tfrac{1}{2}(\gamma_s + \gamma_e)
    \end{equation*}
The segment $1$ is replaced by segment $2$ given between nodes $s$ and $n$ and segment $3$, between $n$ and $e$. The initial \ttt{ds} is split in half so:
\begin{equation*}
    \ttt{ds}_2 =\tfrac{1}{2}\ttt{ds}_1
\end{equation*}
\begin{equation}
    \ttt{ds}_3 = \tfrac{1}{2}\ttt{ds}_1
\end{equation}
and \ttt{iW} of all three nodes are updated. The proper update of \ttt{seg[]} arrays is of course also needed. \ttt{bdry} is inherited by both segments without changes.\\
Below we will check that the above procedure preserves circulation, denoting updated values with primes:
\begin{equation*}
    \ttt{iW}_s' = \ttt{iW}_s - \tfrac{1}{4}\ttt{ds}_1
\end{equation*}
\begin{equation*}
    \ttt{iW}_e' = \ttt{iW}_e - \tfrac{1}{4}\ttt{ds}_1
\end{equation*}
\begin{equation*}
    \ttt{iW}_n = \tfrac{1}{2}\ttt{ds}_1
\end{equation*}
Initially, we have:
\begin{equation}
    \Gamma = \ttt{iW}_s \gamma_s + \ttt{iW}_e \gamma_e
\end{equation}
and after refinement we have:
\begin{equation}
    \Gamma' = \ttt{iW}_s' \gamma_s + \ttt{iW}_e' \gamma_e + \ttt{iW}_n \gamma_n =  
\end{equation}
\begin{equation*}
    = (\ttt{iW}_s-0.25\ttt{ds}_1) \gamma_s + (\ttt{iW}_e-\tfrac{1}{4}\ttt{ds}_1) \gamma_e + \tfrac{1}{2}\ttt{ds}_1 (\gamma_s+\gamma_e)
\end{equation*}
\begin{equation*}
    = \ttt{iW}_s \gamma_s + \ttt{iW}_e \gamma_e
\end{equation*}
The total generation of circulation is also clearly preserved. \\
The refinement due to high curvature is left for future development. Its possible drawback could be the unphysically increased stability of regions of low curvature, due to coarse discretization. Moreover merging segments of significantly different lengths would also be more difficult.

    \subsection{Surgery}
    \label{secSurgery}
\begin{figure}[H]
        \centering
        \begin{subfigure}[b]{0.45\textwidth}
            \centering
            \fbox{\includegraphics[trim={6.5cm 11cm 6cm 10.5cm},clip,width=\textwidth]{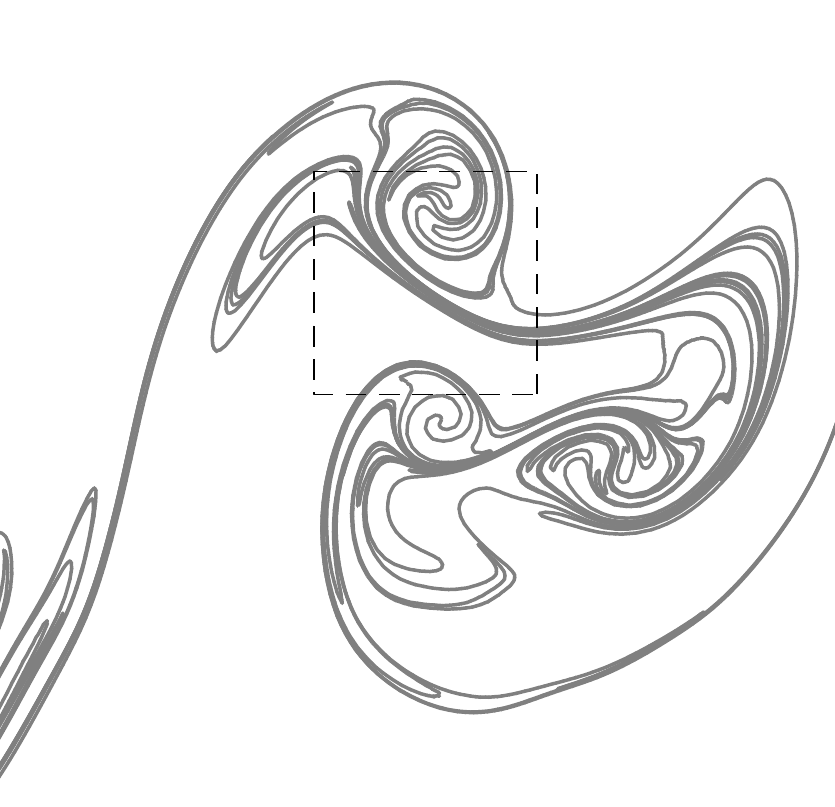}}
            \caption{}
        \end{subfigure}
        \hfill 
        \begin{subfigure}[b]{0.45\textwidth}
            \centering
            \fbox{\includegraphics[trim={6.5cm 11cm 6cm 10.5cm},clip,width=\textwidth]{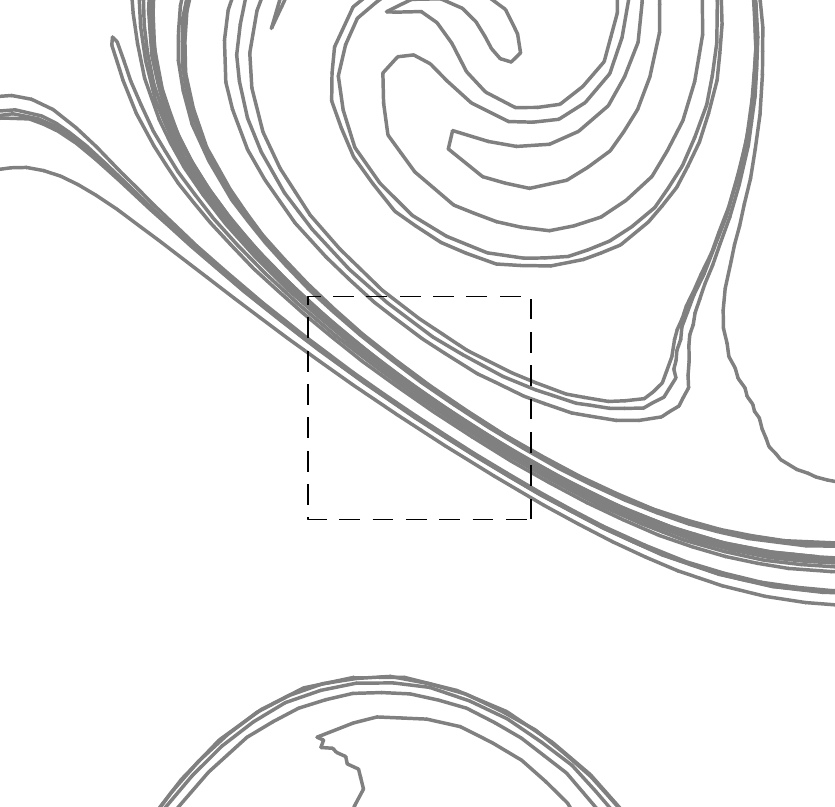}}
            \caption{}
        \end{subfigure}
        \hfill 
        \begin{subfigure}[b]{0.45\textwidth}
            \centering
            \vspace{3mm}
            \fbox{\includegraphics[trim={6.5cm 11cm 6cm 10.5cm},clip,width=\textwidth]{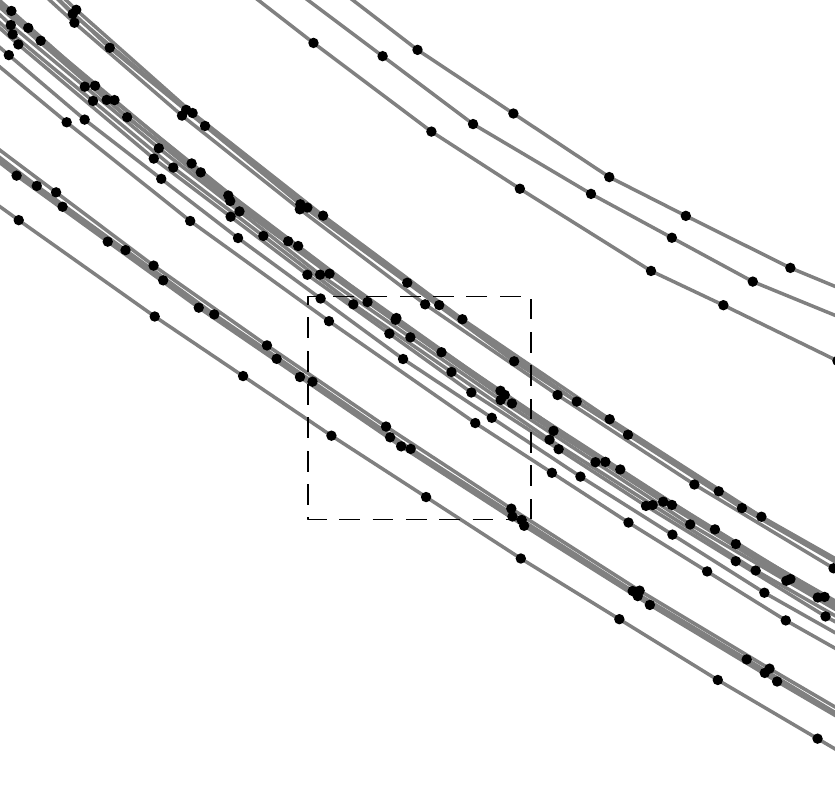}}
            \caption{}
        \end{subfigure}
        \hfill 
        \begin{subfigure}[b]{0.45\textwidth}
            \centering
            \vspace{3mm}
            \fbox{\includegraphics[trim={6.5cm 11cm 6cm 10.5cm},clip,width=\textwidth]{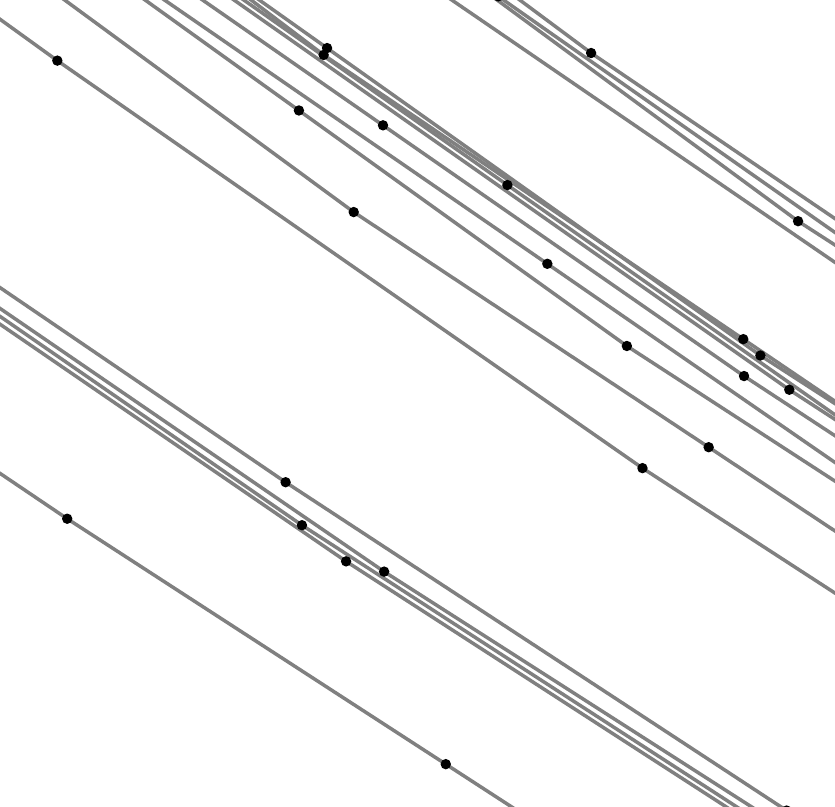}}
            \caption{}
        \end{subfigure}
        \caption{An example of a laminate structure in a series of subfigures of increasing zoom. The dashed square represents a region, enlarged in the next subfigure. In (c) and (d) individual nodes are depicted.}.
        \label{laminate}
    \end{figure}

    Most of the sophistication of the algorithm comes from the need of merging nearby nodes. This idea was inspired by the procedure used with contour dynamics method \cite{surgery}.
    Although it might sound simple, the trade-off between reducing the complexity of the sheet and not degenerating it too much is difficult to balance. We would like to keep the number of nodes in regions of contraction because that could allow resolving the possible development of instability. Such regions are rather rare anyway. The main purpose of the surgery is to simplify laminate structures. By laminate structures, we mean regions of many, relatively straight, parallel, pieces of the vortex sheet, packed tightly, as presented in figure (\ref{laminate}). They are pretty common in the analyzed case, originating in stretching and folding of Kelvin-Helmholtz vortices. Distances between two pieces belonging to such laminate structure tend to get lower and lower, so they could be "glued together" when they are much closer to each other than \ttt{ds0} (discretization scale). This can not only decrease the computational complexity by removing nodes but also avoid problems when pieces are so tightly packed that can cross each other due to numerical inaccuracies. \\\\ 
    The main problem with the process is the change in the sheet topology. What is inside and what is outside the surface is no longer clearly defined. We solve this difficulty by marking the segments that constitute the boundary of the buoyant region with the logical variable \ttt{bdry}. Its value is \ttt{true} when the buoyant fluid is only on one side of the segment. If it is on both or on none, then the value is \ttt{false} and the segment is just a piece of vortex sheet, that does not constitute the interface. This means that it does not generate the vorticity ($\nabla\times\bo{b}=\bo{0}$). All the segments are, of course, initialized with \ttt{bdry}=\ttt{true}. 
    \subsubsection{Case a) merging two interfaces}
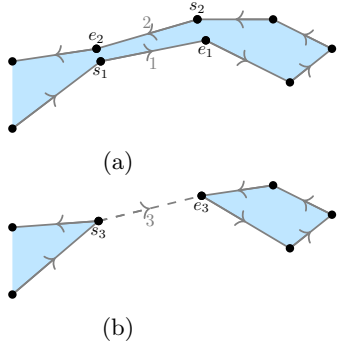
\begin{figure}[h!]
     \centering
     \begin{subfigure}[b]{0.5\textwidth}
         \centering
         \definecolor{myBlue}{rgb}{0.75,0.9,1}
\definecolor{myDarkBlue}{rgb}{0.3,    0.650,    0.8500}

\pgfdeclarelayer{ft2} 
\pgfdeclarelayer{ft}
\pgfdeclarelayer{bg} 
\pgfsetlayers{behind,main,ft,above} 

\newcommand{\vaOne}{60} 
\newcommand{\vaTwo}{115}

\tdplotsetmaincoords{\vaOne}{\vaTwo}
\pgfplotsset{compat=newest}

\begin{tikzpicture}[tdplot_main_coords]

\def\R{0.05cm}
\def\ndsc{0.6}

\coordinate (e) at (2cm,1.25cm);  
\coordinate (s) at (0.75cm,1cm);  
\coordinate (e0) at (3cm,0.75cm);  
\coordinate (s0) at (-0.3cm,0.2cm);
\coordinate (e1) at (2.8cm,1.5cm);
\coordinate (e2) at (1.9cm,1.5cm);
\coordinate (e3) at (0.7cm,1.15cm);
\coordinate (e4) at (-0.3cm,1.0cm);
\coordinate (q) at (3.5cm,1.15cm);
\coordinate (n) at ($0.5*(s)+0.5*(e)$);
\coordinate (o2) at (4cm,0.2cm);

\path[fill, myBlue] (s0)--(s) -- (e)--(e0)--(q) --(e1)--(e2)--(e3)--(e4) --cycle;
\path[draw,gray,semithick] (s0)--(s) -- (e)--(e0)--(q) --(e1)--(e2)--(e3)--(e4);
\draw[->,gray,semithick] (s0)--+($0.5*(s)-0.5*(s0)$);
\draw[->,gray,semithick] (e)--+($0.5*(e0)-0.5*(e)$);
\draw[->,gray,semithick] (e0)--+($0.5*(q)-0.5*(e0)$);
\draw[->,gray,semithick] (q)--+($0.5*(e1)-0.5*(q)$);
\draw[->,gray,semithick] (e1)--+($0.5*(e2)-0.5*(e1)$);
\draw[->,gray,semithick] (e2)--+($0.5*(e3)-0.5*(e2)$)  node [above,scale=\ndsc] {$2$};;
\draw[->,gray,semithick] (e3)--+($0.5*(e4)-0.5*(e3)$);
\draw[->,gray,semithick] (s)--+($0.5*(e)-0.5*(s)$) node [below,scale=\ndsc] {$1$};

\fill[black] (s) circle (\R) node[below,scale=\ndsc] {$s_1$};
\fill[black] (e) circle (\R) node[below,scale=\ndsc] {$e_1$};

\fill[black] (e2) circle (\R) node[above,scale=\ndsc] {$s_2$};
\fill[black] (e3) circle (\R) node[above,scale=\ndsc] {$e_2$};

\fill[black] (e0) circle (\R);
\fill[black] (s0) circle (\R);
\fill[black] (e) circle (\R);
\fill[black] (s) circle (\R);
\fill[black] (q) circle (\R);
\fill[black] (e1) circle (\R);
\fill[black] (e2) circle (\R);
\fill[black] (e3) circle (\R);
\fill[black] (e4) circle (\R);
\end{tikzpicture}
         \caption{}
     \end{subfigure}
     \hfill
     \begin{subfigure}[b]{0.5\textwidth}
         \centering
         \definecolor{myBlue}{rgb}{0.75,0.9,1}
\definecolor{myDarkBlue}{rgb}{0.3,    0.650,    0.8500}

\pgfdeclarelayer{ft2} 
\pgfdeclarelayer{ft}
\pgfdeclarelayer{bg} 
\pgfsetlayers{behind,main,ft,above} 

\newcommand{\vaOne}{60} 
\newcommand{\vaTwo}{115}

\tdplotsetmaincoords{\vaOne}{\vaTwo}
\pgfplotsset{compat=newest}

\begin{tikzpicture}[tdplot_main_coords]

\def\R{0.05cm}
\def\ndsc{0.6}
\coordinate (e) at (2cm,1.25cm);  
\coordinate (s) at (0.75cm,1cm);  
\coordinate (e0) at (3cm,0.75cm);  
\coordinate (s0) at (-0.3cm,0.2cm);
\coordinate (e1) at (2.8cm,1.5cm);
\coordinate (e2) at (1.9cm,1.5cm);
\coordinate (e3) at (0.7cm,1.15cm);
\coordinate (e4) at (-0.3cm,1.0cm);
\coordinate (q) at (3.5cm,1.15cm);
\coordinate (n) at ($0.5*(s)+0.5*(e)$);
\coordinate (s1e2) at ($0.5*(s)+0.5*(e3)$);
\coordinate (e1s2) at ($0.5*(e)+0.5*(e2)$);
\coordinate (o2) at (4cm,0.2cm);

\coordinate (e2) at ($(e1s2)$);
\coordinate (e3) at ($(s1e2)$);
\coordinate (e) at ($(e1s2)$);  
\coordinate (s) at ($(s1e2)$); 

\path[fill, myBlue] (s0)--(s1e2) -- (e1s2)--(e0)--(q) --(e1)--(e1s2)--(s1e2)--(e4) --cycle;
\path[draw,gray,semithick] (s0)--(s1e2);
\path[draw,gray,semithick]  (e1s2)--(e0)--(q) --(e1)--(e1s2);
\path[draw,gray,semithick] (s1e2)--(e4);
\path[draw,semithick,dashed, gray] (s1e2) -- (e1s2);

\draw[->,gray,semithick] (s0)--+($0.5*(s)-0.5*(s0)$);
\draw[->,gray,semithick] (e)--+($0.5*(e0)-0.5*(e)$);
\draw[->,gray,semithick] (e0)--+($0.5*(q)-0.5*(e0)$);
\draw[->,gray,semithick] (q)--+($0.5*(e1)-0.5*(q)$);
\draw[->,gray,semithick] (e1)--+($0.5*(e2)-0.5*(e1)$);
\draw[->,gray,semithick] (e3)--+($0.5*(e4)-0.5*(e3)$);
\draw[->,gray,semithick,dashed] (s1e2)--+($-0.5*(s1e2)+0.5*(e1s2)$) node [below,scale=\ndsc] {$3$};

\fill[black] (s1e2) circle (\R) node[below,scale=\ndsc] {$s_3$};
\fill[black] (e1s2) circle (\R) node[below,scale=\ndsc] {$e_3$};


\fill[black] (e0) circle (\R);
\fill[black] (s0) circle (\R);
\fill[black] (e1s2) circle (\R);
\fill[black] (s1e2) circle (\R);
\fill[black] (q) circle (\R);
\fill[black] (e1) circle (\R);

\fill[black] (e4) circle (\R);
\end{tikzpicture}
         \caption{}
     \end{subfigure}
        \caption{Before and after merging two interface segments. The solid line depicts segments with \ttt{bdry}=\ttt{true}, while the dashed line ones with \ttt{bdry}=\ttt{false}}.
        \label{merging}
\end{figure}

    Let us consider a fragment of vortex sheet, where two segments are close enough to be merged. We will denote them by indices $1$ and $2$. Their starting and ending nodes are denoted by $s$ and $e$ with proper subscripts, as presented in figure (\ref{merging}). Let the curve (the meridional section of the sheet) be positively oriented, so the buoyant region is always on the left side of each segment.
We demand from the procedure not to affect the total circulation and the total generation of the circulation. The latter depends only on the coordinates of the nodes of the segments. Because the segments are assumed to have a buoyant region always on the left side, if two of them are close to each other, they must have (approximately) opposite directions. This means that:
\begin{equation}
    |z_{s1}-z_{e2}| < \varepsilon
\end{equation}
\begin{equation*}
    |z_{e1}-z_{s2}| < \varepsilon
\end{equation*}
where $\varepsilon$ is some small value, determined by the merging criterion. Therefore, the total generation of circulation by these two segments, using eq. (\ref{gammaEq}) is:
\begin{equation}
    \frac{d\Gamma_1}{dt}+\frac{d\Gamma_2}{dt} = b(z_{e1}-z_{s1})+b(z_{e2}-z_{s2}) < 2b\varepsilon
\end{equation}
which is assumed to be negligible. Segments $1$ and $2$, both with \ttt{bdry}=\ttt{true} will be replaced by segment $3$ with \ttt{bdry}=\ttt{false}. We will call it the external vortex sheet. The choice of its direction does not matter. \\ \\
We decide to locate the resulting new nodes at averages weighted by the circulation modulus i.e:
\begin{equation}
    \bo{r}_{s3} = \frac{\bo{r}_{s1}\ttt{iW}_{s1}|\gamma_{s1}|+\bo{r}_{e2}\ttt{iW}_{e2}|\gamma_{e2}|}{\ttt{iW}_{s1}|\gamma_{s1}|+\ttt{iW}_{e2}|\gamma_{e2}|}
    \label{positionWeights}
\end{equation}
\begin{equation*}
    \bo{r}_{e3} = \frac{\bo{r}_{e1}\ttt{iW}_{e1}|\gamma_{e1}|+\bo{r}_{s2}\ttt{iW}_{s2}|\gamma_{s2}|}{\ttt{iW}_{e1}|\gamma_{e1}|+\ttt{iW}_{s2}|\gamma_{s2}|}
\end{equation*}
In the end, we need to update the sheet structure. Segment $3$ might be just modified segment $1$, while $2$ is removed. List of segments linked to the new node will be a union of segments linked to the old ones, without removed segments and with the newly created one.
\begin{equation}
    \ttt{seg[]}_{s3} = \ttt{seg[]}_{s1} \cup \ttt{seg[]}_{e2} \cup \{3\} \backslash \{1,2\}
\end{equation}
\begin{equation*}
    \ttt{seg[]}_{e3} = \ttt{seg[]}_{e1} \cup \ttt{seg[]}_{e2} \cup \{3\} \backslash \{1,2\}
\end{equation*}
There is also an analogical case that results in an internal vortex sheet - surrounded from both sides with a buoyant region. It is qualitatively the same, just colors in fig. (\ref{merging}) are swapped. It is also good to notice that the resulting closed interiors preserve their orientation in both cases. This means that all the positive-oriented cycles in the graph enclose the buoyant fluid, while the negative-oriented enclose non-buoyant one. 
\subsubsection{Case b) merging interface with an external vortex sheet}
In the previous case, we showed, that merging might result in a segment that does not generate circulation. Now, we will analyze how such a segment might be merged with a generating one:
    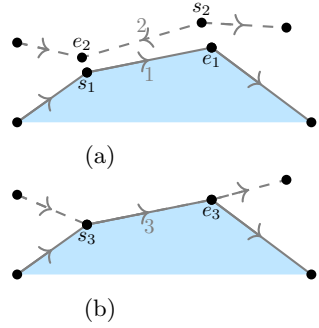
\begin{figure}[h!]
     \centering
     \begin{subfigure}[b]{0.5\textwidth}
         \centering
         \definecolor{myBlue}{rgb}{0.75,0.9,1}
\definecolor{myDarkBlue}{rgb}{0.3,    0.650,    0.8500}

\pgfdeclarelayer{ft2} 
\pgfdeclarelayer{ft}
\pgfdeclarelayer{bg} 
\pgfsetlayers{behind,main,ft,above} 

\newcommand{\vaOne}{60} 
\newcommand{\vaTwo}{115}

\tdplotsetmaincoords{\vaOne}{\vaTwo}
\pgfplotsset{compat=newest}

\begin{tikzpicture}[tdplot_main_coords]

\def\R{0.05cm}
\def\ndsc{0.6}
\coordinate (e1) at (2cm,1.25cm);  
\coordinate (s1) at (0.75cm,1cm);  
\coordinate (ne1) at (3cm,0.5cm);  
\coordinate (ps1) at (0.05cm,0.5cm);
\coordinate (s2) at (1.9cm,1.5cm);
\coordinate (e2) at (0.7cm,1.15cm);
\coordinate (pe2) at (0.05cm,1.3cm);
\coordinate (ns2) at ($(s2)+(0.85cm,-0.05cm)$);
\coordinate (q) at ($2.0*(ne1)-(3.5cm,1.35cm)$);

\coordinate (o2) at (4cm,0.2cm);
\coordinate (p0) at (0.7cm,0.0cm);

\path[fill, myBlue] (ps1)--(s1) -- (e1)--(ne1) --cycle;
\path[draw,gray,semithick] (ps1)--(s1) -- (e1)--(ne1);
\path[draw,gray,semithick,dashed] (pe2)--(e2)--(s2) -- (ns2);

\draw[->,gray,semithick] (ps1)--+($0.5*(s1)-0.5*(ps1)$);
\draw[->,gray,semithick] (e1)--+($0.5*(ne1)-0.5*(e1)$);

\draw[->,gray,semithick,dashed] (s2)--+($-0.5*(s2)+0.5*(ns2)$);

\draw[->,gray,semithick,dashed] ($(s2)+0.49*(e2)-0.49*(s2)$)--($(s2)+0.5*(e2)-0.5*(s2)$)  node [above,scale=\ndsc] {  $2$};
\draw[->,gray,semithick,dashed] (pe2)--+($-0.5*(pe2)+0.5*(e2)$);
\draw[->,gray,semithick] (s1)--+($0.5*(e1)-0.5*(s1)$) node [below,scale=\ndsc] {  $1$};

\fill[black] (s1) circle (\R) node[below,scale=\ndsc] {  $s_1$};
\fill[black] (e1) circle (\R) node[below,scale=\ndsc] {  $e_1$};

\fill[black] (s2) circle (\R) node[above,scale=\ndsc] {  $s_2$};
\fill[black] (e2) circle (\R) node[above,scale=\ndsc] {  $e_2$};

\fill[black] (ne1) circle (\R);
\fill[black] (ps1) circle (\R);
\fill[black] (e1) circle (\R);
\fill[black] (s1) circle (\R);

\fill[black] (pe2) circle (\R);
\fill[black] (ns2) circle (\R);
\end{tikzpicture}
         \caption{}
     \end{subfigure}
     \hfill
     \begin{subfigure}[b]{0.5\textwidth}
         \centering
         \definecolor{myBlue}{rgb}{0.75,0.9,1}
\definecolor{myDarkBlue}{rgb}{0.3,    0.650,    0.8500}

\pgfdeclarelayer{ft2} 
\pgfdeclarelayer{ft}
\pgfdeclarelayer{bg} 
\pgfsetlayers{behind,main,ft,above} 

\newcommand{\vaOne}{60} 
\newcommand{\vaTwo}{115}

\tdplotsetmaincoords{\vaOne}{\vaTwo}
\pgfplotsset{compat=newest}

\begin{tikzpicture}[tdplot_main_coords]

\def\R{0.05cm}
\def\ndsc{0.6}
\coordinate (e1) at (2cm,1.25cm);  
\coordinate (s1) at (0.75cm,1cm);  
\coordinate (ne1) at (3cm,0.5cm);  
\coordinate (ps1) at (0.05cm,0.5cm);
\coordinate (s2) at (1.9cm,1.5cm);
\coordinate (e2) at (0.7cm,1.15cm);
\coordinate (pe2) at (0.05cm,1.3cm);
\coordinate (ns2) at ($(s2)+(0.85cm,-0.05cm)$);
\coordinate (q) at ($2.0*(ne1)-(3.5cm,1.35cm)$);

\coordinate (o2) at (4cm,0.2cm);
\coordinate (p0) at (0.7cm,0.0cm);

\coordinate (e2) at (s1);
\coordinate (s2) at (e1);

\path[fill, myBlue] (ps1)--(s1) -- (e1)--(ne1) --cycle;
\path[draw,gray,semithick] (ps1)--(s1) -- (e1)--(ne1);
\path[draw,gray,semithick,dashed] (pe2)--(e2)--(s2) -- (ns2);

\draw[->,gray,semithick] (ps1)--+($0.5*(s1)-0.5*(ps1)$);
\draw[->,gray,semithick] (e1)--+($0.5*(ne1)-0.5*(e1)$);

\draw[->,gray,semithick,dashed] (s2)--+($-0.5*(s2)+0.5*(ns2)$);

\draw[->,gray,semithick,dashed] (pe2)--+($-0.5*(pe2)+0.5*(e2)$);
\draw[->,gray,semithick] (s1)--+($0.5*(e1)-0.5*(s1)$) node [below,scale=\ndsc] {  $3$};

\fill[black] (s1) circle (\R) node[below,scale=\ndsc] {  $s_3$};
\fill[black] (e1) circle (\R) node[below,scale=\ndsc] {  $e_3$};


\fill[black] (ne1) circle (\R);
\fill[black] (ps1) circle (\R);
\fill[black] (e1) circle (\R);
\fill[black] (s1) circle (\R);

\fill[black] (pe2) circle (\R);
\fill[black] (ns2) circle (\R);
\end{tikzpicture}
         \caption{}
     \end{subfigure}
        \caption{Before and after merging and interface with an external vortex sheet. The solid line depicts segments with \ttt{bdry}=\ttt{true}, while the dashed line ones with \ttt{bdry}=\ttt{false}}.
        \label{mergingB}
\end{figure}
The external vortex sheet is just incorporated into the interior, which keeps its orientation. Because circulation generation depends on the coordinates, keeping the values from the interface seems more reasonable, than using (\ref{positionWeights}). There is also an analogical case with the internal vortex sheet. \\ \\
In all cases, we would like to preserve the total circulation and the total parameter length ($s$), which is important for computing the integrals (section \ref{integralsEvolution}). Due to the latter:
\begin{equation}
    \ttt{ds}_3 = \ttt{ds}_1 + \ttt{ds}_2 
\end{equation}
then, the values of $\ttt{iW}_{s3}$ and $\ttt{iW}_{e3}$ are computed by the formula (\ref{iW}).  Circulation conservation requires:
\begin{equation}
    \gamma_{s3} = \frac{\ttt{iW}_{s1}\gamma_{s1}+\ttt{iW}_{e2}\gamma_{e2}}{\ttt{iW}_{s3}}
\end{equation}
\begin{equation*}
    \gamma_{e3} = \frac{\ttt{iW}_{e1}\gamma_{e1}+\ttt{iW}_{s2}\gamma_{s2}}{\ttt{iW}_{e3}}
\end{equation*}
\subsubsection{Case c) merging two external vortex sheets}
Merging of two external (or two internal) vortex sheets results in a sheet of the same kind as the former two. The coordinates of the nodes are determined according to (\ref{positionWeights}). Because the direction of external (internal) segments is arbitrary, it is not guaranteed that the starting node of one will be merged with the ending node of the other. It is necessary to check for merging also after flipping one of the segments.
\subsubsection{Merging criterion}
As can be seen in fig. (\ref{laminate}), especially (d), nodes of the laminate structure could be shifted along its direction. Therefore their distance might be much bigger (reaching \ttt{ds0}/2) than the actual distance between the two layers of the laminate. For this reason, although straightforward, this might not be the best criterion for merging. We want to keep the process as non-intrusive as possible and 
merging layers that are \ttt{ds0}/2 away, might be too degenerative. This problem will be even amplified if one decides to adjust the discretization to the local curvature, which in regions of interest is generally low. \\ \\
We propose the criterion based on three indicators:
\begin{itemize}
    \item distance between the nodes that could possibly be merged
        \begin{equation}
            (\bo{r}_{e1}-\bo{r}_{s2})^2<\ttt{ds0}^2 {\kappa_{M}}^2 
            \label{surgeryCriterion}
        \end{equation}
        \begin{equation*}
            (\bo{r}_{s1}-\bo{r}_{e2})^2<\ttt{ds0}^2 {\kappa_{M}}^2 
        \end{equation*}
        which should be lower than some percentage of the discretization scale, given by $\kappa_M$. To deal with the problem of shifted nodes it can be set to around 0.5.
    \item "normal distance" between the segments. We demand all the distances between chosen point and an opposite segment to be lower than some percentage of $\ttt{ds0}$, given by $\kappa_{MN}$ ($<\kappa_M$). For convenience, let us denote vectors representing segments by $\bo{a}$ and vectors connecting these segments by $\bo{c}$:
    \begin{equation*}
        \bo{a}_1 = \bo{r}_{e1}-\bo{r}_{s1}
    \end{equation*}
    \begin{equation*}
        \bo{a}_2 = \bo{r}_{e2}-\bo{r}_{s2}
    \end{equation*}
        \begin{equation*}
        \bo{c}_1 = \bo{r}_{s2}-\bo{r}_{e1}
    \end{equation*}
    \begin{equation*}
        \bo{c}_2 = \bo{r}_{s1}-\bo{r}_{e2}
    \end{equation*}
    then the criterion is:
    \begin{equation*}
        |\bo{a}_1 \times \bo{c}_1|\,/\,|\bo{a}_1|<\ttt{ds0}\,\kappa_{MN}
    \end{equation*}
    \begin{equation}
        |\bo{c}_1 \times \bo{a}_2|\,/\,|\bo{a}_2|<\ttt{ds0}\,\kappa_{MN}
        \label{surgeryCriterion2}
    \end{equation}
    \begin{equation*}
        |\bo{a}_2 \times \bo{c}_2|\,/\,|\bo{a}_2|<\ttt{ds0}\,\kappa_{MN}
    \end{equation*}
    \begin{equation*}
        |\bo{c}_2 \times \bo{a}_1|\,/\,|\bo{a}_1|<\ttt{ds0}\,\kappa_{MN}
    \end{equation*}
    \item relative angle (given by its cosine)
    \begin{equation}
        \frac{\bo{a}_1\cdot\bo{a}_2}{|\bo{a}_1|\;|\bo{a}_2|}<\kappa_A
        \label{surgeryCriterion3}
    \end{equation}
    where $\kappa_A$ represents the maximal cosine of the angle and should be negative. This allows targeting approximately parallel layers. Moreover, prevents merging neighboring segments that are in a straight line if they get too short. We do not do this as described at the beginning of the section. This would also require a slightly different procedure to be done in a conservative manner. Nevertheless, sharp enough corners are merged with no problems.
\end{itemize}
    In addition to the above criterion, we need to add a few exceptions to reject some pathological cases.
    \begin{itemize}
        \item we do not allow the converging or diverging segments to be merged. Such a situation could happen because segments with \ttt{bdry}=\ttt{false} have an arbitrary direction and pass the angle criterion. Therefore, if $e_1 = e_2$ or $s_1 = s_2$ merging is not applied.
        \item one-node segments ($s_1 = e_1$) are merged only with each other. Merging such a segment with a normal, two-node one is troublesome to be done in a conservative way. Such a segment might arise in a case presented in fig. (\ref{oneNode}) when segments $1$ and $2$ are merged.
    \end{itemize}
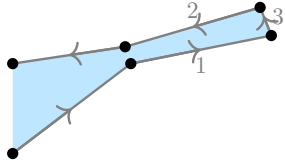
\begin{figure}[H]
         \centering
         \definecolor{myBlue}{rgb}{0.75,0.9,1}
\definecolor{myDarkBlue}{rgb}{0.3,    0.650,    0.8500}

\pgfdeclarelayer{ft2} 
\pgfdeclarelayer{ft}
\pgfdeclarelayer{bg} 
\pgfsetlayers{behind,main,ft,above} 

\newcommand{\vaOne}{60} 
\newcommand{\vaTwo}{115}

\tdplotsetmaincoords{\vaOne}{\vaTwo}
\pgfplotsset{compat=newest}

\begin{tikzpicture}[tdplot_main_coords]

\def\R{0.05cm}
\def\ndsc{0.6}
\coordinate (e) at (2cm,1.25cm);  
\coordinate (s) at (0.75cm,1cm);  
\coordinate (e0) at (3cm,0.75cm);  
\coordinate (s0) at (-0.3cm,0.2cm);
\coordinate (e1) at (2.8cm,1.5cm);
\coordinate (e2) at (1.9cm,1.5cm);
\coordinate (e3) at (0.7cm,1.15cm);
\coordinate (e4) at (-0.3cm,1.0cm);
\coordinate (q) at (3.5cm,1.15cm);
\coordinate (n) at ($0.5*(s)+0.5*(e)$);
\coordinate (o2) at (4cm,0.2cm);

\path[fill, myBlue] (s0)--(s) -- (e)--(e2)--(e3)--(e4) --cycle;
\path[draw,gray,semithick] (s0)--(s) -- (e)--(e2)--(e3)--(e4);
\draw[->,gray,semithick] (s0)--+($0.5*(s)-0.5*(s0)$);
\draw[->,gray,semithick] (e)--+($0.7*(e2)-0.7*(e)$) node [right,scale=\ndsc] {  $3$};
\draw[->,gray,semithick] (e2)--+($0.5*(e3)-0.5*(e2)$)  node [above,scale=\ndsc] {  $2$};;
\draw[->,gray,semithick] (e3)--+($0.5*(e4)-0.5*(e3)$);
\draw[->,gray,semithick] (s)--+($0.5*(e)-0.5*(s)$) node [below,scale=\ndsc] {  $1$};



\fill[black] (s0) circle (\R);
\fill[black] (e) circle (\R);
\fill[black] (s) circle (\R);
\fill[black] (e2) circle (\R);
\fill[black] (e3) circle (\R);
\fill[black] (e4) circle (\R);
\end{tikzpicture}
         \caption{Segment $3$ becomes a one-node segment after merging segments $1$ and $2$}
         \label{oneNode}
\end{figure}
\subsection{Promoting continuity of merging}
If we let segments be tested for merging in ("random") order of their placement in memory, the resulting structures might leave a lot to be desired. E.g. if we consider 3 parallel lines, it might happen that some segments from the middle one will be merged with the left one and others with the right one. This will give rise to a zig-zag reminding shape, which we find unfavorable. Another way in which it might arise is when two lines are balancing on the edge of the merging criterion. We can imagine that the first segment will be merged, the next one will not, the third one will be merged again, the 4th not, etc. We would prefer to merge continuous, possibly long, parts of the sheet. To promote this we test the segments in the order given by the Deep First Search algorithm. We start with a given segment, look for its possible merges, then proceed to its neighbor and do the same. Moreover we order the neighboring segments according to the angle they make with the currently visited segment, to promote traveling over straight lines. \\
One can notice that when the merging starts, we get a node that is connected to (at least) three segments. In addition to that, angles between nearby segments have changed unfavorably for the continuation of the surgery. For that reason, we add one more rule to the criterion: if tested segments have a common node, and if this node is connected to at least 3 segments, then we neglect the angle criterion (\ref{surgeryCriterion3}) and the normal distance criterion (\ref{surgeryCriterion2}). This is not only a correction to the described disadvantage but also actively promotes continuation. If merging just has started, then the next segment has less restrictive criteria and is more likely to also be merged. For that reason, surgery will be continued slightly too far i.e. will include segments that would not be merged if the process was to start at them. As a result, when the process is stopped, segments that are going to be tested, are no longer balancing on the edge of the criterion but rather clearly do not satisfy it. The last advantageous feature that could be added is setting few thresholds of criteria parameters. We would start with the most restrictive one and proceed to the more liberal ones. In the case of a few merging possibilities, this approach would support the best fit, rather than the first found. Nevertheless, we did not use this.

\section{Optimization}
\label{optimization}
    
\subsection{Construction of a tree}
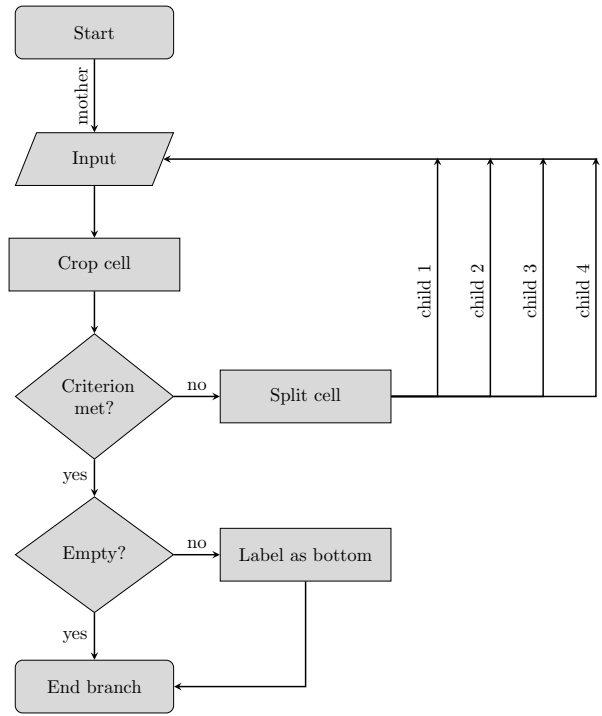
\begin{figure}[h!]
    \centering
    \tikzstyle{startstop} = [rectangle, rounded corners, 
minimum width=3cm, 
minimum height=1cm,
text centered, 
draw=black, 
fill=gray!30]

\tikzstyle{io} = [trapezium, 
trapezium stretches=true, 
trapezium left angle=70, 
trapezium right angle=110, 
minimum width=3cm, 
minimum height=1cm, text centered, 
draw=black, fill=gray!30]

\tikzstyle{process} = [rectangle, 
minimum width=3cm, 
minimum height=1cm, 
text centered, 
text width=3cm, 
draw=black, 
fill=gray!30]

\tikzstyle{decision} = [diamond, 
minimum width=3cm, 
minimum height=2.2cm, 
text width=1.25cm,
text centered, 
draw=black, 
fill=gray!30]
\tikzstyle{arrow} = [thick,->,>=stealth]

\begin{tikzpicture}[node distance=2cm]

\node (start) [startstop] {Start};
\node (in1) [io, below of=start, yshift=-0.4cm] {Input};
\node (crop) [process, below of=in1] {Crop cell};

\node (crit) [decision, below of=crop, yshift=-0.5cm] {Criterion met?};

\node (Empty) [decision, below of=crit, yshift=-1.0cm] {Empty?};

\node (split) [process, right of=crit, xshift=2cm] {Split cell};
\node (stop) [startstop, below of=Empty,yshift=-0.5cm] {End branch};

\node (emptyNode1) [right of=in1, xshift=4.5cm,yshift=0.15cm] {};
\node (emptyNode2) [right of=in1, xshift=5.5cm,,yshift=0.15cm] {};
\node (emptyNode3) [right of=in1, xshift=6.5cm,,yshift=0.15cm] {};
\node (emptyNode4) [right of=in1, xshift=7.5cm,yshift=0.15cm] {};
\node (emptyNode5) [right of=in1, xshift=7.65cm] {};
\node (labelB) [process,right of=Empty, xshift=2cm] {Label as bottom};
\node (emptyNode7) [right of=Empty, xshift=1.87cm,yshift=0.13cm] {};

\draw [arrow] (start) -- node[above,rotate=90, xshift=0cm] {mother}(in1);
\draw [arrow] (in1) -- (crop);
\draw [arrow] (crop) -- (crit);
\draw [arrow] (crit) -- node[anchor=east] {yes} (Empty);
\draw [arrow] (crit) -- node[anchor=south] {no} (split);
\draw [arrow] (split) -| node[above,rotate=90, xshift=2cm] {child 1} (emptyNode1);
\draw [arrow] (split) -| node[above,rotate=90, xshift=2cm] {child 2} (emptyNode2);
\draw [arrow] (split) -| node[above,rotate=90, xshift=2cm] {child 3} (emptyNode3);
\draw [arrow] (split) -| node[above,rotate=90, xshift=2cm] {child 4} (emptyNode4);
\draw [arrow] (emptyNode5) -- (in1);

\draw [arrow] (Empty) -- node[anchor=south] {no}(labelB);
\draw [arrow] (labelB) |- (stop);
\draw [arrow] (Empty) -- node[anchor=east] {yes}(stop);

\end{tikzpicture}
    \caption{Construction of a quadtree}
    \label{flowchart}
\end{figure}

\begin{figure}[H]
        \centering
        \begin{subfigure}[b]{0.34\textwidth}
            \centering
            {\includegraphics[trim={8cm 9.5cm 7cm 9cm},clip,width=\textwidth]{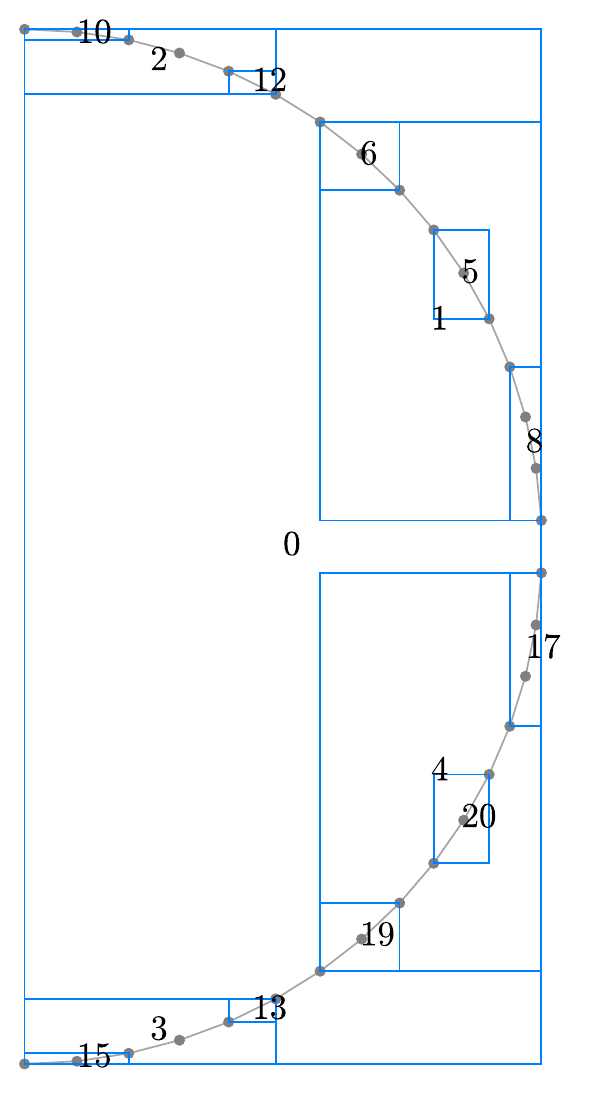}}
            \caption{Geometrical strucutre}
        \end{subfigure}
    \hfill
        \begin{subfigure}[b]{0.65\textwidth}
            \centering
            {\includegraphics[trim={6cm 11.0cm 5cm 11cm},clip,width=\textwidth]{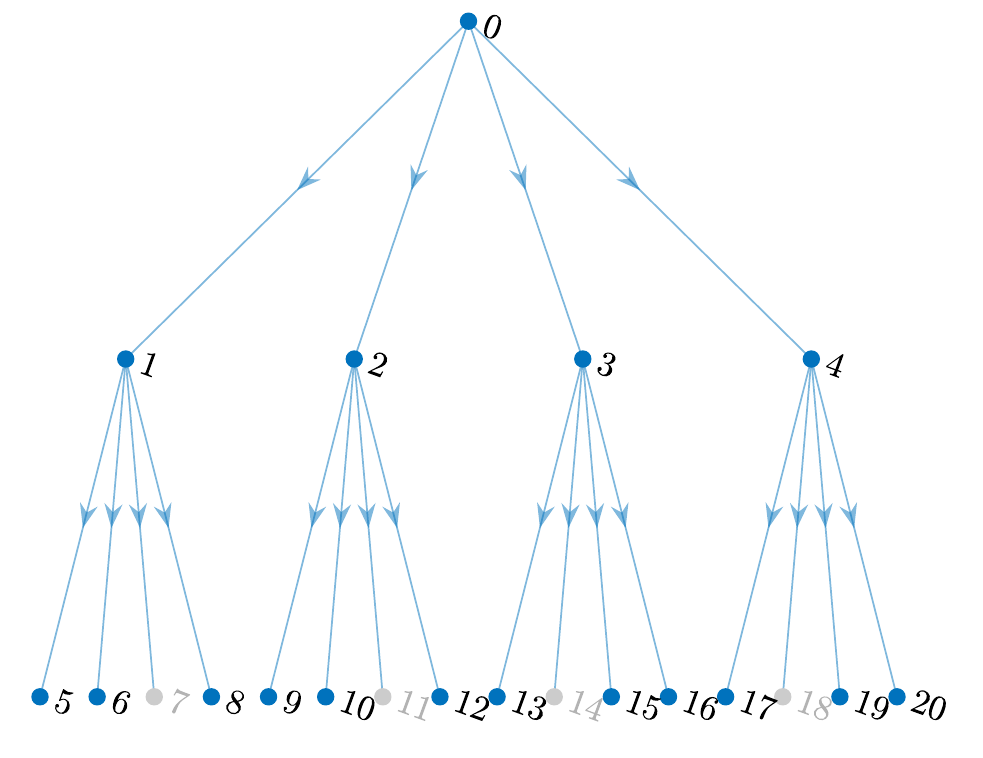}}
            \caption{Hierarchical structure (empty cells colored gray)}
        \end{subfigure}
        \caption{Quad tree constructed for the system of 32 nodes with $h=0.4$ at initial condition}
        \label{fig_treeBoth}
\end{figure}

Next to nodes and segments, we add to the vortex sheet structure the third element - the quad-tree. It consists of:
\begin{itemize}
    \item \ttt{nodeMap[]} - array that redefines nodes' indices. Used for sorting.
    \item \ttt{cells[]} - an array of cells
    \item \ttt{bottoms[]} - list of cells that are at the bottom - are childless
\end{itemize}
Each cell consists of:
\begin{itemize}
    \item \ttt{i0}, \ttt{i1} - first and last index in \ttt{nodeMap} of the nodes that belong to the cell
    \item $\rho_0$, $\rho_1$, $z_0$, $z_1$ - radial and vertical bounds of the cell
    \item \ttt{children[4]} - 4 indices of the child-cells of that cell
\end{itemize}
and some other variables of secondary importance for the general algorithm. The hierarchical structure of cells is constructed according to the flow chart in fig. (\ref{flowchart}). We choose the desired maximal bottom-cell size $h$ (understand as a diagonal). Starting with one, mother cell that includes the whole sheet, we search for the extremal coordinates of the nodes. Then we use them to adjust the boundaries ($\rho_0$, $\rho_1$, $z_0$, $z_1$) of the cell, and check the criterion:
\begin{equation}
    (\rho_1-\rho_0)^2 + (z_1-z_0)^2 < h^2    
\end{equation}
If it is not met, the cell is split into four child cells coincident with the quadrants. Nodes are distributed among the children by properly sorting the piece of \ttt{nodeMap} that belongs to the initial cell. As a result, nodes inherited by a child are in a continuous interval of indices. Then coordinates of all four children are cropped and the process continues till the creation of small enough cells. Those of them, that are not empty, will be called bottom cells and their indices are stored in \ttt{bottoms}. An exemplary tree is presented in fig. (\ref{fig_treeBoth}).

    \subsection{Fast velocity induction}
        
\label{section_dynnikova}
\begin{figure}[h!]
    \centering
    \tikzstyle{startstop} = [rectangle, rounded corners, 
minimum width=3cm, 
minimum height=1cm,
text centered, 
draw=black, 
fill=gray!30]

\tikzstyle{io} = [trapezium, 
trapezium stretches=true, 
trapezium left angle=70, 
trapezium right angle=110, 
minimum width=3cm, 
minimum height=1cm, text centered, 
draw=black, fill=gray!30]

\tikzstyle{process} = [rectangle, 
minimum width=3cm, 
minimum height=1cm, 
text centered, 
text width=3cm, 
draw=black, 
fill=gray!30]

\tikzstyle{decision} = [diamond, 
minimum width=3cm, 
minimum height=2.2cm, 
text width=1.25cm,
text centered, 
draw=black, 
fill=gray!30]
\tikzstyle{arrow} = [thick,->,>=stealth]

\begin{tikzpicture}[node distance=2cm]

\node (start) [startstop] {Start};
\node (in1) [io, below of=start, yshift=-0.4cm] {Input};

\node (crit) [decision, below of=in1, yshift=-2.5cm] {Far enough?};

\node (bot) [decision, right of=crit, xshift=1.75cm] {Bottom?};

\node (Fast) [process, below of=crit, yshift=-1.0cm] {Approximate induction};

\node (split) [process, above of=bot, yshift=0.5cm] {Go down the tree};
\node (add) [process, below of=Fast,yshift=-0.5cm,xshift=1.75cm] {Add to velocity};

\node (stop) [startstop, below of=add,yshift=-0.25cm] {Branch done};

\node (FastNode1) [right of=in1, xshift=4.5cm,yshift=0.15cm] {};
\node (FastNode2) [right of=in1, xshift=5.5cm,,yshift=0.15cm] {};
\node (FastNode3) [right of=in1, xshift=6.5cm,,yshift=0.15cm] {};
\node (FastNode4) [right of=in1, xshift=7.5cm,yshift=0.15cm] {};
\node (FastNode5) [right of=in1, xshift=7.65cm] {};
\node (naive) [process,below of=bot, yshift=-1cm] {Naive induction};
\node (FastNode7) [right of=Fast, xshift=1.87cm,yshift=0.13cm] {};

\draw [arrow] (start) -- node[above,rotate=90, xshift=0cm] {mother}(in1);
\draw [arrow] (in1) -- (crit);
\draw [arrow] (crit) -- node[anchor=east] {yes} (Fast);
\draw [arrow] (crit) -- node[anchor=south] {no} (bot);
\draw [arrow] (split) -| node[above,rotate=90, xshift=1cm] {child 1} (FastNode1);
\draw [arrow] (split) -| node[above,rotate=90, xshift=1cm] {child 2} (FastNode2);
\draw [arrow] (split) -| node[above,rotate=90, xshift=1cm] {child 3} (FastNode3);
\draw [arrow] (split) -| node[above,rotate=90, xshift=1cm] {child 4} (FastNode4);
\draw [arrow] (FastNode5) -- (in1);

\draw [arrow] (naive) -- (add);
\draw [arrow] (bot) -- node[anchor=east] {yes}(naive);
\draw [arrow] (bot) -- node[anchor=east] {no}(split);
\draw [arrow] (Fast) -- (add);
\draw [arrow] (add) -- (stop);
\end{tikzpicture}
    \caption{Fast induction of velocity for a given probing point}
    \label{flowchart_vel}
\end{figure}
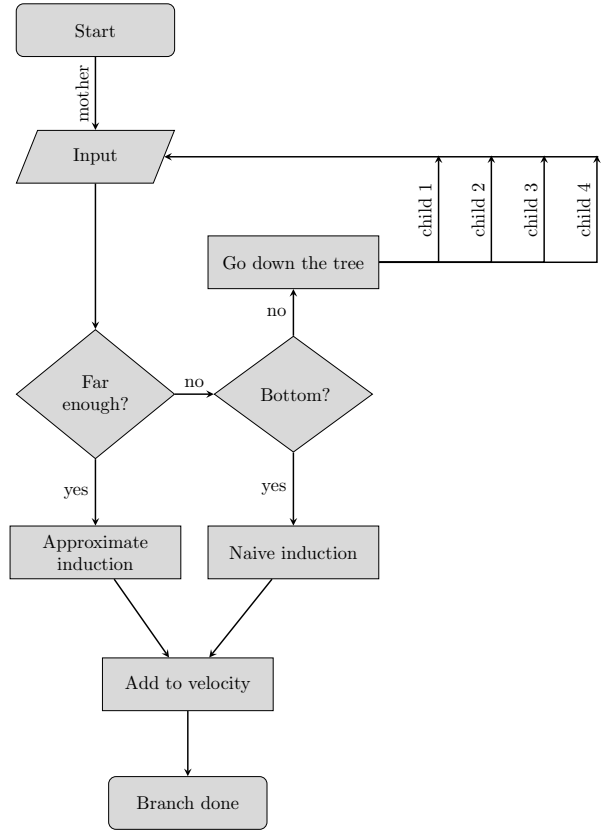

This section presents an application of the method described in \cite{dynnikova} for two dimensions, to the three-dimensional, axisymmetric case.\\ \\
The main idea comes from the N-body problem in the context of celestial mechanics. Interactions of far-away clusters of bodies are simplified to the interaction between their barycenters, which is then "distributed" over particular bodies. Clustering is done with the help of a hierarchical structure like the one described in the previous section.
Although we are dealing with a continuous problem, due to discretization it becomes an N-body problem as well. \\
In the original context, mass is always positive, therefore barycenters lie within the clusters. It is not the case in the N-vortex problem, where its analog - circulation, might also be negative, making the location of barycenters unbounded. The solution is to split the vortices into two groups - these with positive circulation and the rest with negative. Then barycenters are determined separately for each group. \\ \\
The method is summarized in three steps:
\begin{enumerate}
    \item build a tree
    \item compute circulations and barycenters
    \item induce velocities
\end{enumerate}
A more detailed flowchart for the last one is presented in fig. (\ref{flowchart_vel}).
\subsubsection{Approximation of the induced field}
\begin{figure}
    \centering
    \definecolor{myBlue}{rgb}{0.75,0.9,1}
\definecolor{myDarkBlue}{rgb}{0.3,    0.650,    0.8500}
\definecolor{myRed}{rgb}{0.9,0.2,0}

\pgfdeclarelayer{ft2} 
\pgfdeclarelayer{ft}
\pgfdeclarelayer{bg} 
\pgfsetlayers{behind,main,ft,above} 

\newcommand{\vaOne}{60} 
\newcommand{\vaTwo}{115}

\tdplotsetmaincoords{\vaOne}{\vaTwo}
\pgfplotsset{compat=newest}

\begin{tikzpicture}[tdplot_main_coords]

\def\R{0.02cm}
\def\op{0.0}
\def\opp{0.3}

\input{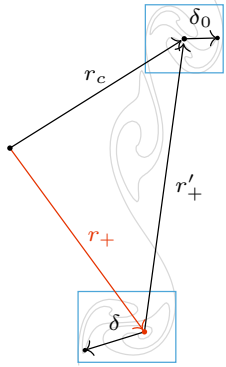}

\draw[myDarkBlue, thin](29.26 cm, 1.86 cm)
--(29.26 cm, 2.65 cm)
--(30.35 cm, 2.65 cm)
--(30.35 cm, 1.86 cm) --cycle;

\coordinate (rP) at (30.0 cm, 2.2 cm);  
\coordinate (sourceN) at (29.335980 cm, 1.995000 cm);

\draw[->] (rP)-- (sourceN) node [midway, above] {\scriptsize $\delta$};

\coordinate (rC) at (30.4440 cm, 5.4697 cm);  
\coordinate (probingN) at (30.815070 cm, 5.478840 cm);  
\coordinate (O) at (28.5 cm, 4.25 cm);

\draw[myDarkBlue, thin] (30.877980 cm, 5.849430 cm)
--(30.877980 cm, 5.09 cm)
--(30.01 cm, 5.09 cm)
--(30.01 cm, 5.849430 cm) --cycle;

\fill[black] (rC) circle ({1.5*\R});
\draw[->] (rC)-- (probingN) node [midway, above] {\scriptsize $\delta_0$};

\draw[->] (O)--+ ($0.985*(rC)-0.985*(O)$) node [midway, above] {\scriptsize $r_c$};
\draw[->,myRed] (O)--+ ($0.99*(rP)-0.99*(O)$) node [midway, right] {\scriptsize $r_+$};

\draw[->] (rP)--+ ($0.985*(rC)-0.985*(rP)$) node [midway, right] {\scriptsize $r_+'$};

\fill[black] (probingN) circle ({1.25*\R});
\fill[black] (sourceN) circle ({1.25*\R});
\fill[myRed] (rP) circle ({1.5*\R});
\fill[black] (O) circle ({1.5*\R});
\end{tikzpicture}
    \caption{Two exemplary nodes belonging to far-away clusters, captured by the cells}
    \label{fig_dynnikova_vecs}
\end{figure}

Although it is an abuse of nomenclature, let us refer to the following as the positive and negative parts of the circulation in the axisymmetric case:
\begin{subequations}
    \begin{equation}
        \Gamma_+ = \int_{C_+} \gamma\rho\, ds
        \label{vort+}
    \end{equation}
    \begin{equation}
        \Gamma_{-} = \int_{C_-} \gamma\rho\, ds
        \label{vort-}
    \end{equation}
\end{subequations}
We associate it with a group of nodes gathered in a cell ($C$).
Let us now define the corresponding, positive and negative barycenters of that cell:
\begin{subequations}
\begin{equation}
    \bo{r_+} = \frac{1}{\Gamma_+}\int_{C_+} \bo{r} \gamma\rho \,ds
    \label{center+}
\end{equation}
\begin{equation}
    \bo{r_-} = \frac{1}{\Gamma_+}\int_{C_-} \bo{r} \gamma\rho \,ds
    \label{center-}
\end{equation}
\end{subequations}
although for brevity, we will discuss only the positive one. The negative will be completely analogical. \\
Finally, let us consider two nodes from far away cells. The first one we will call the probing node (located in $\bo{r_0}$) and the second the source node (located in $\bo{r}$). The source cell has barycenter in $\bo{r_+}$ and let $\bo{r_c}$ be some reference point in the probing cell - e.g. its centroid. Then we introduce the following decomposition:
\begin{subequations}
\begin{equation}
    \bo{r_0} = \bo{r_c} + \bo{\delta_0}
\end{equation}
\begin{equation}
    \bo{r} = \bo{r_+} + \bo{\delta}
\end{equation}
\label{sub2}
\end{subequations}
Vectors $\bo{\delta} = \delta_\rho\hat{\rho}+z_\rho\hat{z}$ and $\bo{\delta_0} = \delta_{0\rho}\hat{\rho_0}+z_\rho\hat{z}$ should not be confused with the smoothing parameter from section \ref{section_delta}.
We would like to substitute this into the Biot-Savart formula (\ref{BS2}). For simplicity, let us, now, consider just the following expression:
\begin{equation}
     \frac{\bo{r_0}-\bo{r}}{(\bo{r_0}-\bo{r})^3} = \frac{\bo{r_c}+\bo{\delta_0}-(\bo{r_+}+\bo{\delta}) }{[\bo{r_c}+\bo{\delta_0}-(\bo{r_+}+\bo{\delta})]^{3/2}}
\end{equation}
To take an advantage of the assumption that the cells are far away, we will introduce also:
\begin{subequations}
\begin{equation}
    \bo{\delta'}=\bo{\delta_0}-\bo{\delta}
    \label{sub1}
\end{equation}
\begin{equation}
    \bo{r_+'} = \bo{r_c}-\bo{r_+}
    \label{r+'def}
\end{equation}
\label{sub_1}
\end{subequations}
The considered case, together with the nomenclature was presented in fig. (\ref{fig_dynnikova_vecs}).
We now can write:
\begin{equation}
     \frac{\bo{r_0}-\bo{r}}{(\bo{r_0}-\bo{r})^3} = \frac{\bo{r_+'}+\bo{\delta'}}{(r_+'^2 + \delta'^2 + 2\bo{r_+'}\cdot \bo{\delta'})^{3/2}} = \frac{\bo{r_+'}+\bo{\delta'}}{r_+'^3} \bigg[1+\big(\frac{\delta'}{r_+'}\big)^2+2\frac{\bo{r_+'}\cdot\bo{\delta'}}{r_+'^2}   \bigg]^{-3/2}
    \label{preTaylor}
\end{equation}
The fraction in parentheses is by assumption (clusters far away from each other) a small parameter that we will denote $\varepsilon$, so:
\begin{equation}
    \varepsilon = \frac{\delta'}{r_+'}
\end{equation}
In practice, in numerical computation it will be estimated as:
\begin{equation}
    \varepsilon' = \frac{h+h_0}{r_+'}
    \label{estimateEpsilon}
\end{equation}
where $h$ and $h_0$ are diagonals of probing and source cells.
It is also good to notice that
\begin{equation}
    \frac{\bo{r_+'}\cdot\bo{\delta'}}{r_+'^2} = \frac{\hat{r_+'}r_+'\cdot \hat{\delta'}\delta'}{r_+'^2} = (\hat{r_+'}\cdot \hat{\delta'})\varepsilon 
\end{equation}
Reminding the following Taylor expansion around 0:
\begin{equation}
    (1+x)^{-3/2} = 1 - \frac{3}{2}x + \frac{15}{8}x^2 + ...
\end{equation}
we will expand the square bracket in (\ref{preTaylor}), neglect the terms of order higher than first in $\varepsilon$, obtaining:
\begin{equation}
     \frac{\bo{r_0}-\bo{r}}{(\bo{r_0}-\bo{r})^3} = \bigg( \frac{\hat{r_+'}+\hat{\delta'}\varepsilon}{r_+'^2} \bigg) 
    \bigg[  1 - \frac{3}{2}\big(  \varepsilon^2 + 2\hat{r_+'}\cdot \hat{\delta'}\varepsilon  \big) + \frac{15}{8} \big(  \varepsilon^2 + 2\hat{r_+'}\cdot \hat{\delta'}\varepsilon  \big)^2 \bigg .... \bigg]
\end{equation} 
\begin{equation*}
    \approx
     \frac{\hat{r_+'}}{r_+'^2} \bigg[  1 - 3\hat{r_+'}\cdot \hat{\delta'}\varepsilon  \bigg]  +\frac{\hat{\delta'}\varepsilon}{r_+'^2}
\end{equation*}
Now, when we truncated the series, it would be convenient to substitute back for $\bo{\delta'}$ from (\ref{sub1}) and (\ref{sub2}): $\hat{\delta'}\varepsilon = \bo{\delta'}/r_+' = (\bo{\delta_0}+\bo{r_+}-\bo{r})/r_+'$ in hope of extracting the integral quantities defined at the very beginning (\ref{vort+}), (\ref{center+}). Also for that purpose, we will present it in full Biot-Savart context:
\begin{equation}
    \frac{1}{4\pi}\int_0 ^{2\pi}\int_{C_+}\hat{\phi}\times \frac{\bo{r_0}-\bo{r}}{(\bo{r_0}-\bo{r})^3}\; \gamma\;ds\rho d\phi \approx 
\end{equation}
\begin{equation*}
   \approx
   \frac{1}{4\pi} \int_0^{2\pi} \hat{\phi}\times \Bigg(  \int_{C_+}  \frac{\bo{r_+'}}{r_+'^3}
    \bigg[  1 - 3\frac{\bo{r_+'}\cdot    (\bo{\delta_0}+\bo{r_+}-\bo{r}) }{r_+'^2}  \bigg]  \;\gamma\; ds 
    +\int_{C_+} \frac{(\bo{\delta_0}+\bo{r_+}-\bo{r}) }{r_+'^3}\;\gamma\; ds \Bigg) \rho d\phi
\end{equation*}
Noticing that $\bo{r'_+}$, $\bo{r_+}$, $\bo{\delta_0}$ do not depend on $s$ (at least for a given pair of clusters) and the only $s$-dependent quantities are $\bo{r}$ and $\gamma$, we will rearrange the integration: 
\begin{equation*}
=
\frac{1}{4\pi} \int_0^{2\pi} \hat{\phi}\times \Bigg( 
  \frac{\bo{r_+'}}{r_+'^3}\int_{C_+} \;\gamma\rho\; ds
    -3\frac{\bo{r_+'}}{r_+'^5} \bigg[ \bo{r_+'}\cdot 
         (\bo{\delta_0}+\bo{r_+})\int_{C_+}\gamma\rho\; ds -\bo{r_+'}\cdot\int_{C_+}\bo{r}\;g\rho\; ds   \bigg] +
\end{equation*}
\begin{equation*}
    + \frac{\bo{\delta_0}+\bo{r_+}}{r_+'^3}\int_{C_+}\;\gamma \rho \; ds-\frac{1}{r_+'^3}\int_{C_+}\bo{r} \;\gamma\rho\; ds
    \Bigg)  d\phi
\end{equation*}
Substituting integral quantities from (\ref{vort+}) and (\ref{center+}):
\begin{equation}
 =
 \frac{1}{4\pi} \int_0 ^{2\pi}\int_{C_+}\hat{\phi}\times \Bigg(
     \frac{\bo{r_+'}}{r_+'^3}\Gamma_+
    -3\frac{\bo{r_+'}}{r_+'^5} \bigg[ \bo{r_+'}\cdot 
         (\bo{\delta_0}+\bo{r_+})\Gamma_+ -\bo{r_+'}\cdot\Gamma_+\bo{r_+}   \bigg] 
    + \frac{\bo{\delta_0}+\bo{r_+}}{r_+'^3}\Gamma_+ -\frac{\Gamma_+ \bo{r_+}}{r_+'^3} \Bigg)  d\phi
\end{equation}
and canceling out terms, we finally get the formula for an approximate velocity in point $\bo{r_0}$, induced by the nodes of positive circulation from cell $C$:
\begin{equation}
     \bo{u}(\bo{r_0})_{C+} = 
     \frac{1}{4\pi}\int_0^{2\pi}  \Gamma_+\hat{\phi} \times \Bigg(
     \underbrace{
     \frac{\bo{r_+'}}{r_+'^3}
     }_{A}
    -3 \underbrace{
    \frac{\bo{r_+'}}{r_+'^5} \bigg[ \bo{r_+'}\cdot 
         \bo{\delta_0}    \bigg]
         }_{B}
    + \underbrace{
    \frac{\bo{\delta_0}}{r_+'^3}}_{C}
    \Bigg)  d\phi + \bigO(\varepsilon^2)
    \label{ABC}
\end{equation}
The result is identical to one obtained in \cite{dynnikova} with exception of the coefficient in front of term B. That difference comes from three-dimensional space instead of two and arises is Taylor expansion of $(1+x)^{-3/2}$ instead of $(1+x)^{-1}$.
\subsubsection{Integration over the azimuth} 
It can be noticed that the term A from equation (\ref{ABC}) is analogical to (\ref{BS2}), so we will write down the corresponding quantities (denoting them by $\to$):
\begin{equation*}
    \rho_c\to\rho_0
\end{equation*}
\begin{equation}
    z_c \to z_0
    \label{limits}
\end{equation}
\begin{equation*}
    z_+ \to z
\end{equation*}
\begin{equation*}
    \Gamma_+ \to \gamma\rho\;ds
\end{equation*}
and proceed in exactly the same way as before (including smoothing), obtaining:
\begin{equation}
    u_{\rho_0 A}^\delta = \frac{1}{4\pi} \frac{\Gamma_+(z_c-z_+)}{{R_{2A}^\delta}^3}\bigg(I_{1A}^\delta-I_{2A}^\delta\bigg)
    \label{urA}
\end{equation}
\begin{equation}
    u_{z A}^\delta = \frac{1}{4\pi} \frac{\Gamma_+}{{R_{2A}^\delta}^3}\bigg((\rho_+-\rho_c)I_{1A}^\delta+\rho_c I_{2A}^\delta\bigg)
    \label{uzA}
\end{equation}
where:
\begin{align}
    {k_A^\delta}^2 = 1-\bigg(\frac{R_{1A}^\delta}{R_{2A}^\delta}\bigg)^2   \label{kA}\\ 
    {R_{1A}^\delta}^2 = (\rho_c-\rho_+)^2+(z_c-z_+)^2 + \delta^2\\
    {R_{2A}^\delta}^2 = (\rho_c+\rho_+)^2+(z_c-z_+)^2 + \delta^2\\
\end{align}
Resulting integrals have exactly the same form as (\ref{I1}) and (\ref{I2}). The only difference is that $k_\delta$ is replaced with $k_A^\delta$ and $R_2^\delta$ with $R_{2A}^\delta$. We will denote them as $I_{1A}^\delta$ and $I_{2A}^\delta$. It is good to notice, that (\ref{limits}) represent also behavior in a limit as dimensions of both cells ($\bo{\delta}$ and $\bo{\delta_0}$) approach zero. Thus (\ref{uzA}) and (\ref{urA}) approach the integrands of (\ref{uz}) and (\ref{ur}).\\ \\
Let us now proceed to the term C:
\begin{equation}
    C = \frac{1}{4\pi}\int_0^{2\pi}  \frac{\Gamma_+\hat{\phi}\times\bo{\delta_0}}{r_+'^3} d\phi
\end{equation}
its denominator is exactly the same as for term A. Let $\bo{\delta_0} = \delta_{0\rho} \hat{\rho_0} + \delta_{0z} \hat{z}$, and consider the numerator:
\begin{equation}
    \hat{\phi} \times (\delta_{0\rho} \hat{\rho_0} + \delta_{0z} \hat{z}) = -\delta_{0\rho}\cos(\Delta\phi)\hat{z} + \delta_{0z}\hat{\rho}
\end{equation}
Projecting it with dot product onto directions $z$ and $\rho_0$ gives:
\begin{equation}
        \hat{\phi} \times (\delta_{0\rho} \hat{\rho_0} + \delta_{0z} \hat{z}) = -\delta_{0\rho}\cos(\Delta\phi)\hat{z} + \delta_{0z}\cos(\Delta\phi)\hat{\rho_0}
\end{equation}
We switch to $\theta$ via (\ref{phi2theta}), remembering that $d\phi = 2d\theta$ and to half the integration upper limit. We obtain:
\begin{equation}
    u_{\rho 0 C}^\delta = \frac{1}{4\pi}\int_0^{\pi}\frac{\Gamma_+\delta_{0z}(2-4\cos^2\theta)}{{R_{2A}^\delta}^2(1-{k_A^\delta}^2\sin^2\theta)}\,d\theta = 
    \frac{1}{4\pi}\frac{\Gamma_+\delta_{0z}}{{R_{2A}^\delta}^3}\bigg( I_{1A}^\delta-I_{2A}^\delta \bigg) 
\end{equation}
\begin{equation}
    u_{z C}^\delta = \frac{1}{4\pi}\int_0^{\pi}\frac{\Gamma_+\delta_{0z}(2-4\cos^2\theta)}{{R_{2A}^\delta}^2(1-{k_A^\delta}^2\sin^2\theta)}\,d\theta = 
    \frac{-1}{4\pi}\frac{\Gamma_+\delta_{0\rho}}{{R_{2A}^\delta}^3}\bigg( I_{1A}^\delta-I_{2A}^\delta \bigg) 
\end{equation}
where quantities with subscript $A$ are defined near (\ref{kA}).
\begin{equation}
    B = \frac{3}{4\pi}\int_0^{2\pi}\Gamma_+ \hat{\phi}\times \frac{\bo{r'_+}(  \bo{r_+'}\cdot\bo{\delta_0})}{r_+'^5}\,d\phi
\end{equation}
The denominator can be transformed using the formula (\ref{denom}). In the first part of the numerator - the cross product, we can use (\ref{num}).
The remainder is the dot product in the numerator. It is convenient to decompose $\bo{r_+'}$ with (\ref{r+'def}).
\begin{equation}
    \bo{r_+'} \cdot \bo{\delta_0} = (\bo{r_c}-\bo{r_+})\cdot \bo{\delta_0} = (\rho_c \hat{\rho_0} + z_c\hat{z}-\rho_+\hat{\rho}-z_+\hat{z}) \cdot (\delta_{0\rho}\hat{\rho_0}+\delta_{0z}\hat{z})
\end{equation}
\begin{equation*}
    = \rho_c\,\delta_{0\rho}-\rho_+\delta_{0\rho}\cos(\Delta\phi)+(z_c-z_+)\delta_{0z}
\end{equation*}
then we can substitute $\theta$ using (\ref{phi2theta}) and overall, we obtain:
\begin{equation}
    v_{B\rho0} = \frac{3}{4\pi}\int_0^{\pi}\Gamma_+ \frac{\rho_c\,\delta_{0\rho}-\rho_+\delta_{0\rho}(1-2\cos^2\theta)+(z_c-z_+)\delta_{0z}}{{R_{2A}^\delta}^5(1-{k_A^\delta}^2\sin^2\theta)^{5/2}}(z_c-z_+)(2-4\cos^2\theta) \,d\theta
\end{equation}
\begin{equation*}
    =\frac{6}{4\pi}\int_0^\pi \Gamma_+ \frac{\delta_{0\rho}(\rho_c-\rho_+)+(z_c-z_+)\delta_{0z}}{{{R_{2A}^\delta}^5(1-{k_A^\delta}^2\sin^2\theta)^{5/2}}}(z_c-z_+)\,d\theta
\end{equation*}
\begin{equation*}
    +\frac{12}{4\pi}\int_0^\pi\Gamma_+\frac{-(\rho_c\,\delta_{0\rho}-\rho_+\delta_{0\rho}+(z_c-z_+)\delta_{0z})+\rho_+\delta_{0\rho}}{{R_{2A}^\delta}^5(1-{k_A^\delta}^2\sin^2\theta)^{5/2}}(z_c-z_+)\cos^2\theta\, d\theta
\end{equation*}
 \begin{equation*}
     +\frac{24}{4\pi}\int_0^\pi\Gamma_+ \frac{-\rho_+\delta_{0\rho}}{{R_{2A}^\delta}^5(1-{k_A^\delta}^2\sin^2\theta)^{5/2}}(z_c-z_+)\cos^4\theta\, d\theta
 \end{equation*}
 what finally leads to:
\begin{equation}
    v_{B\rho0}^\delta = \frac{3(z_c-z_+)\Gamma_+}{4\pi {R_{2A}^\delta}^5}\Bigg[  \bigg(\delta_{0\rho}(\rho_c-\rho_+)+(z_c-z_+)\delta_{0z}\bigg)(I_{1B}^\delta -I_{2B}^\delta)
\end{equation}
\begin{equation*}
    +\rho_+\delta_{0\rho} (I_{2B}^\delta-I_{3B}^\delta) \Bigg]
\end{equation*}
and analogically for $z$ component:
\begin{equation}
    v_{Bz}^\delta = \frac{3}{4\pi}\int_0^{\pi}\Gamma_+ \frac{\rho_c\,\delta_{0\rho}-\rho_+\delta_{0\rho}(1-2\cos^2\theta)+(z_c-z_+)\delta_{0z}}{{R_{2A}^\delta}^5(1-{k_A^\delta}^2\sin^2\theta)^2}(2\rho_+-\rho_c(2-4\cos^2\theta)) \,d\theta
\end{equation}
\begin{equation*}
    =\frac{3}{4\pi \, {R_{2A}^\delta}^5} \Gamma_+  \Bigg(
    [(\rho_c -\rho_+)^2\delta_{0\rho}+(z_c-z_+)\delta_{0z}\rho_c]( I_{2B}^\delta -I_{1B}^\delta)
\end{equation*}
\begin{equation*}
 +\rho_+ [(z_c-z_+)\delta_{0z} I_{1B}^\delta
    + \delta_{0\rho} \rho_c I_{3B}^\delta]\Bigg)
 \end{equation*}
where
\begin{equation}
    I_{1B}^\delta = \int_0^\pi \frac{2\,d\theta}{(1-{k_A^\delta}^2\sin^2\theta)^{5/2}}
    = \frac{4}{3}\Bigg(  \frac{2E({k_A^\delta}^2)-K({k_A^\delta}^2)}{1-{k_A^\delta}^2}+\frac{2E({k_A^\delta}^2)}{(1-{k_A^\delta}^2)^2} \Bigg)
\end{equation}
\begin{equation}
    I_{2B}^\delta = \int_0^\pi \frac{4\cos^2\theta}{(1-k_A^2\sin^2\theta)^{5/2}}\,d\theta = \frac{1}{3}\Big( I_{2A} + \frac{8 E({k_A^\delta}^2)}{1-{k_A^\delta}^2}\Big)
\end{equation}
\begin{equation}
    I_{3B}^\delta = \int_0^\pi \frac{8\cos^4\theta}{(1-{k_A^\delta}^2\sin^2\theta)^{5/2}}d\theta = 
        \frac{2}{3{k_A^\delta}^2}\Bigg(  (2+{k_A^\delta}^2)I_{2A}^\delta - 8E({k_A^\delta}^2)  \Bigg)
\end{equation}
and the total velocity induced by the cluster is:
\begin{flalign}
    & u_{\rho 0}^\delta = u_{\rho 0A}^\delta -u_{\rho 0B}^\delta+ u_{\rho 0C}^\delta +\bigO(\varepsilon^2)\\
    & u_{z 0}^\delta = u_{z 0A}^\delta -v_{z 0B}^\delta + v_{z 0C}^\delta  +\bigO(\varepsilon^2)
\end{flalign}
\subsubsection{Final formulas}
One can notice that in equations derived in the previous subsection, only $\delta_{0\rho}$ and $\delta_{0z}$ depend on the specific probing point. Other quantities depend on the locations of the centers of interacting clusters exclusively. It means, that they can be computed once for a cluster-cluster interaction. Thus, we express the velocity induced by the nodes of positive circulation, from a given cluster as follows: 
\begin{flalign}
    & u_{\rho 0}^\delta = \alpha_{\rho}\; \delta_{0\rho} + \beta_{\rho} \; \delta_{0z } + \gamma_\rho +\bigO(\varepsilon^2)\\
    & u_{z 0}^\delta = \alpha_{z}\; \delta_{0\rho } + \beta_{z}\; \delta_{0z} +\gamma_z +\bigO(\varepsilon^2)
    \label{urho0}
\end{flalign}
\begin{equation}
    \alpha_\rho = \frac{3\Gamma_+}{4\pi {R_{2A}^\delta}^5} \Bigg(
     (\rho_c-\rho_+)(I_{2B}^\delta-I_{1B}^\delta) + \rho_+(I_{3B}^\delta-I_{2B}^\delta)
    \Bigg) (z_c - z_+)
\end{equation}
\begin{equation}
    \beta_\rho = \frac{\Gamma_+}{4\pi {R_{2A}^\delta}^3} \Bigg(
    I_{1A}^\delta-I_{2A}^\delta + \frac{3}{{R_{2A}^\delta}^2}(z_c-z_+)^2(I_{2B}^\delta-I_{1B}^\delta)
    \Bigg)
\end{equation}
\begin{equation}
    \gamma_\rho = \frac{\Gamma_+}{4\pi {R_{2A}^\delta}^3}
    (z_c-z_+)(I_{1A}^\delta-I_{2A}^\delta)
\end{equation}
\begin{equation}
    \alpha_z = \frac{\Gamma_+}{4\pi {R_{2A}^\delta}^3}\Bigg(
    (I_{2A}^\delta-I_{1A}^\delta) +\frac{3}{{R_{2A}^\delta}^2}\bigg[    
    (\rho_c-\rho_+)^2(I_{1B}^\delta-I_{2B}^\delta)
    -\rho_+\rho_c I_{3B}^\delta
    \bigg]
    \Bigg)
\end{equation}
\begin{equation}
    \beta_z = \frac{3\Gamma_+}{4\pi {R_{2A}^\delta}^5} \Bigg(
    \rho_c (I_{1B}^\delta-I_{2B}^\delta) -\rho_+ I_{1B}^\delta
    \Bigg)(z_c-z_+)
\end{equation}
\begin{equation}
    \gamma_z = \frac{\Gamma_+}{4\pi {R_{2A}^\delta}^3}\bigg( (\rho_+-\rho_c)I_{1A}^\delta +\rho_c I_{2A}^\delta\bigg)
    \label{gamma_z}
\end{equation}
The above equations work also for the negative part of circulation if we replace $\Gamma_+$ with $\Gamma_-$ and $\bo{r_+}$ with $\bo{r_-}$. The total induced velocity is obtained by summing both contributions from all the far enough cells. Contributions from nearby cells are determined naively, according to (\ref{ur_delta}) and (\ref{uz_delta}).
\subsubsection{Test cases}
\begin{figure}[H]
        \centering
        \begin{subfigure}[b]{0.49\textwidth}
            \centering
            {\includegraphics[trim={3.6cm 9.3cm 4cm 9.5cm},clip,width=\textwidth]{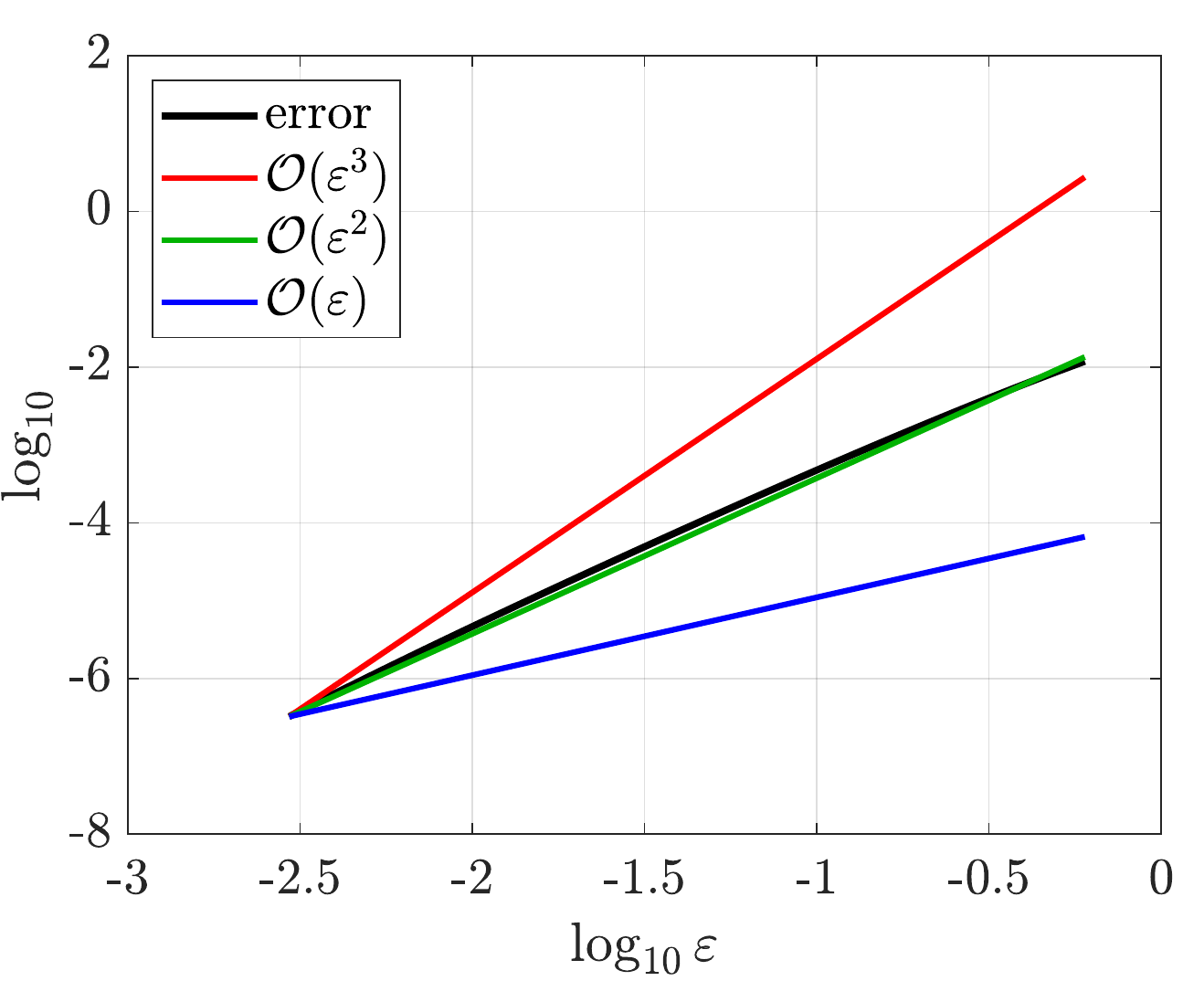}}
            \caption{far away from the $z$ axis}
        \end{subfigure}
    \hfill
        \begin{subfigure}[b]{0.49\textwidth}
            \centering
            {\includegraphics[trim={3.6cm 9.3cm 4cm 9.5cm},clip,width=\textwidth]{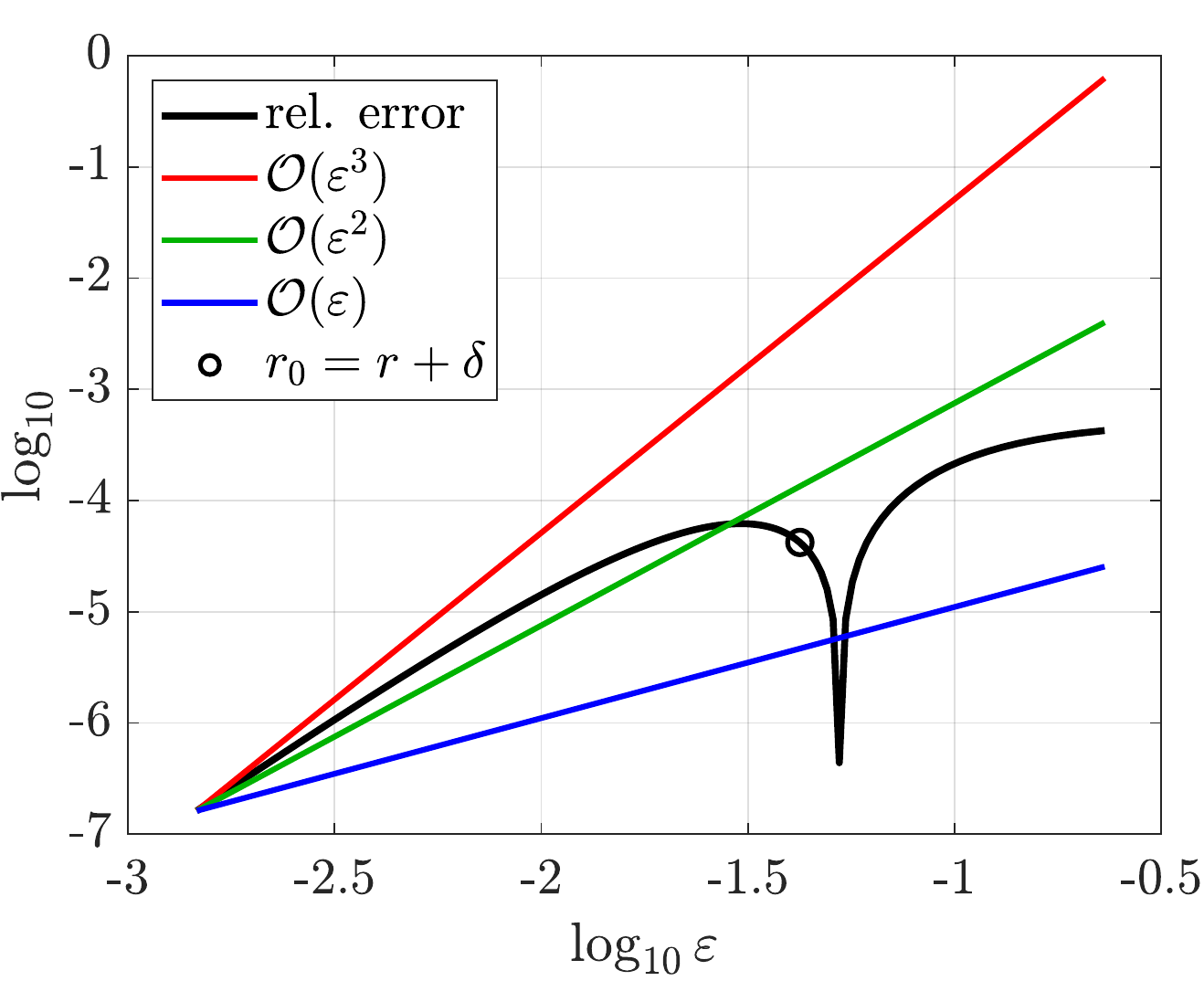}}
            \caption{at the typical distance}
        \end{subfigure}
        \caption{Velocity (absolute) error of the fast method compared with respect to the naive one. Axes are logarithmic (base 10). A circle marks measurement from point at a smoothing distance from the closest source. Therefore, nodes to the right lay in a small viscous region, and to the left, cover the rest of the sheet. On the horizontal axis, we used $\varepsilon$ computed using the left node.}  
        \label{dynnikovaConv}
\end{figure}

\begin{figure}[H]
        \centering
        \begin{subfigure}[b]{0.49\textwidth}
            \centering
            {\includegraphics[trim={3.6cm 9.3cm 4cm 9.5cm},clip,width=\textwidth]{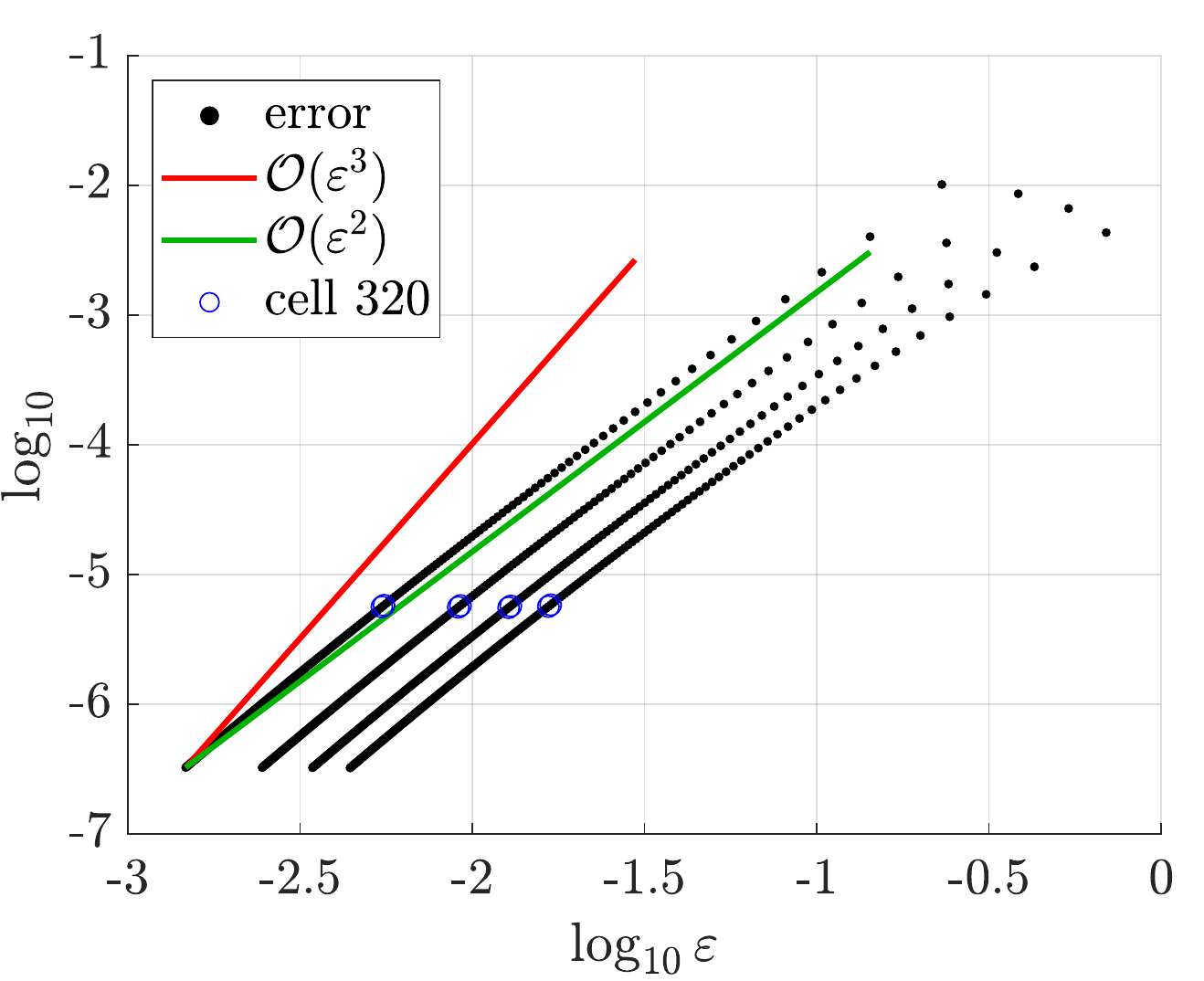}}
            \caption{far away from the $z$ axis}
        \end{subfigure}
    \hfill
        \begin{subfigure}[b]{0.49\textwidth}
            \centering
            {\includegraphics[trim={3.6cm 9.3cm 4cm 9.5cm},clip,width=\textwidth]{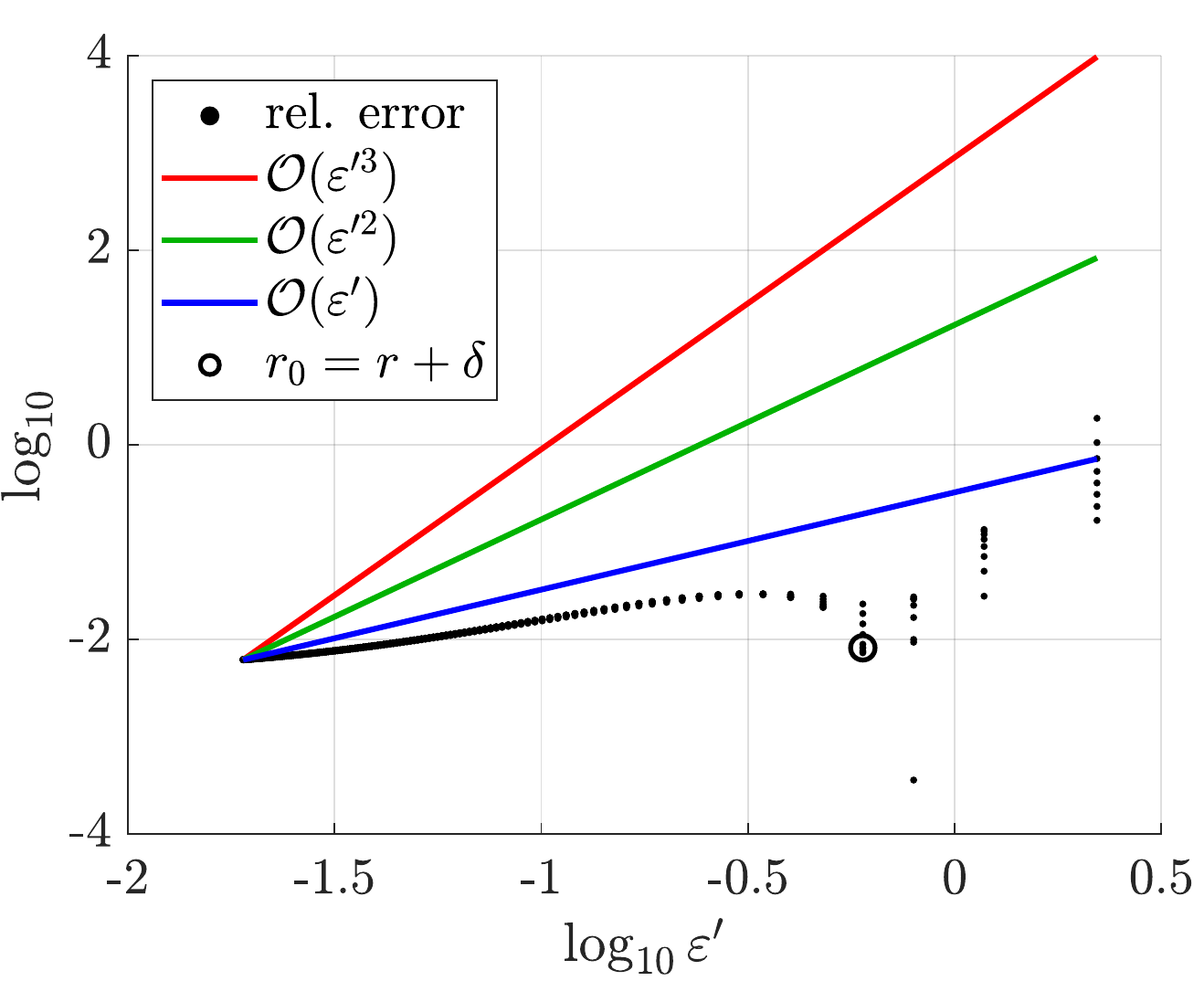}}
            \caption{at the typical distance}
        \end{subfigure}
        \caption{}  
        \label{dynnikovaConv2}
\end{figure}

\begin{figure}[H]
        \centering
        \begin{subfigure}[b]{0.48\textwidth}
            \centering
            {\includegraphics[trim={3.6cm 9.3cm 4cm 9.5cm},clip,width=\textwidth]{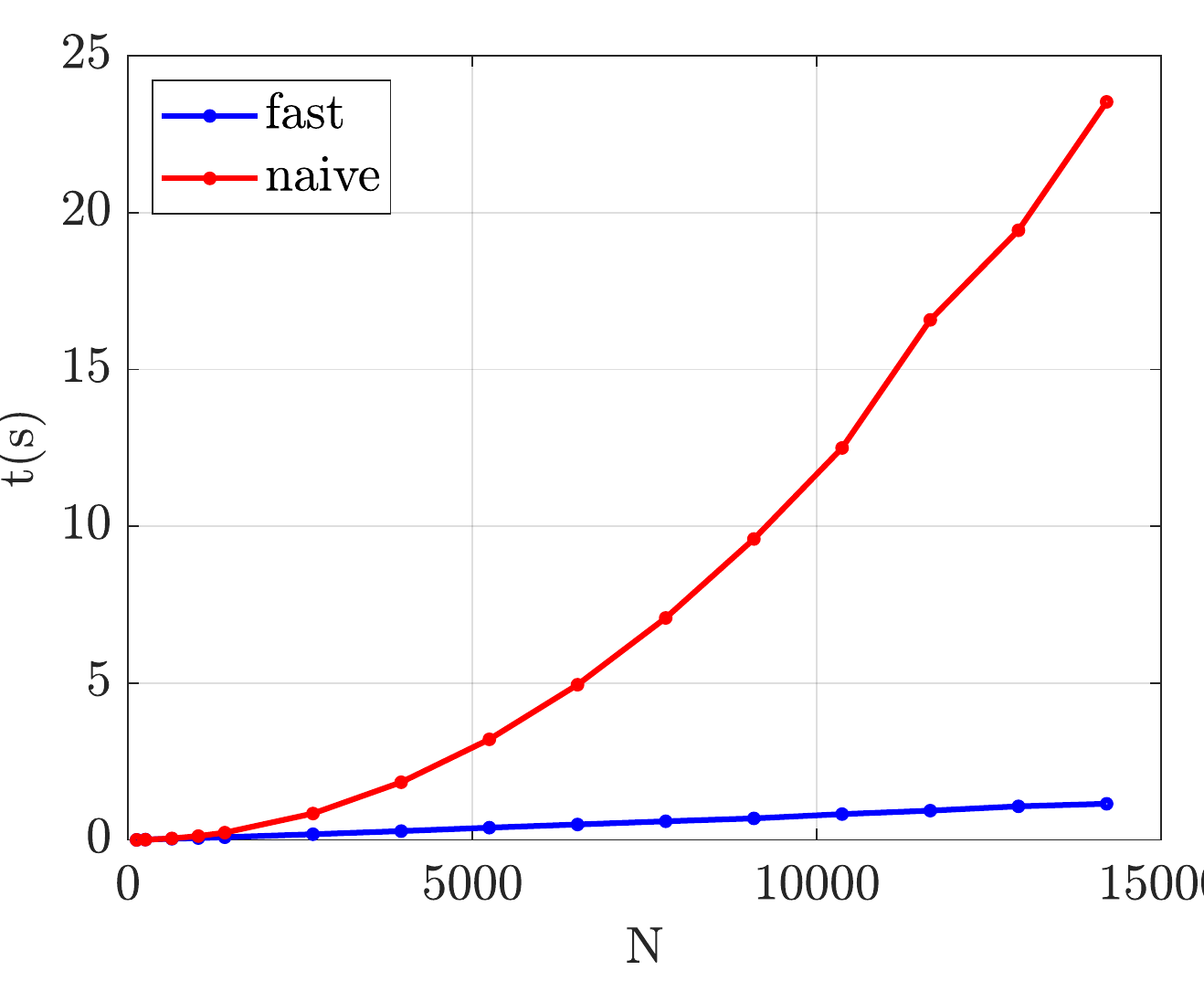}}
            \caption{}
        \end{subfigure}
    \hfill
        \begin{subfigure}[b]{0.51\textwidth}
            \centering
            {\includegraphics[trim={3.6cm 9.3cm 3.2cm 9.3cm},clip,width=\textwidth]{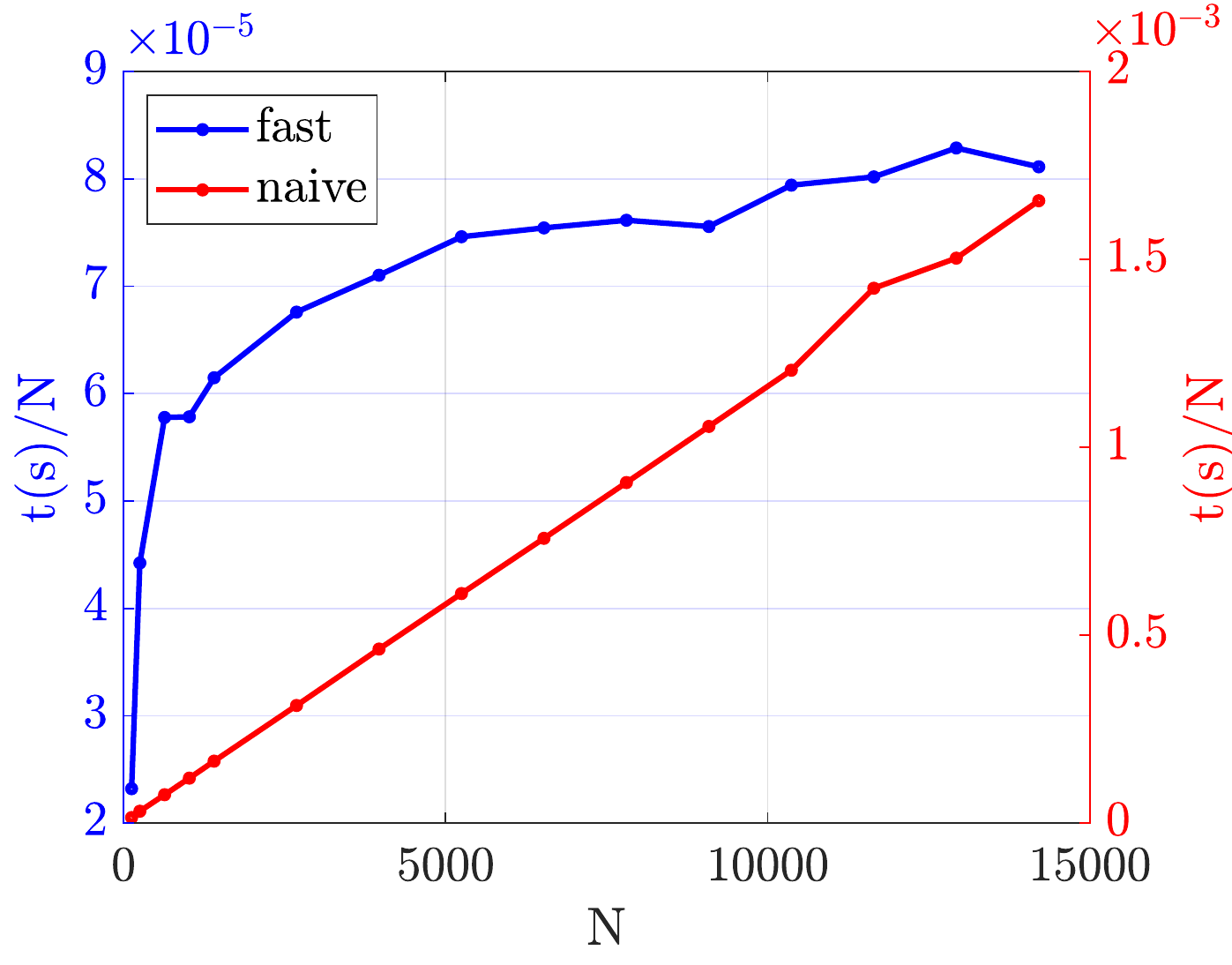}}
            \caption{}
        \end{subfigure}
        \caption{Average processing time for velocity induction against the number of the nodes. Notice different scaling of the axes at (b)}  
        \label{speedTest}
\end{figure}

The above formulas are rather complicated so to be sure of their correctness a few simple test cases were designed. We initialize a straight vortex sheet, with 1024 nodes, stretching radially between $10$ and $10+\pi$ (a) or $0.5$ and $1.5$ (b). We let it have $z=0$ and $\gamma=0$ everywhere. We build a tree such that each bottom cell has one node inside. We pick a non-bottom cell from the left edge, including 8 nodes. We assign $\gamma=2$ and $\gamma=1$ to one at the left cell boundary and one at the right respectively. Additionally, we shift the latter in $z$ to make the cell a square (practice shows that otherwise, we get a special case with $\bigO(\varepsilon^3)$). The exact values are of course arbitrary. We choose (a) to show the general convergence and (b) to have rather typical values. We induce velocity with both, naive and fast method (which we apply no matter of $\varepsilon$), using smoothing parameter $\delta=0.0001$ (a) or $\delta=0.03$ (b). Figure (\ref{dynnikovaConv}) presents the behavior of the error understood as the absolute value of the exact and the approximated velocity.
The close proximity of the smoothed source seems to decrease the convergence rate to linear while the effects of strong curvature increase it to nearly $\bigO(\varepsilon^3)$.\\
To also check terms $\alpha$ and $\beta$ (eq. \ref{urho0}-\ref{gamma_z}), we modify the tree from former cases to have 8 nodes per bottom cell. The rest of the setup stays the same and the results are in the figure (\ref{dynnikovaConv2}).
The value of $\varepsilon$ varies non-monotonically, causing the stripes visible in (a). We depicted probes from a chosen cell with blue circles (that overlay in pairs). We see that although per each value of $\varepsilon$, we get a range of possible errors, the convergence is proper. Case (b) we will use to determine the acceptable value of $\varepsilon$. We plotted the relative error against the estimated value - $\varepsilon'$ (constant in a cell), according to formula \ref{estimateEpsilon}.
The convergence is weaker than linear, due to the influence of the viscous core, which additionally decreases an error for relatively high values of $\varepsilon'$.
For low values of smoothing parameter $\delta$, the error is higher, approaching the clear linear convergence. We can notice, that for relative error to be lower than 1\%, $\varepsilon'< 10^{-1.3} \approx 0.05$. For lower values of $\delta$ it rises up to $\varepsilon'< 10^{-1.5} \approx 0.03$ \\ \\
Finally, we will test the efficiency of the method, by measuring the time of computing the induced velocity. For that purpose, we used an initial condition, with $\gamma = \cos(s)$ and 128 nodes. We computed the velocity five times per each number of nodes (without taking the actual step), took the average and then refined the discretization. We used $\delta = 0.03$ and $\varepsilon'=0.05$, even in the fast method, some of the interactions were handled naively. For processing we used one thread, not to disrupt the measurements with parallelization performance. We compare the results for the fast method with the naive approach in figure (\ref{speedTest})
The algorithm seems to reduce $N^2$ computational complexity to $N\log N$.

    \subsection{Fast surgery}
        
The constructed tree can also easily be used to highly optimize the process of surgery, reducing its complexity from $N^2$ to $N\log N$.
The idea is similar to the case of velocity induction, although we determine which cells are far away with a different criterion. We start by creating a list of centers of the segments. Then, we treat them as nodes and build a tree. Each bottom cell, we test with other cells (starting with the mother cell) for a long enough relative distance. We define it as the shortest line connecting the boundaries of the cells. In other words, for cells $A$ and $B$, having boundaries at $z_0^A$, $z_1^A$ and $z_0^B$, $z_1^B$ we can define a one-dimensional distance:
\begin{equation}
d_z(A,B) =
\left\{ 
  \begin{array}{ c l }
    0 & \quad \textrm{if } (z_0^A - z_1^B)<0 \textrm{ and } (z_1^A - z_0^B)<0 \\
    \\
    \min(|z_0^A - z_1^B|, |z_1^A - z_0^B|)                 & \quad \textrm{otherwise}
  \end{array}
\right. 
\end{equation}
where the first case captures an overlap. We define analogical quantity for $\rho$ coordinate. The shortest line connecting cell boundaries is then:
\begin{equation}
    d(A,B) = \sqrt{d_z(A,B)^2+d_\rho(A,B)^2}
\end{equation}
We can state, using a rather conservative bound, that if for two cells
\begin{equation}
    d(A,B)>2\ttt{ds0}
\end{equation}
then surgery criterion (\ref{surgeryCriterion}) cannot be satisfied. Otherwise we check the same for children of the cell given. If there are no more children (both cells are bottom cells) then we naively check for surgery possibilities.

    \subsection{Notes on parallelization and scaling}
        
\begin{figure}[H]
    \centering
    \includegraphics[trim={3cm 8.75cm 4cm 9cm},clip,width=0.75\textwidth]{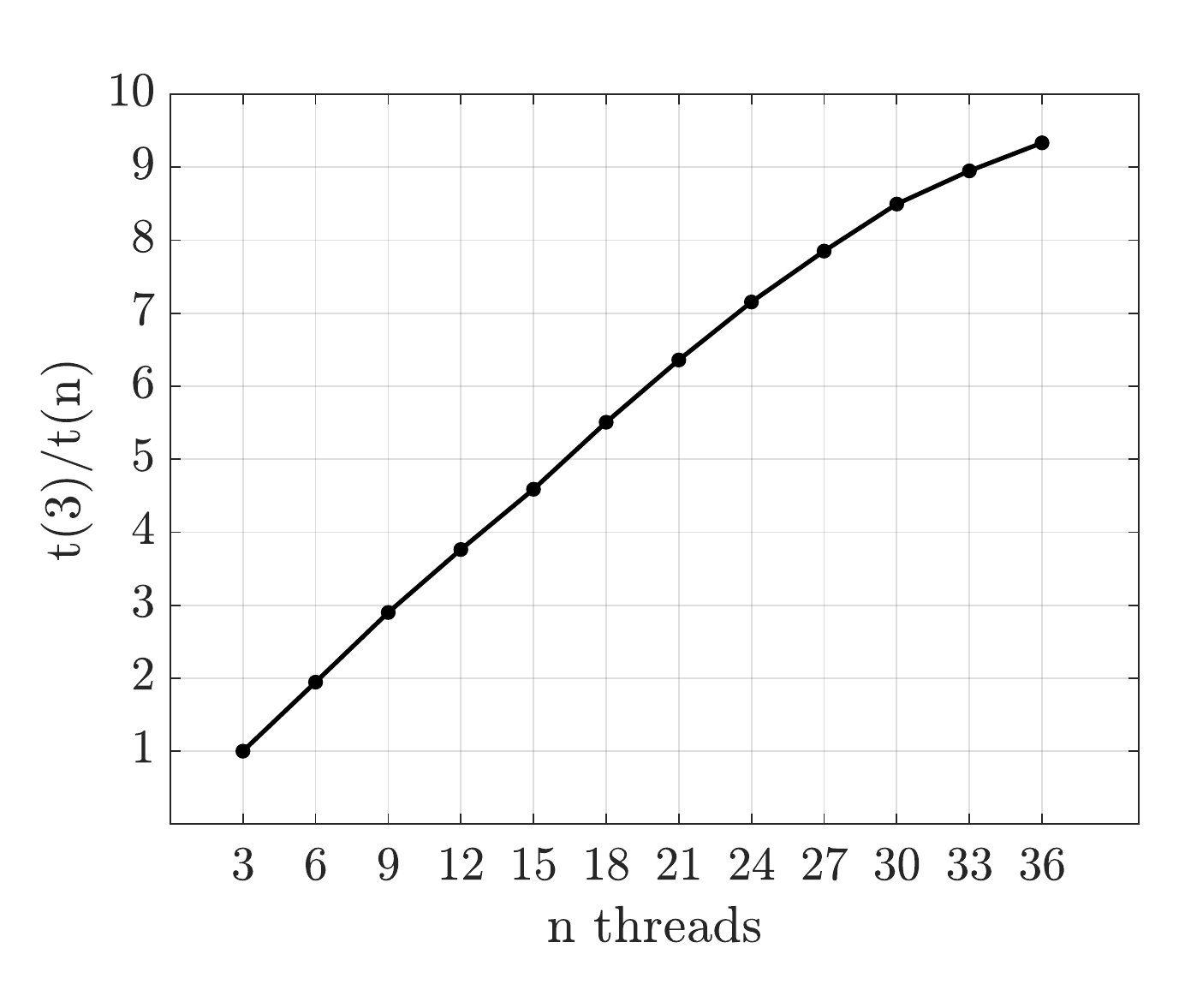}
    \caption{Speedup}
    \label{speedup}
\end{figure}

The code was parallelized with OpenMP. Among all the operations that have to be done every timestep, the velocity induction takes far the most time (difference of a few orders of magnitude). The loop over bottom cells was distributed among the threads so that each thread processes the interactions of its bottom cell with all other cells.
Dynamic scheduling was used because the time required for a given bottom cell might differ a lot. Additionally, tree construction was parallelized in the aspect of distributing nodes among the cells. Computation of all the integrals was done in parallel as well. \\
To test the scaling of the code we used one of the timesteps from $b=1$, $\delta=0.008$ case to initialize the sheet, with around 85 000 nodes. We made 6 complete timesteps with each number of threads $n$, measured the time $t(n)$, and presented the results in figure \ref{speedup}.
Nearly all the simulations were performed on a 36-core node of Rysy cluster in the Interdisciplinary Centre for Mathematical and Computational Modelling UW.

    \subsection{Numerical parameters}
        
\subsubsection{Discretization}
Parameter $\ttt{ds0}$ sets the accuracy of spatial discretization (including surgery), by defining the desired length of segments. We want it to be sufficiently small to solve all the scales that might arise in the simulation. Their size is dependent on the smoothing parameter $\delta$. From experience, we find that
\begin{equation}
    \ttt{ds0} = \frac{\delta}{10}
\end{equation}
is enough. \\
There are two mechanisms that bind the length of the time step $\ttt{dt}$ - movement of the nodes and generation of their circulation density $\gamma$. We estimate the scale of induced velocity as: $U = \Gamma/R$ where $\Gamma$ is the total circulation at a given time. Then, to simulate structures of scale $\delta$ moving with $U$, we let:
\begin{equation}
    \ttt{dt1} = 2\frac{\delta}{\Gamma R}
\end{equation}
where for 4th order Runge-Kutta, 2 seems to be rather a conservative choice. \\
At the same time, to properly solve the production of circulation (especially at the very beginning), we do not want it to increase by more than 10\% between the time steps. This leads to:
\begin{equation}
    \ttt{dt2} = 0.1\frac{\Gamma}{(z_2-z_1)b}
\end{equation}
where $z_2$ and $z_1$ are coordinates of the points at the $z$ axis - the top and the bottom (see formula (\ref{totalCircEvolution})). The time step size is then determined by:
\begin{equation}
    \ttt{dt} = \min(\ttt{dt1},\ttt{dt2})
\end{equation}
The initial condition for $\gamma$ was set with the initial tendency rather than exact zero, i.e.:
\begin{equation}
    \gamma(s)\Big|_{t=0} = 0.1\cos(s)
\end{equation}
\subsubsection{Refinement and surgery}
The exact values of the following parameters were worked out mostly by trial and error as a tradeoff between efficiency and accuracy.\\
The maximal length of a segment before splitting (as multiple of $\ttt{ds0}$):
\begin{equation}
    \kappa_E = 1.25
\end{equation}
The maximal distance between the nodes for merging (as multiple of $\ttt{ds0}$):
\begin{equation}
    \kappa_M = 0.65
\end{equation}
The maximal "normal distance" between the segments for merging (as multiple of $\ttt{ds0}$):
\begin{equation}
    \kappa_{MN} = 0.2
\end{equation}
The maximal cosine of the angle between the segments for merging:
\begin{equation}
    \kappa_{A} = -0.985
\end{equation}
what corresponds to the angle of 170 degrees.
The maximal tree cell diameter:
\begin{equation}
    h = 3\ttt{ds0}
\end{equation}
the algorithm seems to have the best efficiency when there are about 3-5 nodes per cell. \\ \\
We judged the quality of the simulation by monitoring mass conservation, and the evolution of the moment of vorticity, according to formula (\ref{zMoment}). Of course, both were affected by surgery.

\section{Results}
    
We performed a series of simulations with different values of $\delta$. In each scenario, we investigated the case with buoyant vorticity generation, and as a reference - the case with $b=0$. Cases vary in time length due to different behavior of the system and increasing computational complexity.
\subsection{The general evolution of the vortex ring}
\begin{figure}[h]
    \centering
    \vspace{-0.5cm}
 \includegraphics[trim={3.5cm 9.5cm 4cm 9.5cm},clip,width=0.5\textwidth]{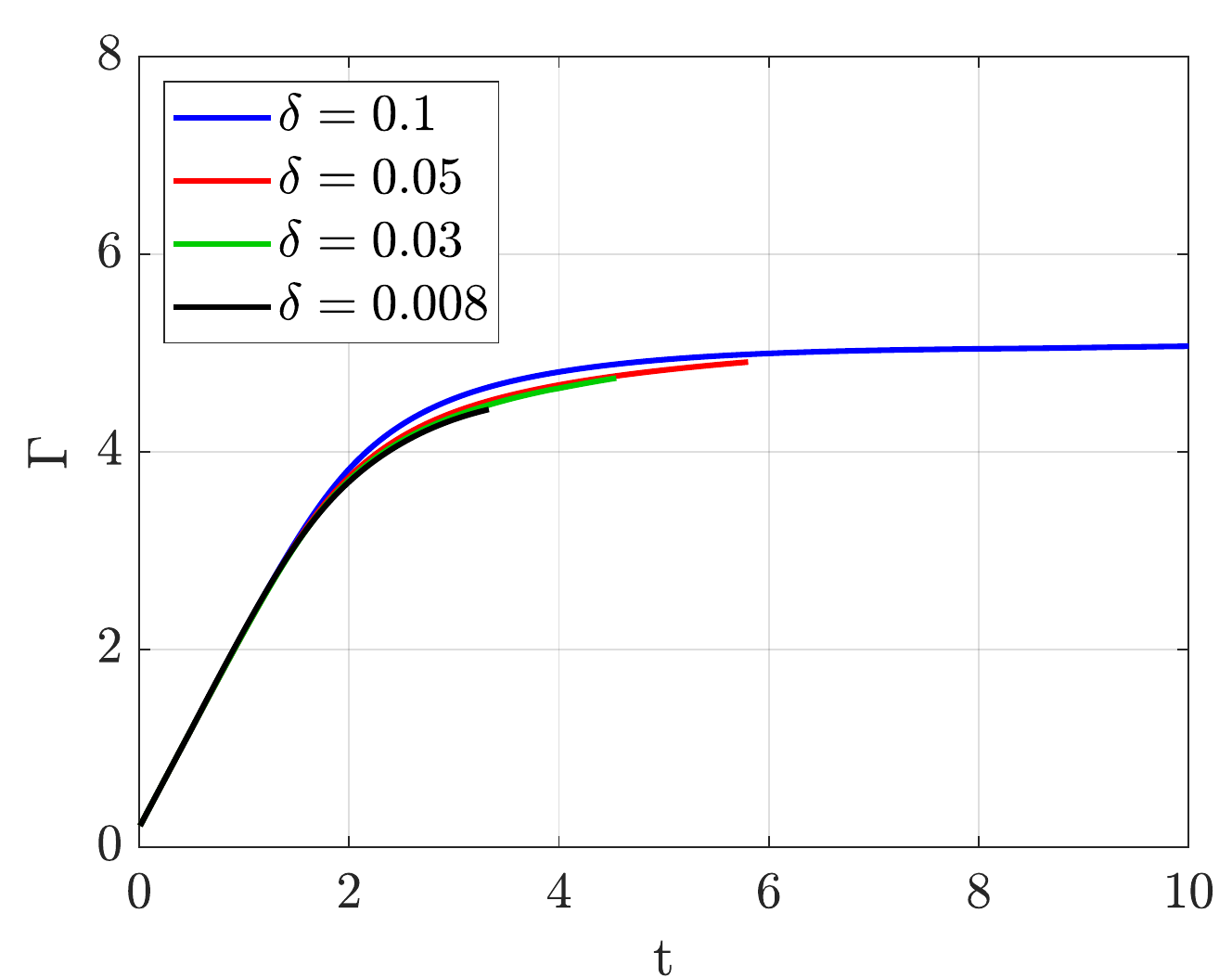}
    \caption{Evolution of total circulation, which reaches 5.07 after 10 time units. The system is generating circulation till the limiting value of around 5.07.}
    \label{meanCirc}
\end{figure}

Although the detailed behavior of the system is strongly $\delta$-dependent, as will be shown in further sections, the evolution of integral quantities seems to be more stable. For all investigated values of $\delta$, the system is rising in a similar manner. This also applies to radial expansion and accumulation of total circulation. All tendencies were presented in figures (\ref{meanCoords}) and (\ref{meanCirc}). Relative differences of $\langle z \rangle$ (volume-averaged coordinate) from the most accurate case ($\delta=0.008$), at its last step, were: 5.2\%, 2.5\%, 1.6\%, for decreasing $\delta$ respectively. In case of $\langle \rho \rangle$: 0.6\%, 1.1\%, 0.8\% and for $\Gamma$: 4.9\%, 1.9\%, 1.0\%, both in the same order. Although $\Gamma$ might not seem to be directly $\delta$-dependend, it depends on the coordinates at the $z$-axis. The bottom coordinate seems to be highly sensitive to the amount of smoothing.
\begin{figure}[H]
        \centering
        \begin{subfigure}[b]{0.49\textwidth}
            \centering
            \includegraphics[trim={4.5cm 9.5cm 4cm 9.5cm},clip,width=\textwidth]{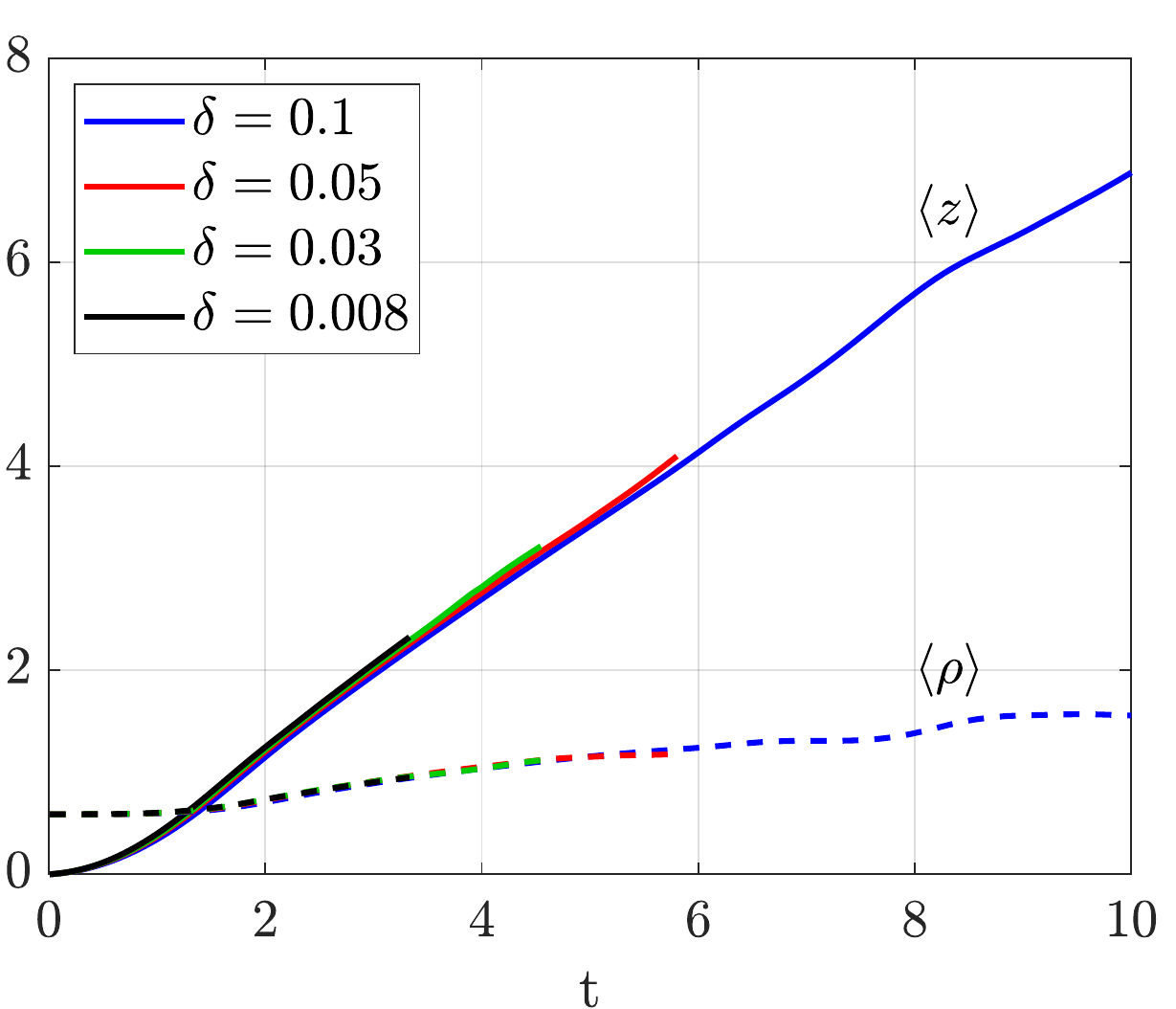}
            \caption{Mean height and radius, case b=1.}
        \end{subfigure}
        \hfill 
            \begin{subfigure}[b]{0.49\textwidth}
            \centering
           \includegraphics[trim={4.5cm 9.5cm 4cm 9.5cm},clip,width=\textwidth]{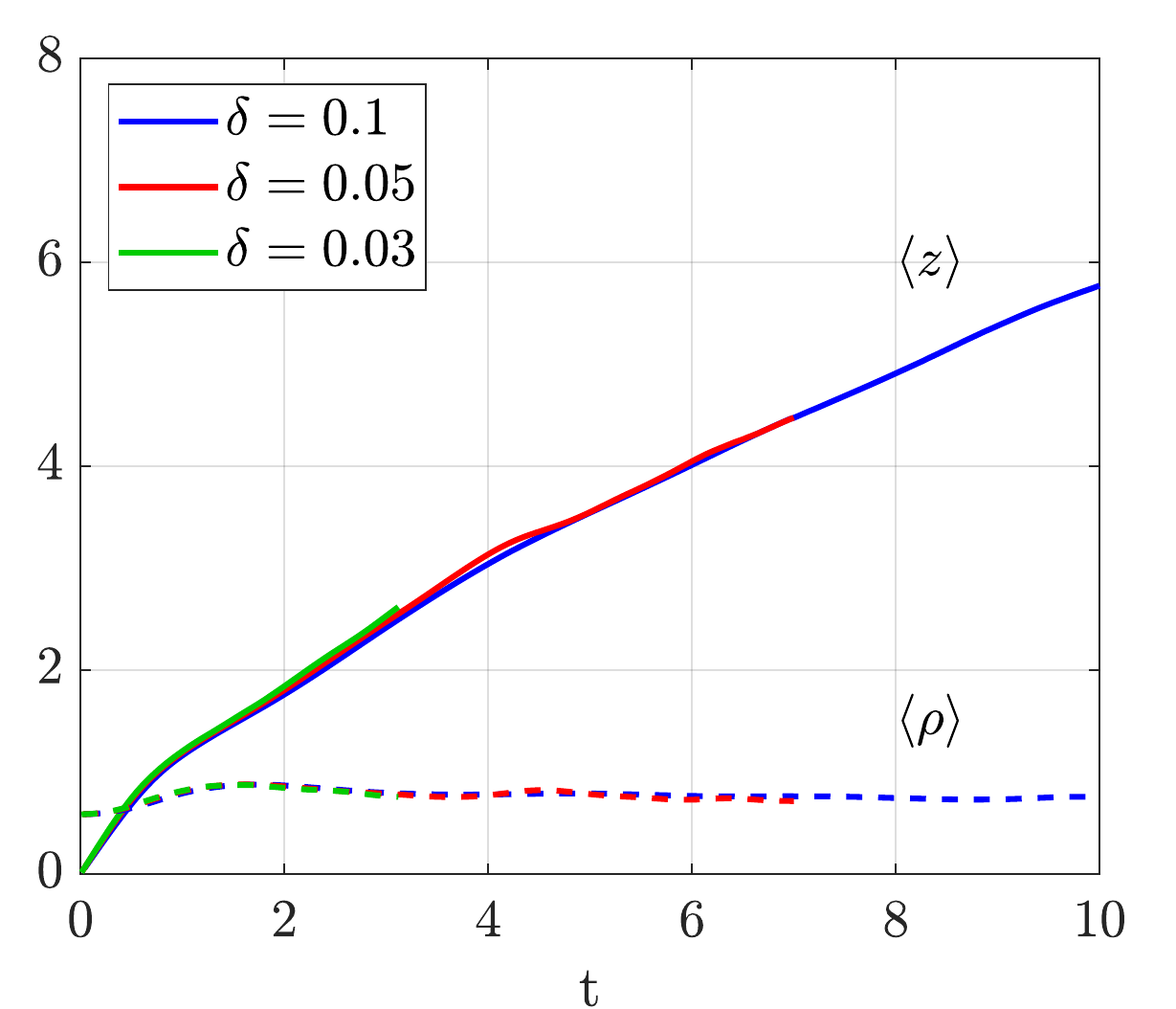}
            \caption{Mean height and radius, case b=0.}
        \end{subfigure}
 \end{figure}
 \begin{figure}\ContinuedFloat
     \centering
            \begin{subfigure}[b]{0.49\textwidth}
            \centering
           \includegraphics[trim={4cm 9.5cm 4cm 10cm},clip,width=\textwidth]{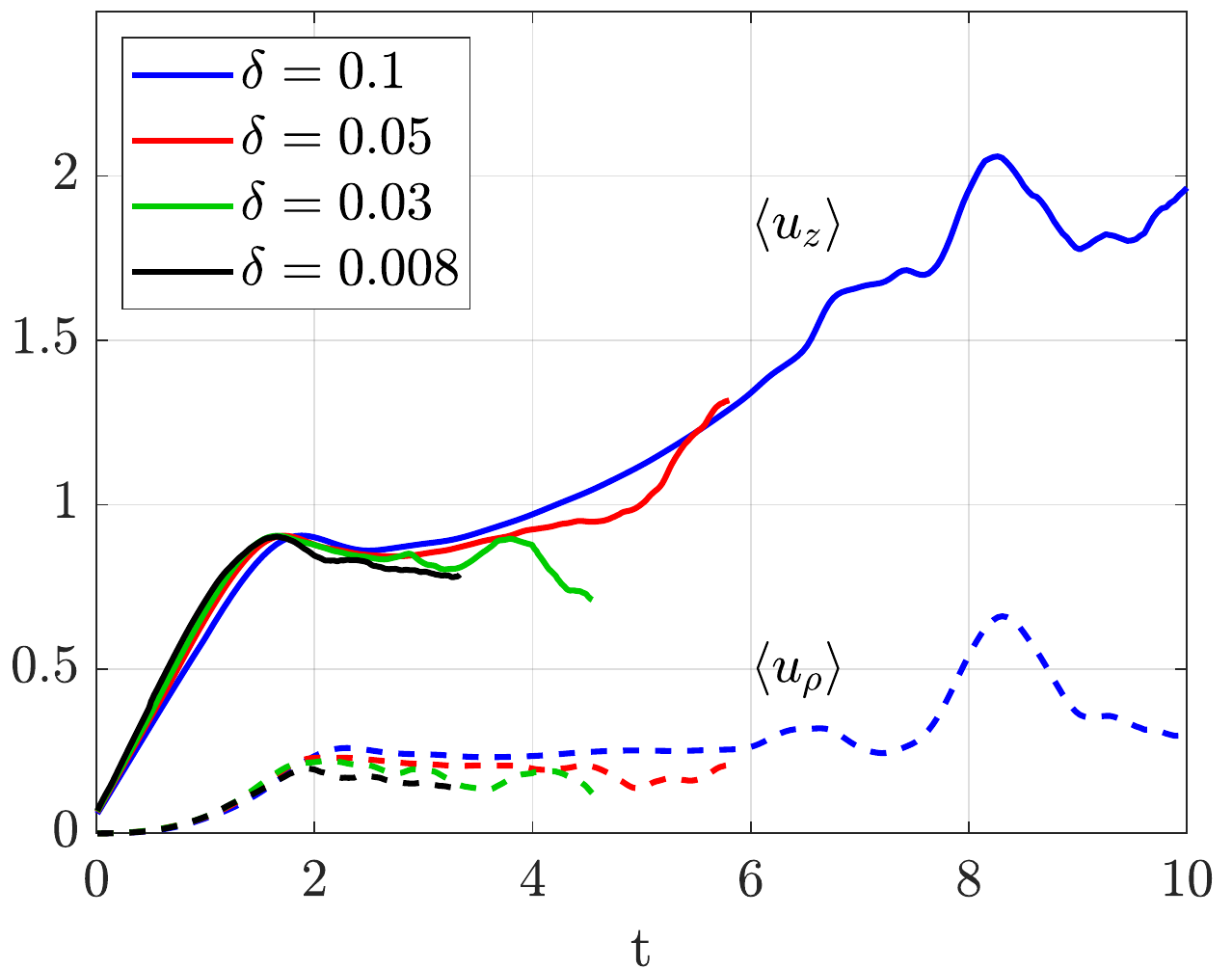}
            \caption{Mean velocities, case b=1.}
        \end{subfigure}
        \hfill 
            \begin{subfigure}[b]{0.49\textwidth}
            \centering
           \includegraphics[trim={4cm 9.5cm 4cm 10cm},clip,width=\textwidth]{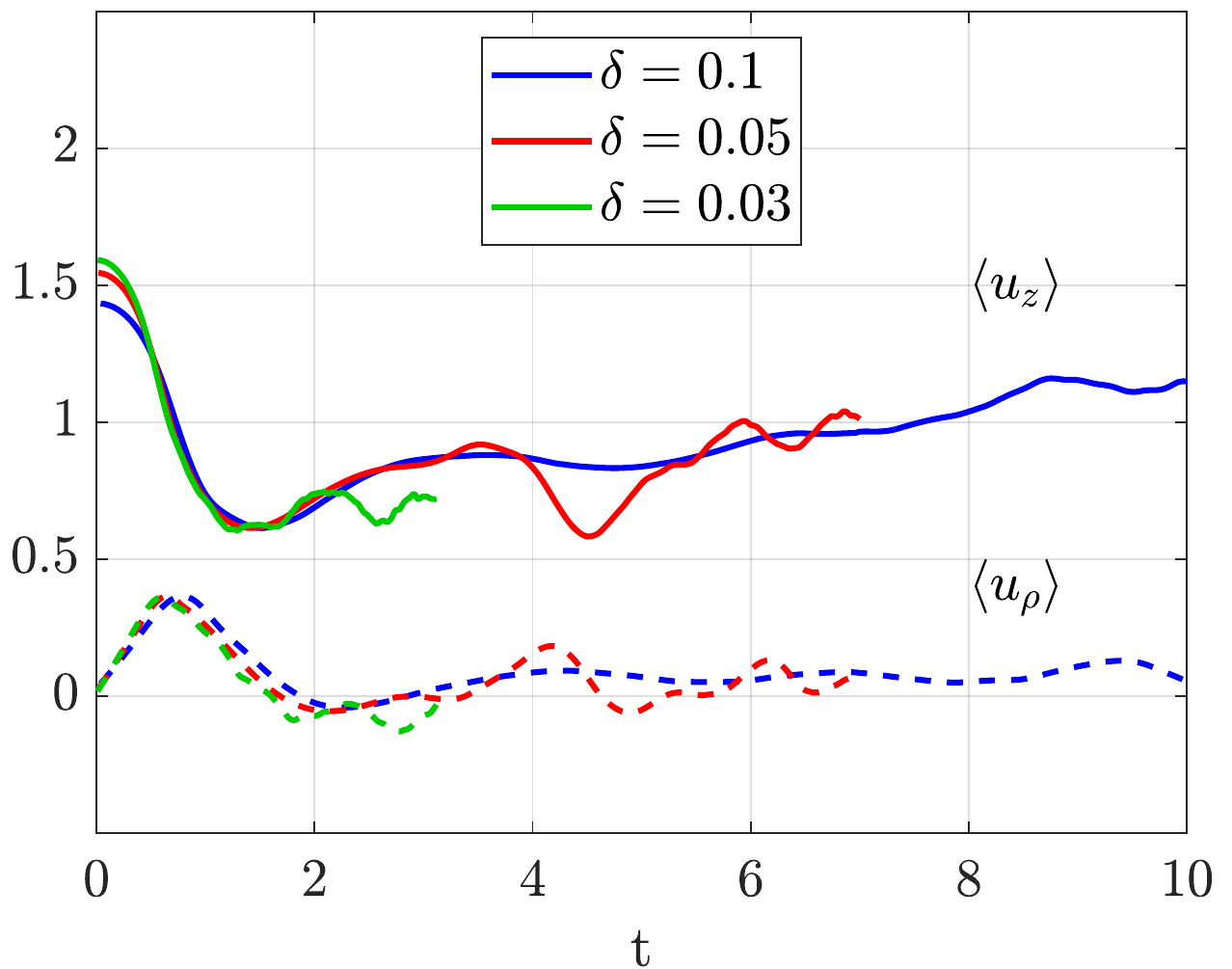}
            \caption{Mean velocities, b=0}
        \end{subfigure}
        \caption{Rising (solid line) and radial expansion (dashed) of the ring. Characterized by volume-mean coordinates and velocities of the region enclosed.}
        \label{meanCoords}
\end{figure} 
\subsection{The effect of buoyancy on the qualitative behavior}
\begin{figure}[H]
\vspace{-3cm}
        \centering
        \begin{subfigure}[b]{0.94\textwidth}
            \centering
            \noindent\makebox[\textwidth]{
            \includegraphics[trim={0.0cm 9.3cm 0.0cm 10.0cm},clip,width=1.6\textwidth]{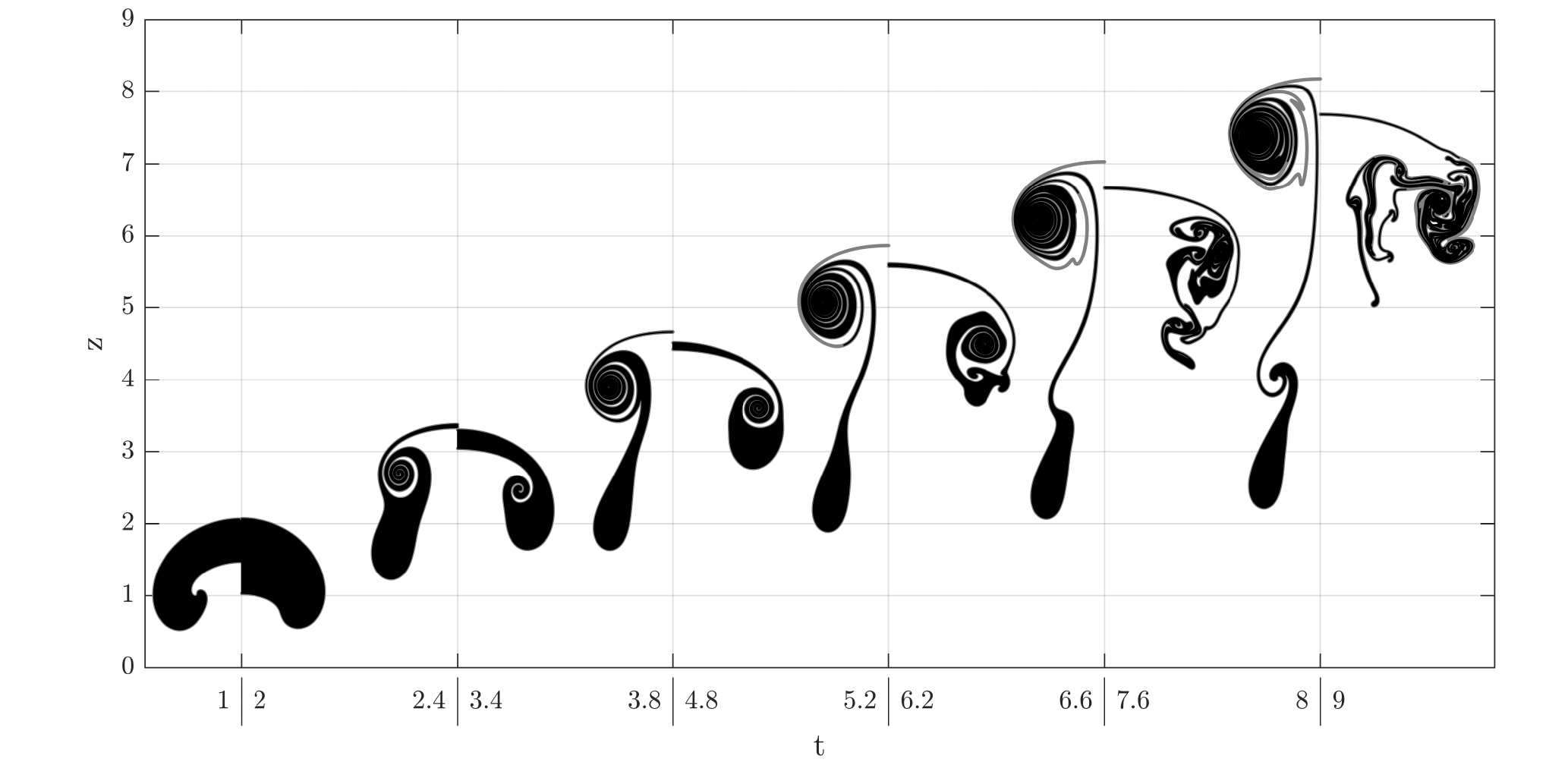}}
            \put(-351,235){(a) $\delta=0.1$}
        \end{subfigure}
        \hfill 
            \begin{subfigure}[b]{0.885\textwidth}
            \centering
            \noindent\makebox[\textwidth]{
            \includegraphics[trim={0.0cm 10.1cm 0.0cm 10.7cm},clip,width=1.6\textwidth]{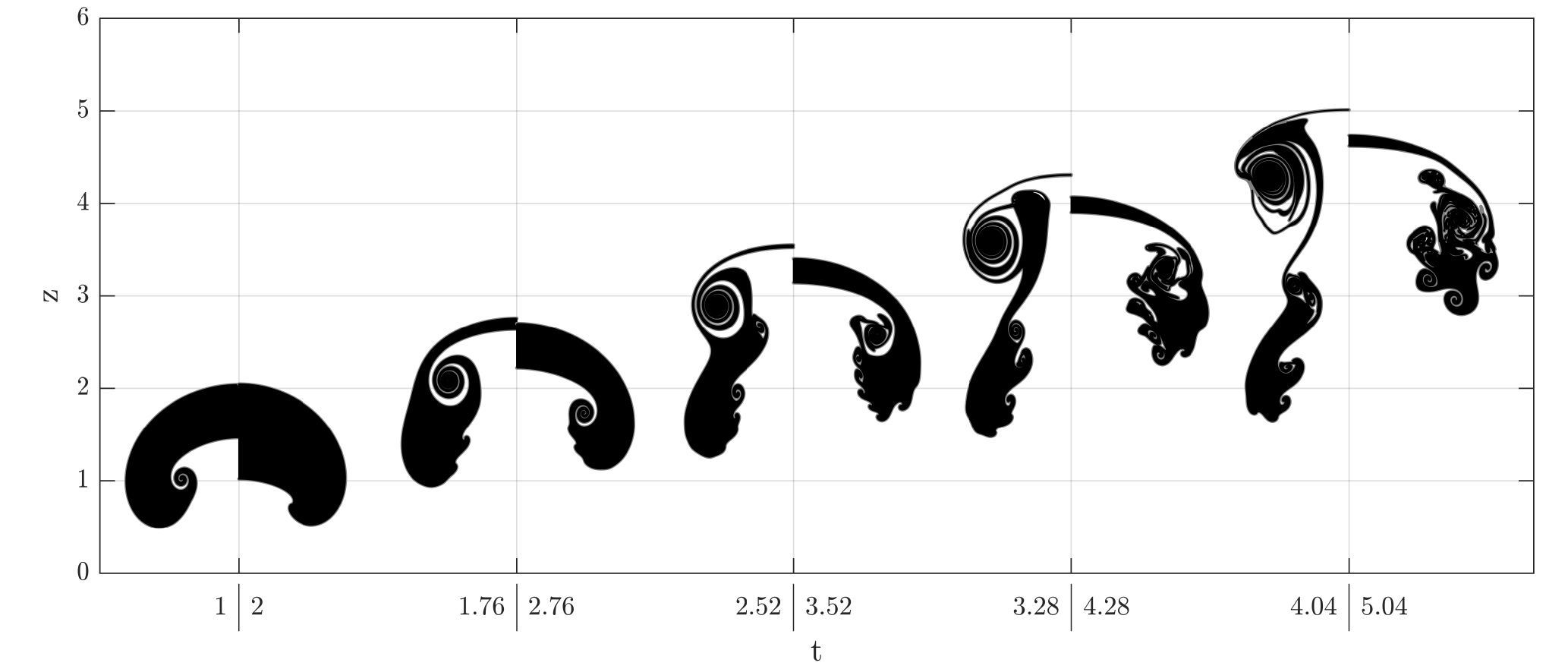}}
            \put(-340,180){(b) $\delta=0.05$}
        \end{subfigure}
        \hfill 
            \begin{subfigure}[b]{0.9\textwidth}
            \centering
            \noindent\makebox[\textwidth]{
            \includegraphics[trim={0.0cm 10.4cm 0.0cm 11.0cm},clip,width=1.6\textwidth]{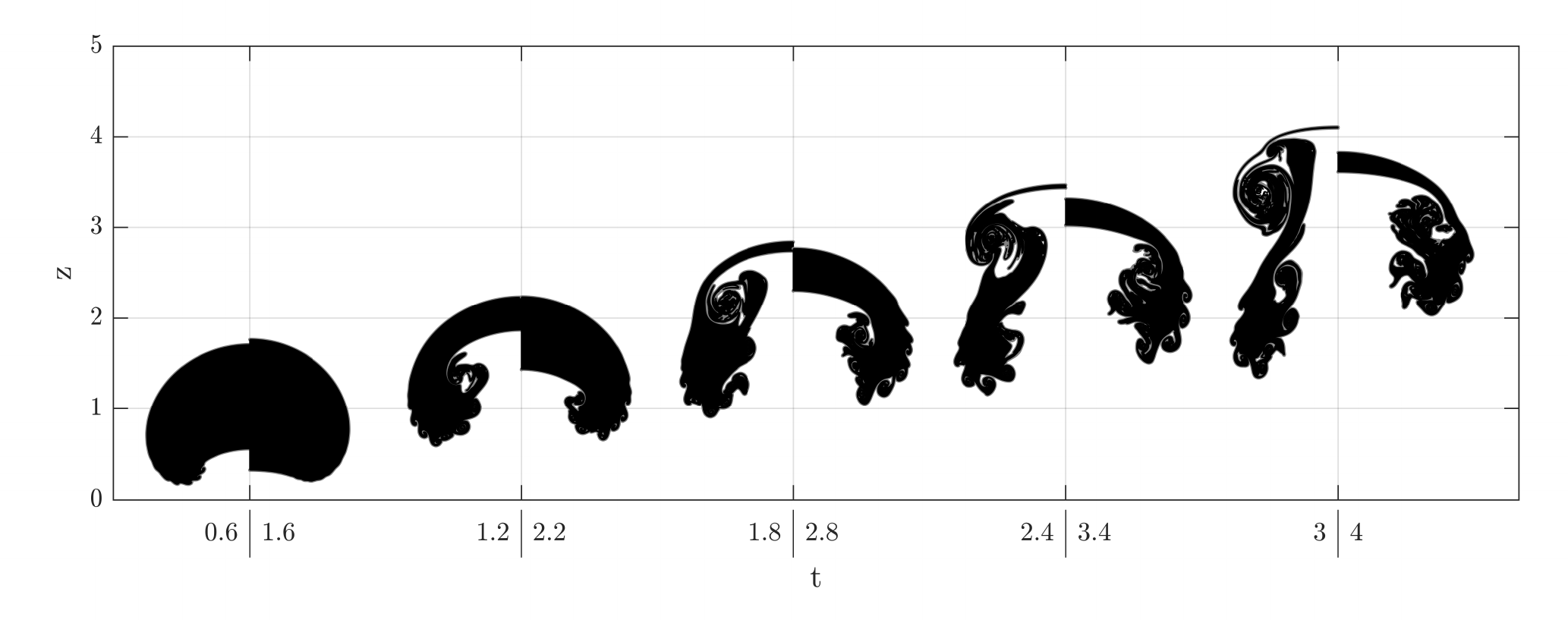}}
            \put(-342,160){(c) $\delta=0.03$}
        \end{subfigure}
        \caption{Development of the vortex ring. Left from the axis - constant vorticity case, right - buoyancy driven. The same convention for time labels. Subfigures represent $\delta=0.1$, $0.05$, and $0.03$ respectively from the top.}
        \label{b01}
    \end{figure}

\begin{figure}[H]
        \centering
        \begin{subfigure}[b]{0.45\textwidth}
            \centering
            \fbox{\includegraphics[trim={5cm 10.5cm 6.5cm 10cm},clip,width=\textwidth]{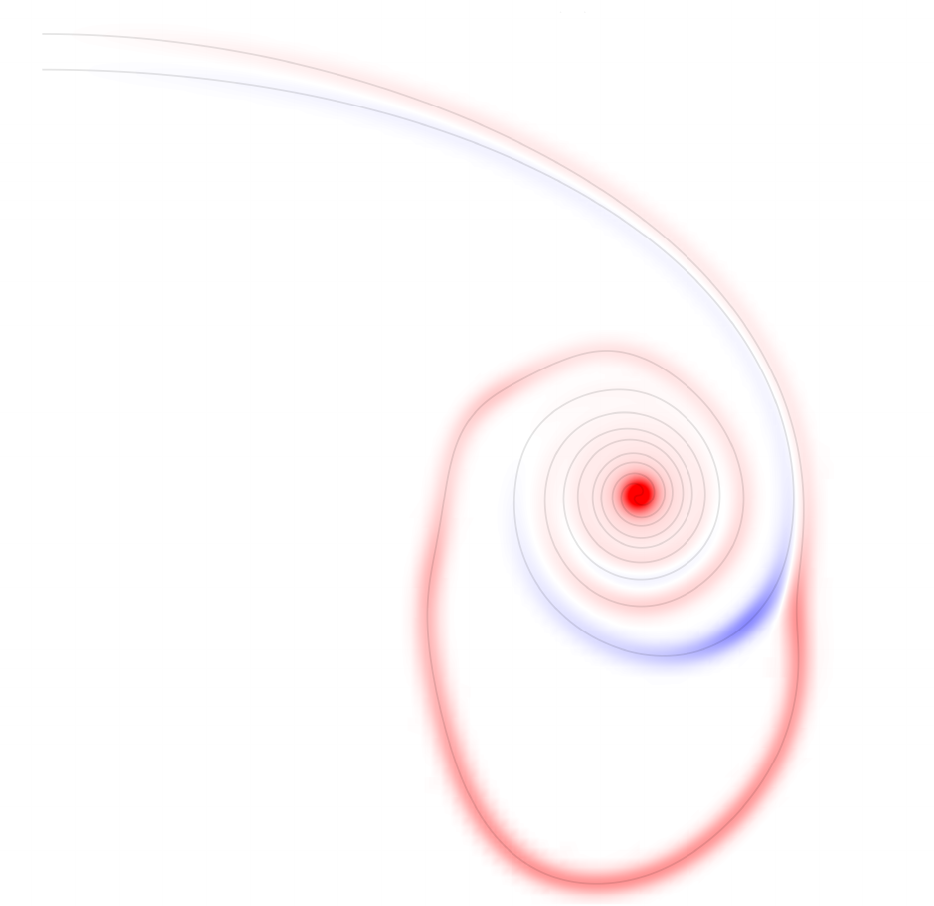}}
            \caption{}
        \end{subfigure}
        \hfill 
            \begin{subfigure}[b]{0.45\textwidth}
            \centering
            \fbox{\includegraphics[trim={5cm 10.5cm 6.5cm 10cm},clip,width=\textwidth]{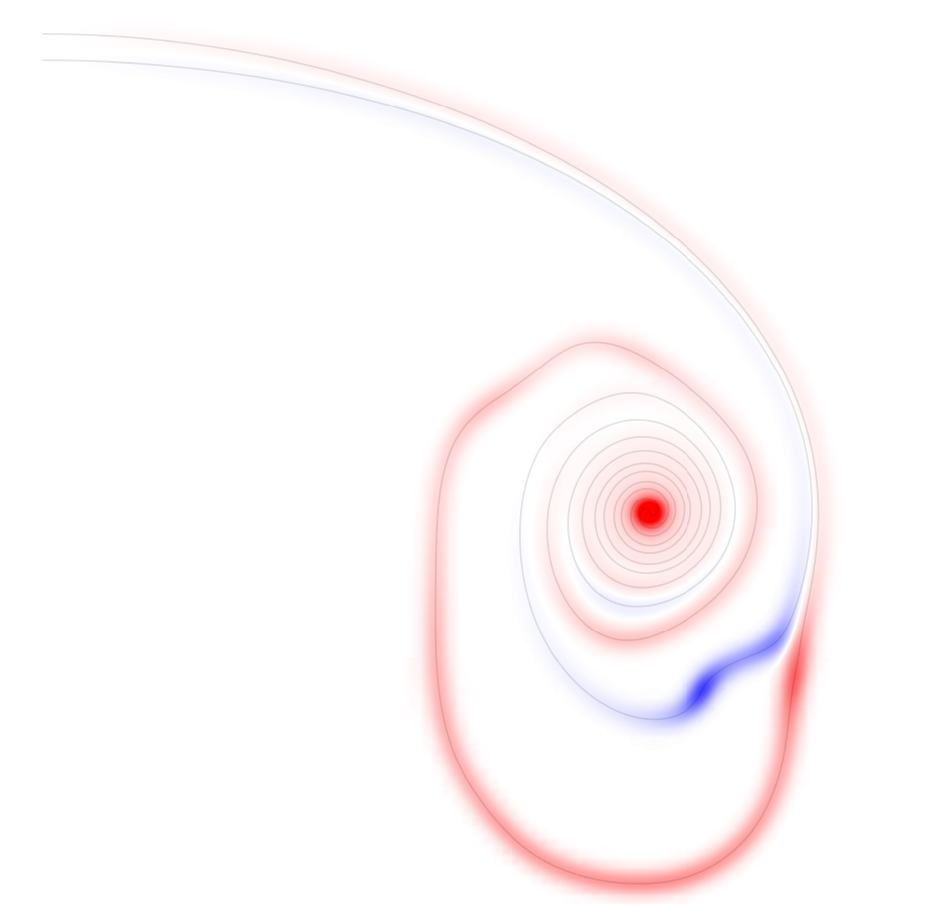}}
            \caption{}
        \end{subfigure}
        \hfill 
        \begin{subfigure}[b]{0.45\textwidth}
            \centering
            \fbox{\includegraphics[trim={5cm 10.5cm 6.5cm 10cm},clip,width=\textwidth]{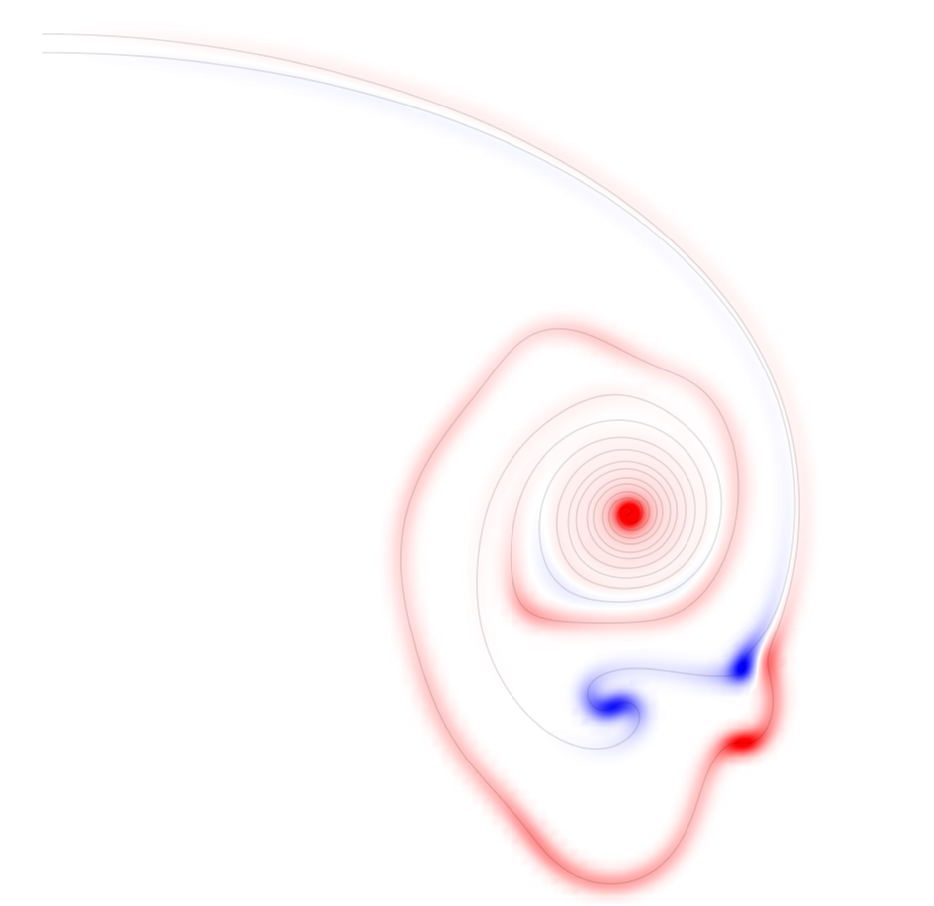}}
            \caption{}
        \end{subfigure}
        \hfill 
           \begin{subfigure}[b]{0.45\textwidth}
            \centering
            \fbox{\includegraphics[trim={5cm 10.5cm 6.5cm 10cm},clip,width=\textwidth]{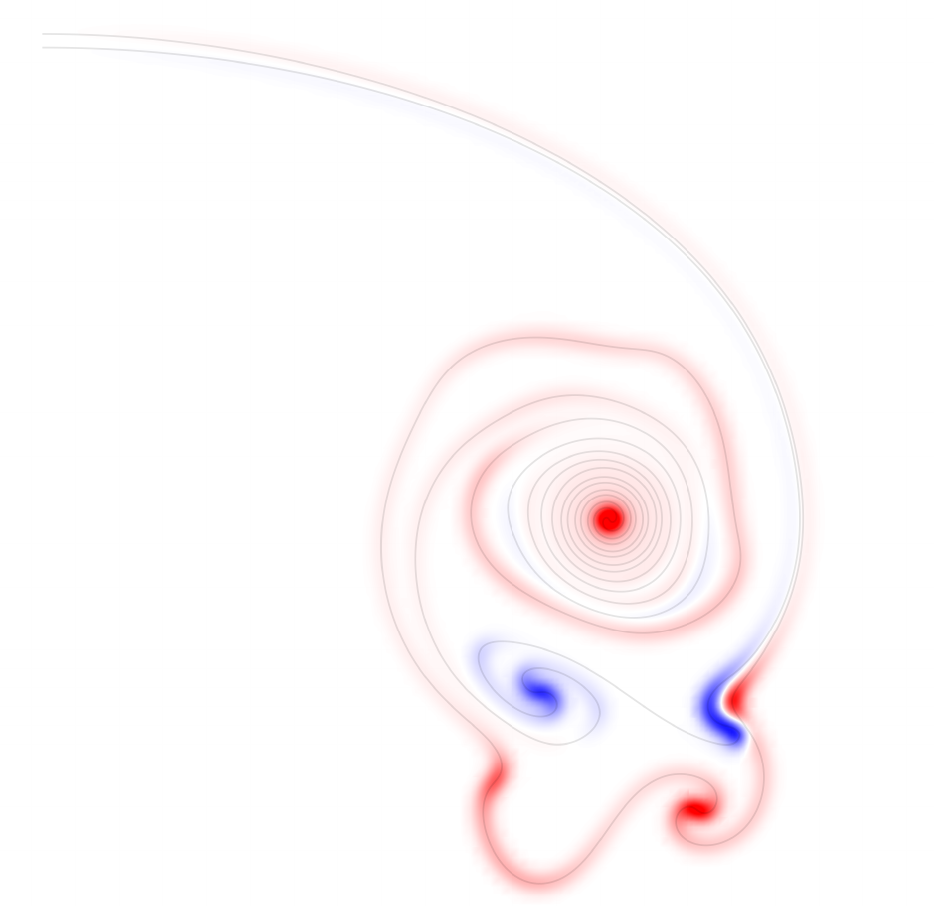}}
            \caption{}
        \end{subfigure}
        \includegraphics[trim={1.5cm 10.7cm 1.5cm 16.7cm},clip,width=\textwidth]{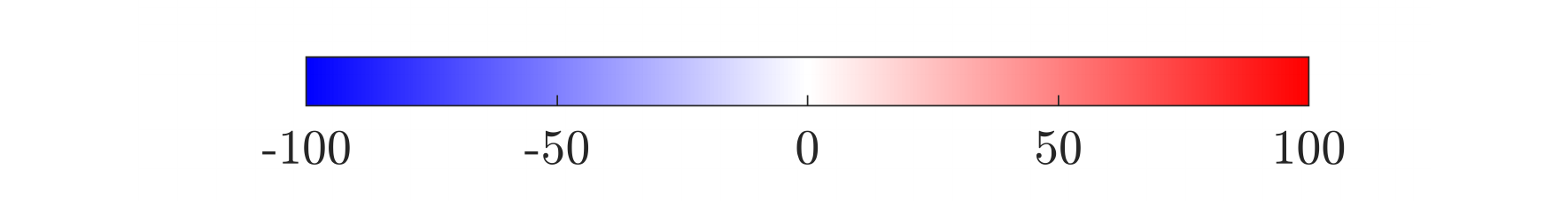}
        \caption{Vorticity distribution in the development of the buoyancy-driven ring instability ($\delta=0.1$).}
        \label{bInstab}
    \end{figure}

In the figure (\ref{b01}) we presented a comparison between the evolution of a buoyancy-driven vortex ring (with $\gamma$ changing according to eq. (\ref{gammaEq})) and a vortex ring with fixed strength. The latter might correspond to a ring generated by shearing stresses in a nozzle (e.g. \cite{understanding}). It was initialized with $\gamma(s) = 2.5\cos(s)$ which gives a total circulation approximately equal to the limiting value from the buoyancy-driven case (see fig. \ref{meanCirc}). At the left-sided labels, we denoted the time of the fixed-strength ring, which is exactly one unit lower than the time of the buoyancy-driven one (right). In this initial unit, the former was nearly in place accumulating vorticity.
\newpage 
\thispagestyle{empty}
We notice that while the fixed-vorticity ring is stable, buoyancy introduces an instability mechanism that disintegrates the ring. The phenomenon is the clearest in case of $\delta=0.1$, where Kelvin-Helmholtz instability does not develop. The collapse of the initial sphere generates a concave region that slowly accumulates negative vorticity. During the development of the ring, this region is attracted to the orbit of the main vortex and as a highly concentrated, counter-rotating vortex, launches the breakdown. The process was depicted in figure (\ref{bInstab}).
\subsection{The ironing of Kelvin-Helmholtz vortices}
\begin{figure}[H]
        \centering
        \begin{subfigure}[b]{0.49\textwidth}
            \centering
            \includegraphics[trim={3.8cm 9.5cm 4.5cm 9.7cm},clip,width=\textwidth]{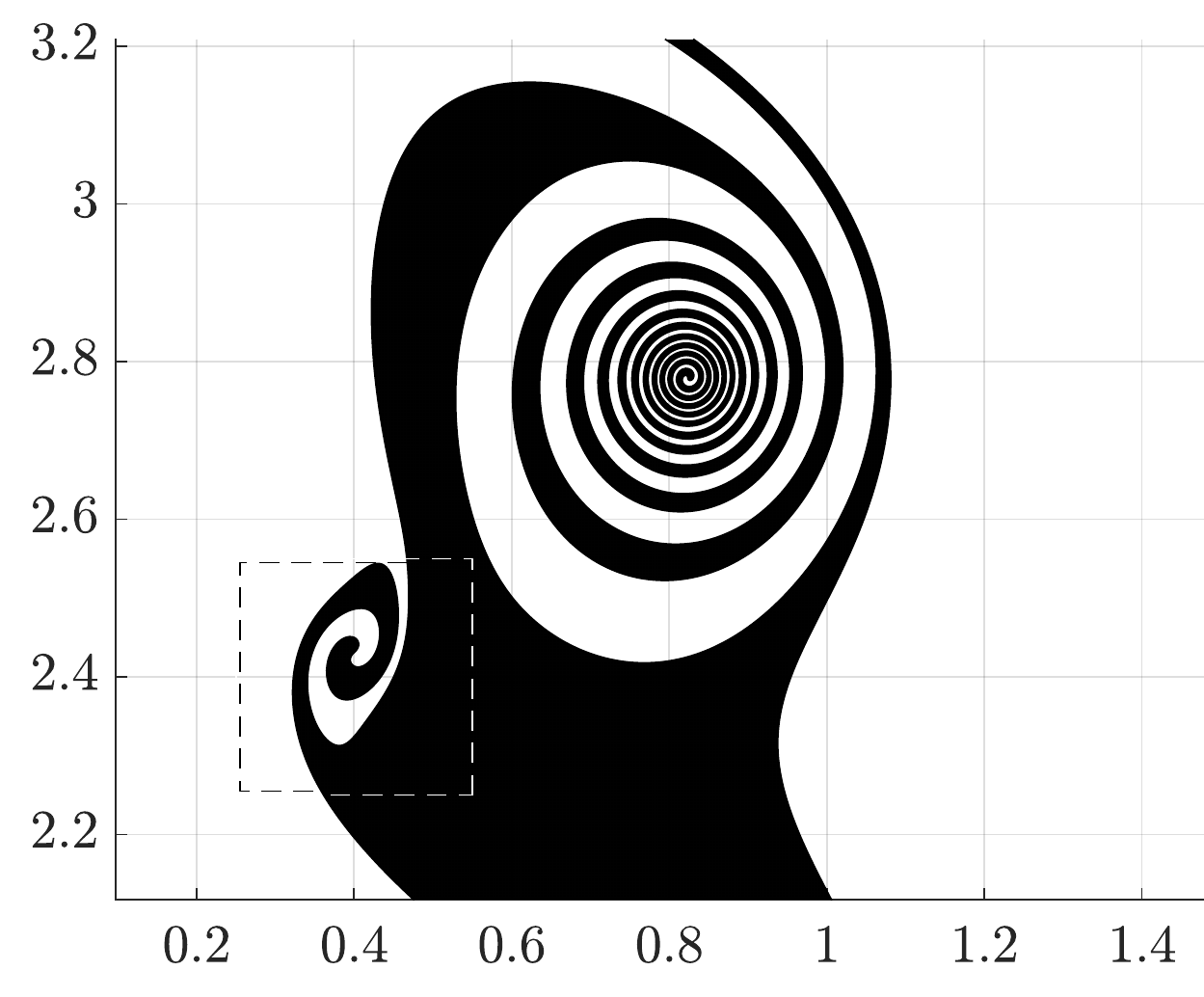}
            \caption{t=2.42}
        \end{subfigure}
        \hfill 
            \begin{subfigure}[b]{0.49\textwidth}
            \centering
          \includegraphics[trim={3.8cm 9.5cm 4.5cm 9.7cm},clip,width=\textwidth]{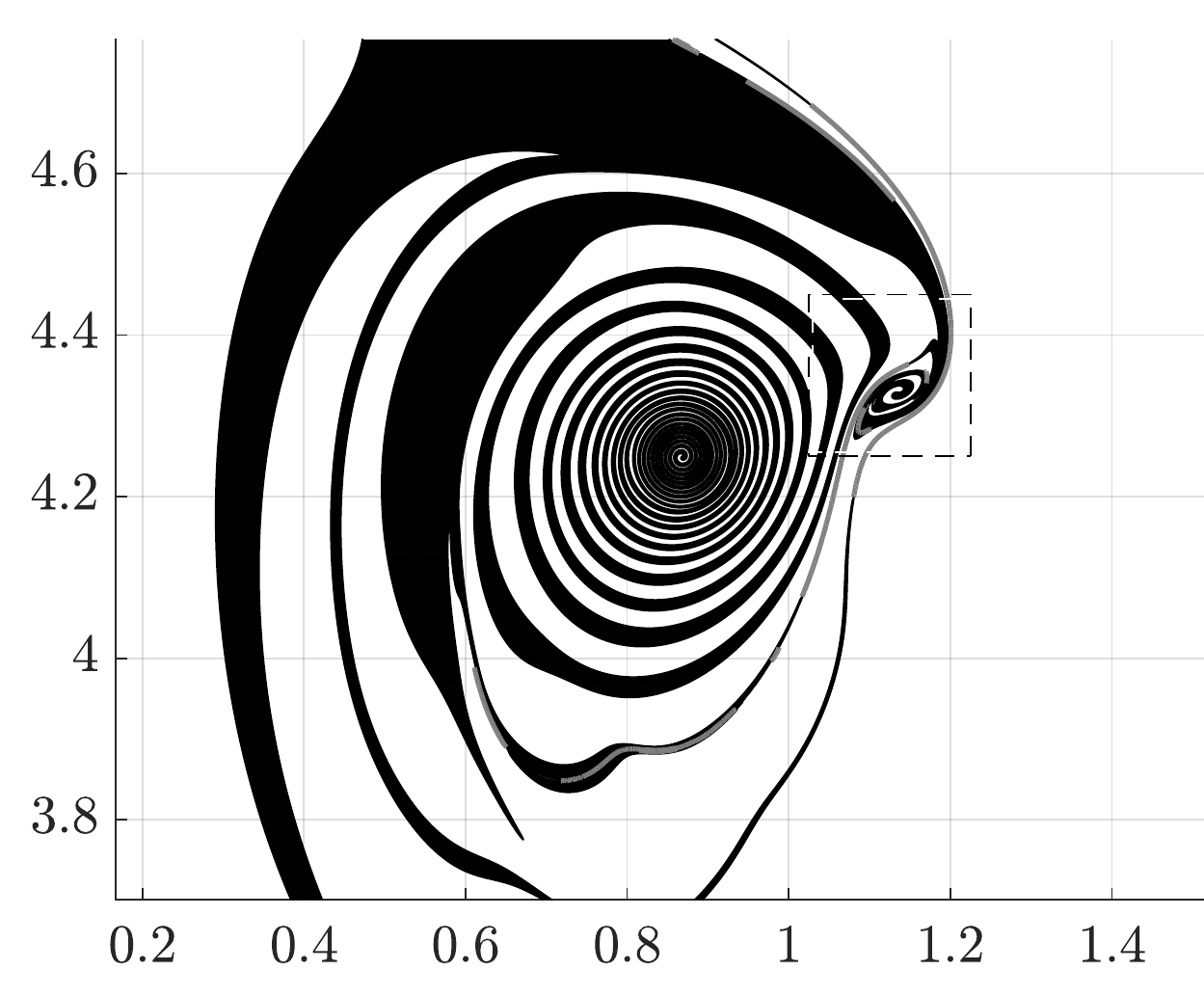}
            \caption{t=4.02}
        \end{subfigure}
        \hfill 
        \begin{subfigure}[b]{0.49\textwidth}
            \centering
          \includegraphics[trim={3.8cm 9.5cm 4.5cm 9.7cm},clip,width=\textwidth]{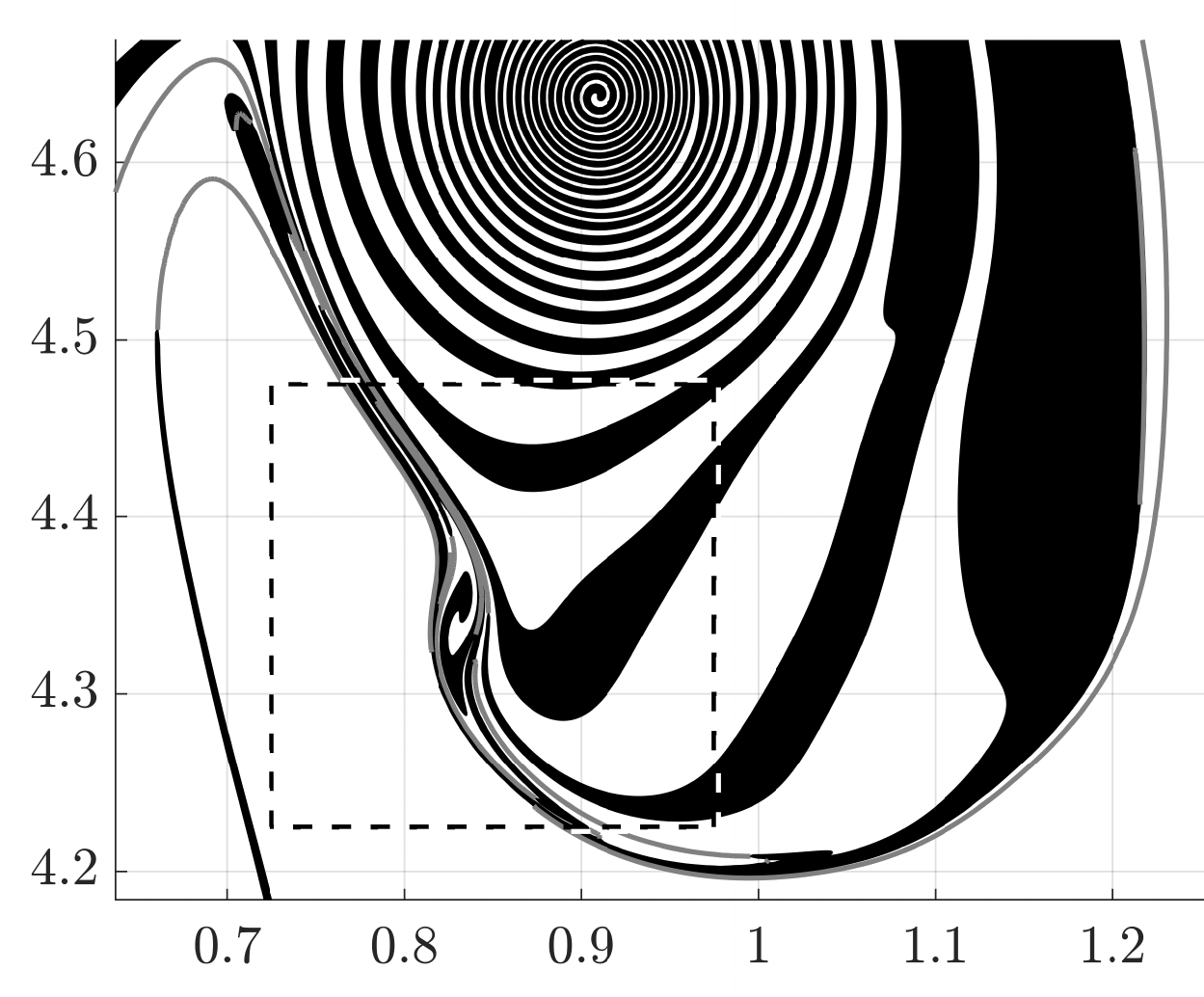}
            \caption{t=4.42}
        \end{subfigure}
        \hfill 
           \begin{subfigure}[b]{0.49\textwidth}
            \centering
          \includegraphics[trim={3.8cm 9.5cm 4.5cm 9.7cm},clip,width=\textwidth]{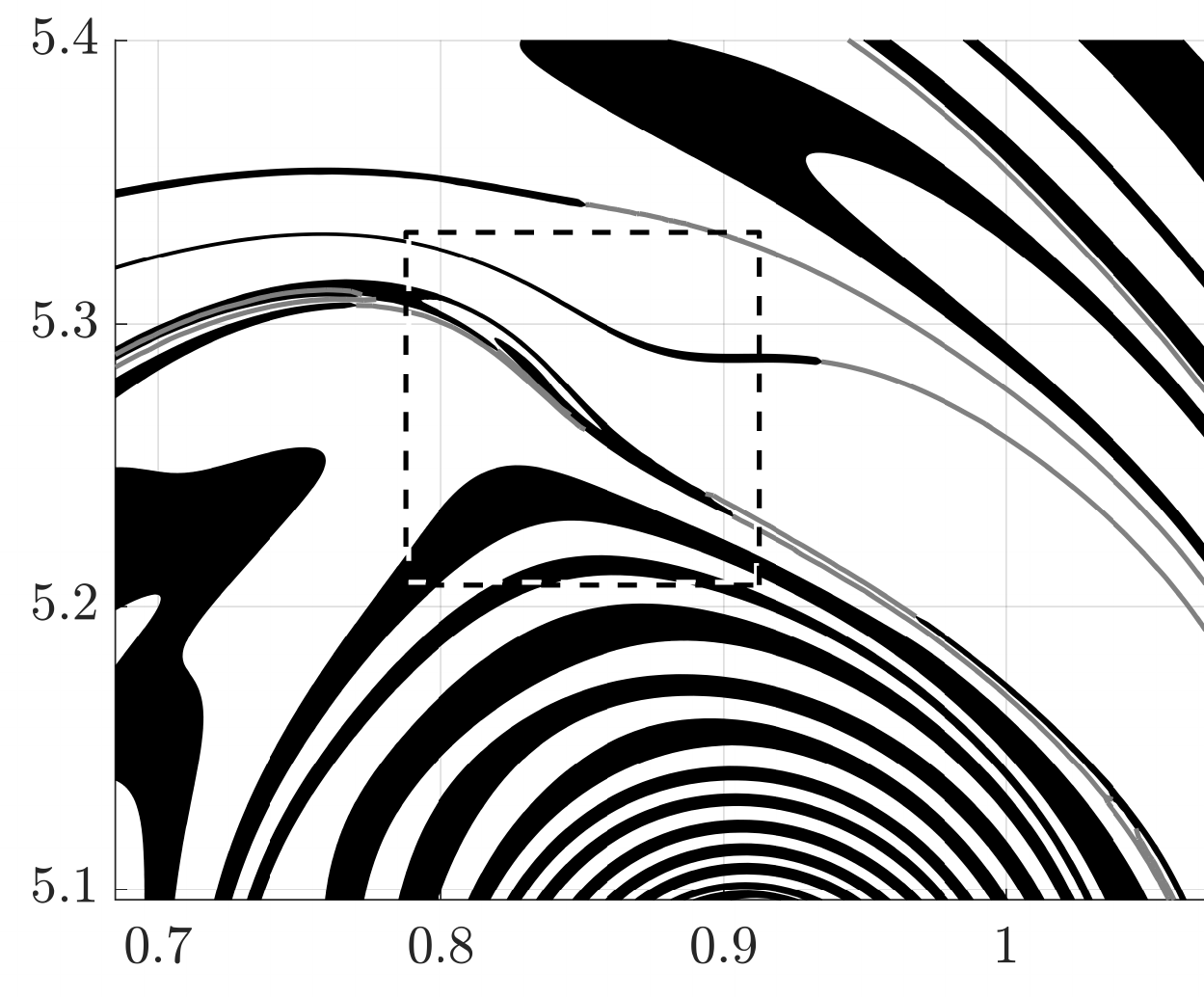}
        \caption{t=4.92}
        \label{ironing_d}
        \end{subfigure}
        \caption{Ironing of the wake vortex into a regular sheet. $\delta=0.05$, $b=0$}  
        \label{ironing}
    \end{figure}

\begin{figure}
    \centering
    \includegraphics[trim={3.8cm 9.5cm 4.5cm 9.7cm},clip,width=0.7\textwidth]{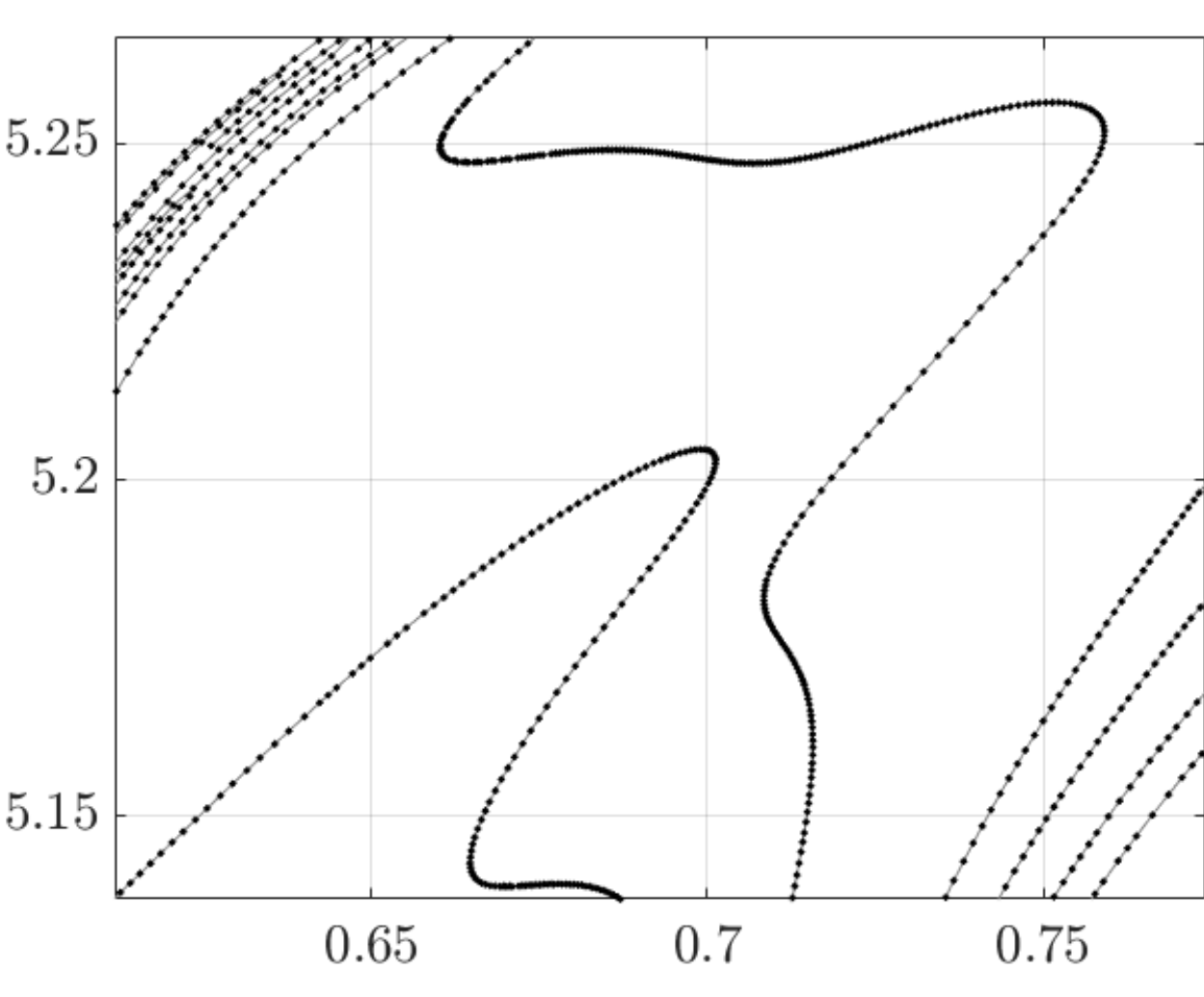}
    \caption{Example of the discretization. Structure visible in fig. (\ref{ironing_d}). The same case and time.}
    \label{ironing_discretization}
\end{figure}

\begin{figure}[H]
        \centering
        \begin{subfigure}[b]{0.49\textwidth}
            \centering
            \includegraphics[trim={5cm 9.5cm 5.5cm 9.5cm},clip,width=\textwidth]{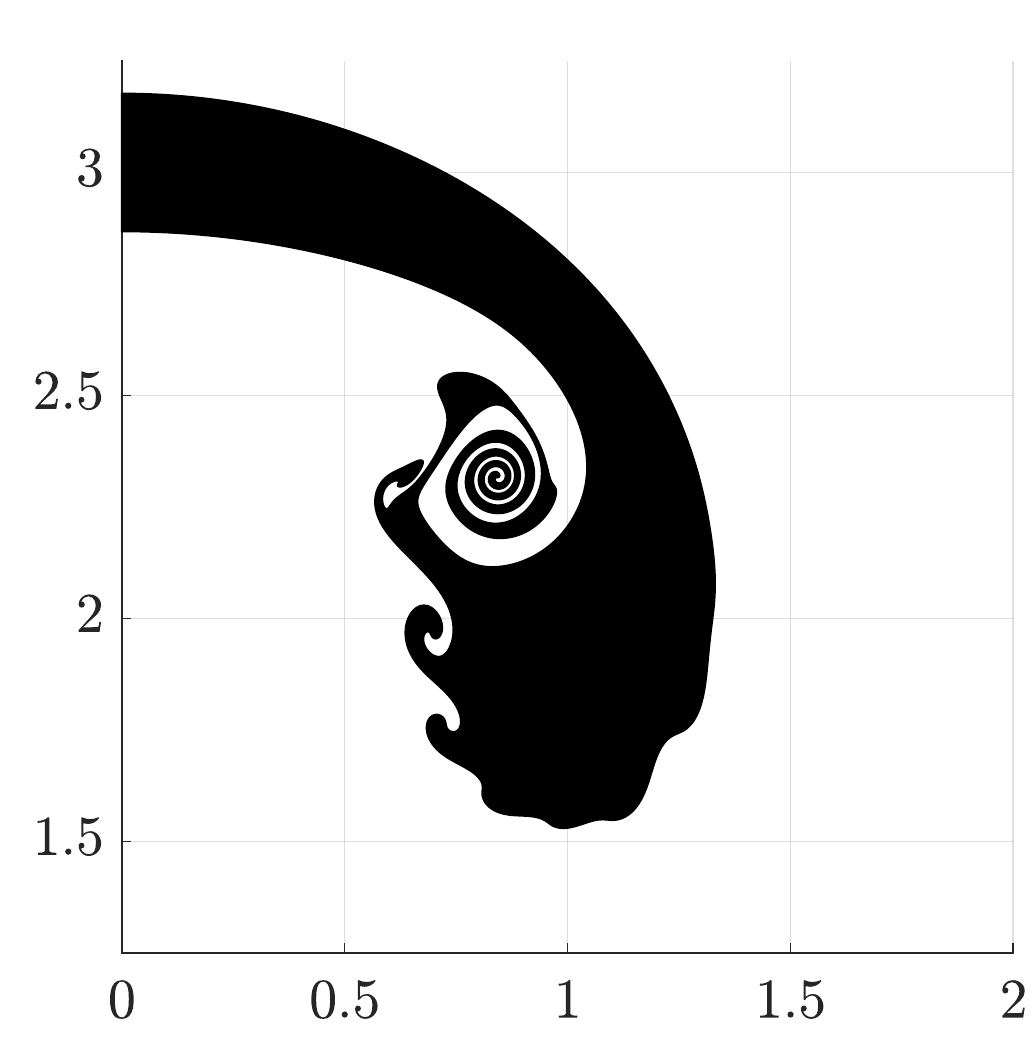}
            \caption{t=3.30}
        \end{subfigure}
        \hfill 
            \begin{subfigure}[b]{0.49\textwidth}
            \centering
            \includegraphics[trim={5cm 9.5cm 5.5cm 9.5cm},clip,width=\textwidth]{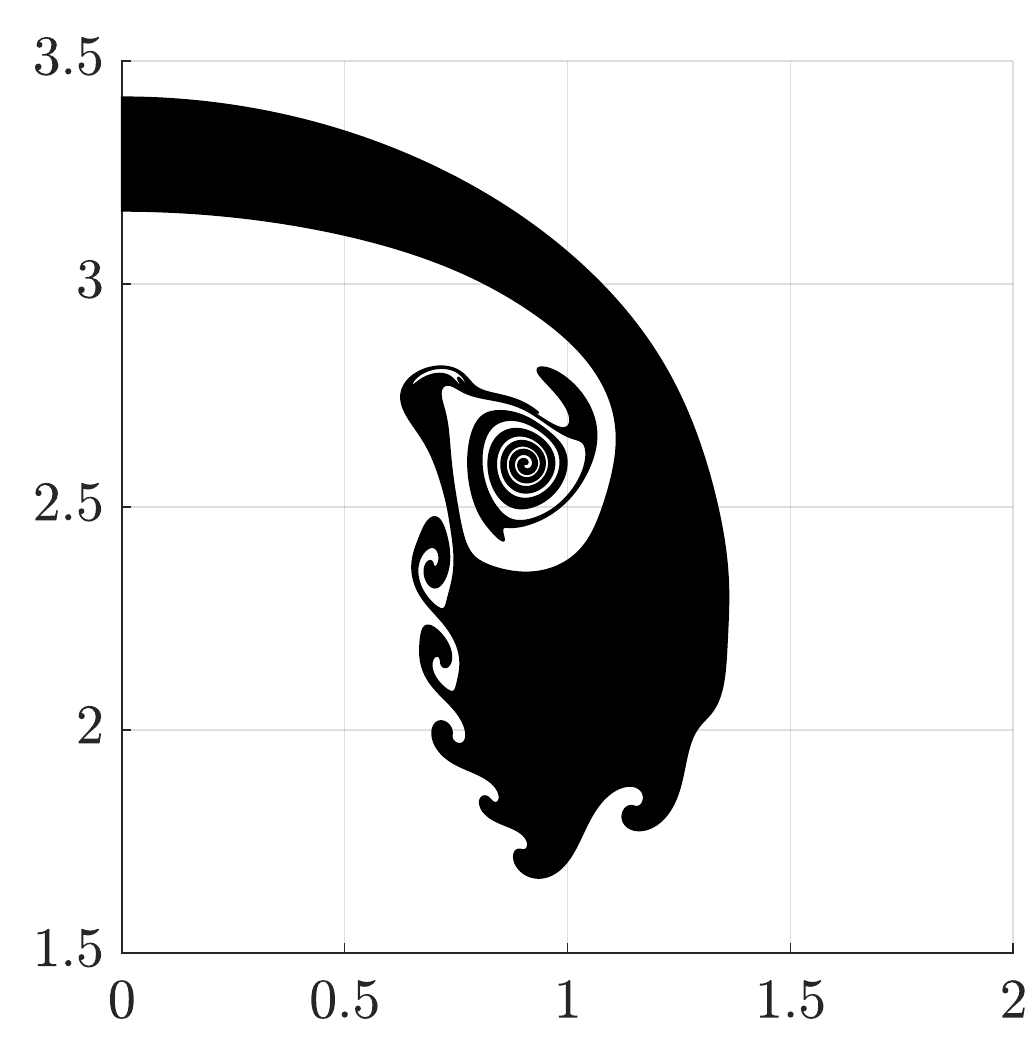}
            \caption{t=3.56}
        \end{subfigure}
        \hfill 
        \begin{subfigure}[b]{0.49\textwidth}
            \centering
            \includegraphics[trim={5cm 9.5cm 5.5cm 9.5cm},clip,width=\textwidth]{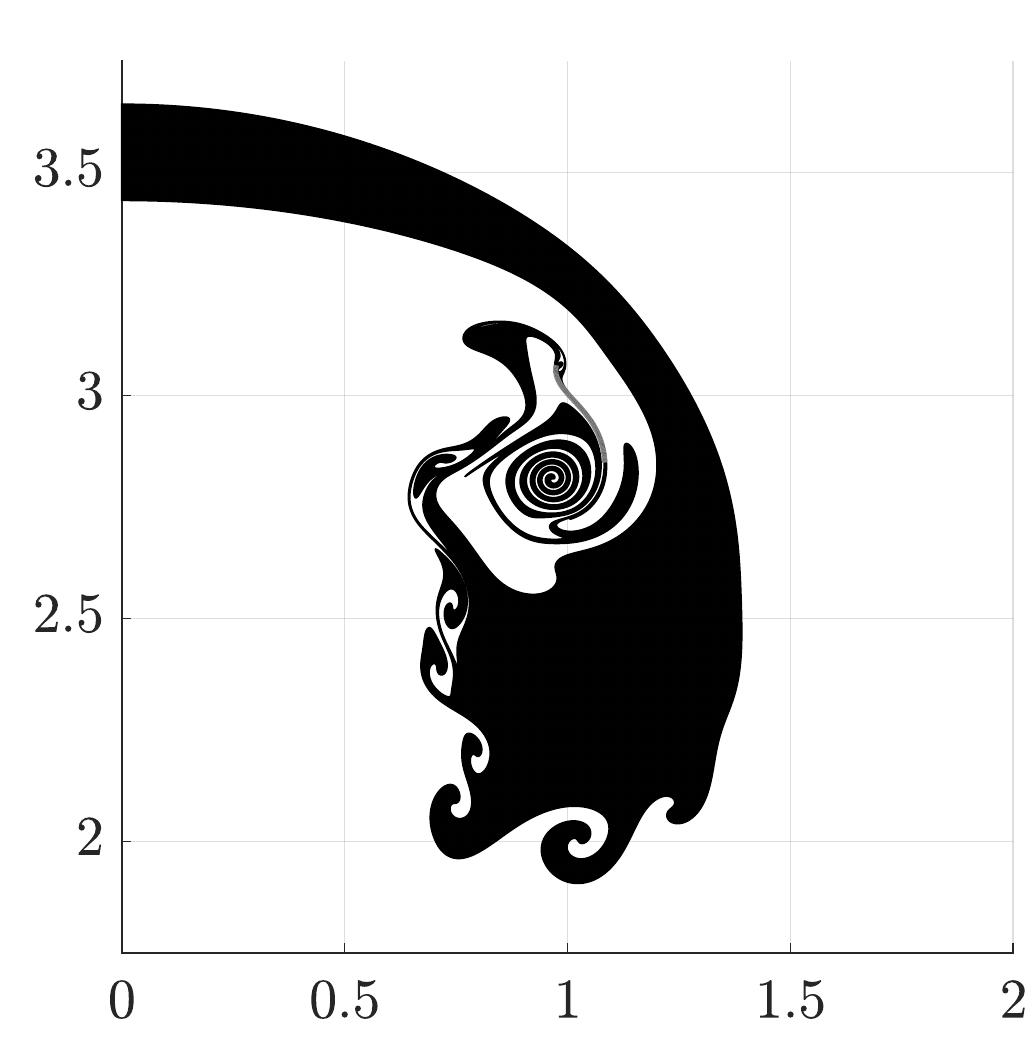}
            \caption{t=4.08}
        \end{subfigure}
        \hfill 
           \begin{subfigure}[b]{0.49\textwidth}
            \centering
            \includegraphics[trim={5cm 9.5cm 5.5cm 9.5cm},clip,width=\textwidth]{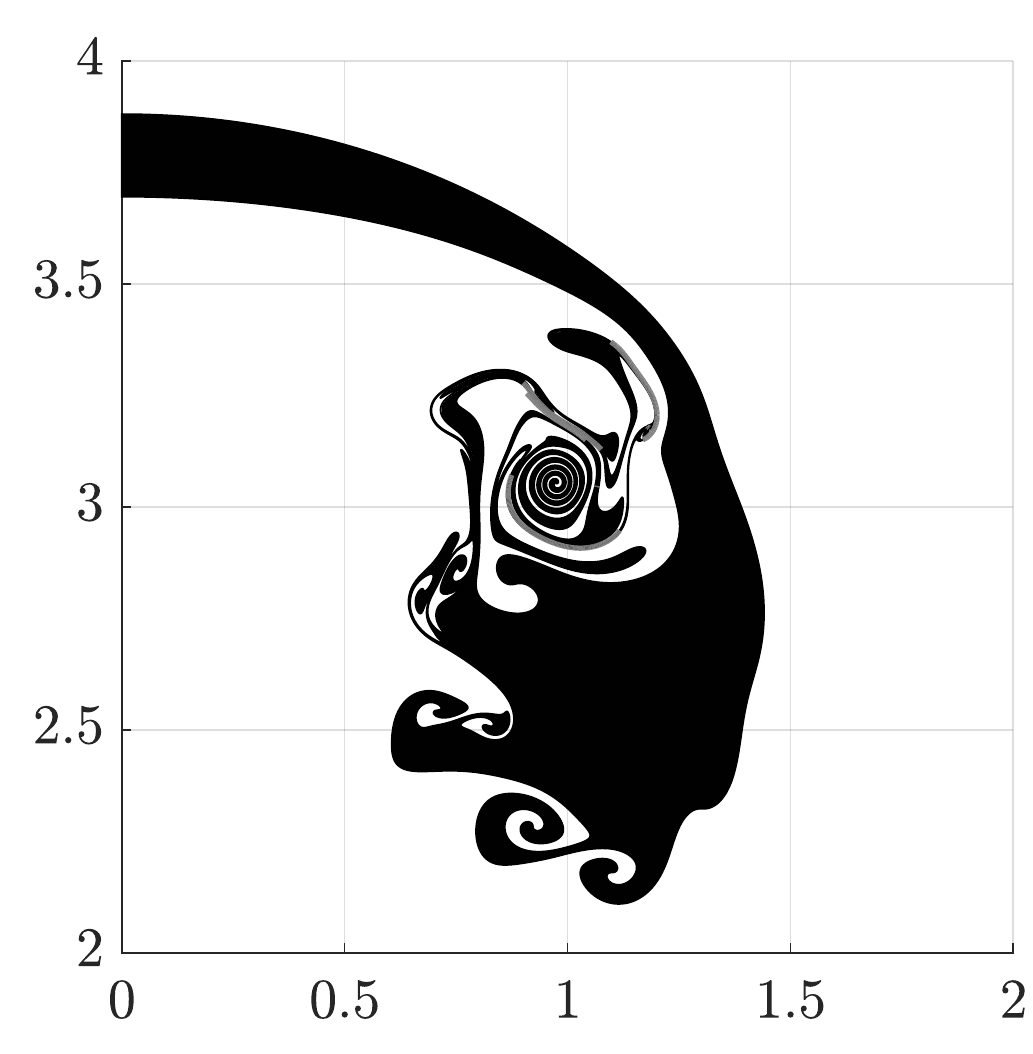}
        \caption{t=4.92}
        \end{subfigure}
        \caption{Winding of a sequence of Kelvin-Helmholtz vortices around the core, $\delta=0.05$, $b=1$}  
        \label{winding}
    \end{figure}

\begin{figure}[h]
    \centering
    \includegraphics[trim={5cm 9.5cm 5.0cm 9.5cm},clip,width=0.7\textwidth]{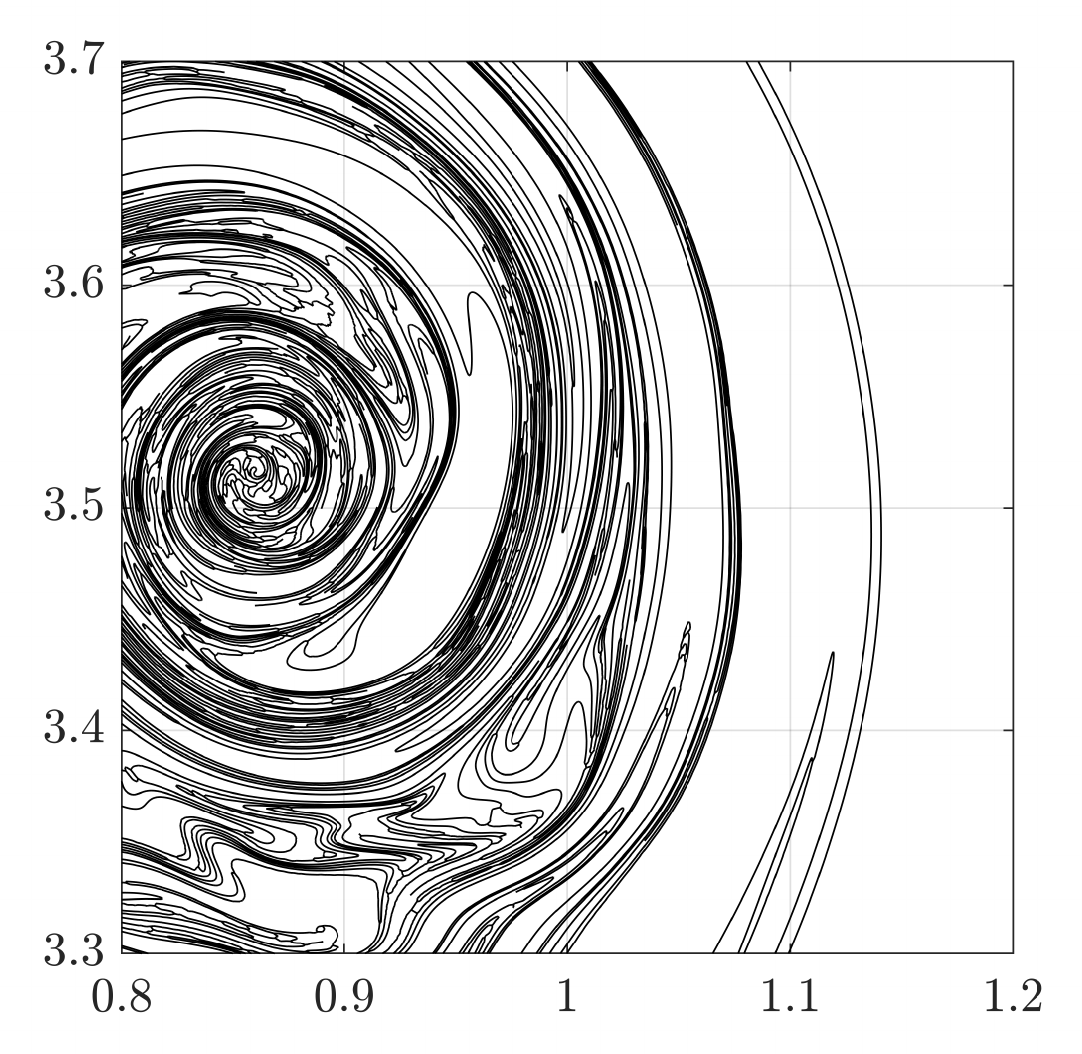}
    \caption{Core of the main vortex. $\delta=0.03$, $b=0$, $t=3.12$. The internal region was not marked black to highlight the concentration of the vortex sheet.}
    \label{tight}
\end{figure}

Decreasing $\delta$ launches the Kelvin-Helmholtz instability in both cases ($b=0$ and $b=1$). A broad discussion of its nature might be found e.g. in (\cite{vallis}, chapter 6). The lower the $\delta$, the less stable the sheet is, and, generally, the lower the size of vortices arising. Nevertheless, their size does not seem to scale with $\delta$ in a simple way - in tests, we were able to obtain the same wavenumber for two significantly different $\delta$ with the same setup. \\
The presence of additional vortices highlights another interesting phenomenon. Case of $\delta=0.05$ captures it in a well-separated form. We can observe an evolution of a vortex, that arose in the wake and is advected towards the center of the ring. During this time it is subjected to intense stretching in direction of the nearby vortex sheet and contraction in the normal. This leads to the ``ironing" of the vortex, which is further incorporated into the sheet and, indistinguishable, gets susceptible to instability again. The whole process reminds an iteration of Smale's horseshoe map and is illustrated in figure (\ref{ironing}).
As the simulation proceeds, some structures evolve into bulky, sharp-edged shapes, which might suggest too coarse discretization. Although, figure (\ref{ironing_discretization}) shows that they are well resolved in space. We also repeated part of the simulation when they arise with time step halved (with RK4 this should decrease the error 16 times) and noticed no significant change.
In the case of buoyant production of vorticity (same $\delta$ - equal 0.05), there is no clear wake and marked fluid, understandably, moves in more compact manner. The Kelvin-Helmholtz instability is also more intense. In fig. (\ref{winding}) we can observe how vortices arise, are ironed, and winded around the core, one after another. The core grows until it is torn apart by the same mechanism as in the $\delta=0.1$ case. 
Further decreasing the smoothing parameter to $\delta=0.03$ intensifies previously described mechanisms. Nevertheless $b=0$ case remains a coherent vortex ring, not being destabilized by negative vorticity. Due to winding, it quickly becomes very tightly packed, still exhibiting sheet structure, but thicker - see fig. (\ref{tight}). Such a sheet characterizes by repetitive strands of internal and external volume
building a Damascus-like structure. 
\subsection{The hierarchy of Kelvin-Helmholtz instabilities}
\begin{figure}[h]
    \centering
    \includegraphics[trim={5cm 9cm 4.0cm 10cm},clip,width=0.7\textwidth]{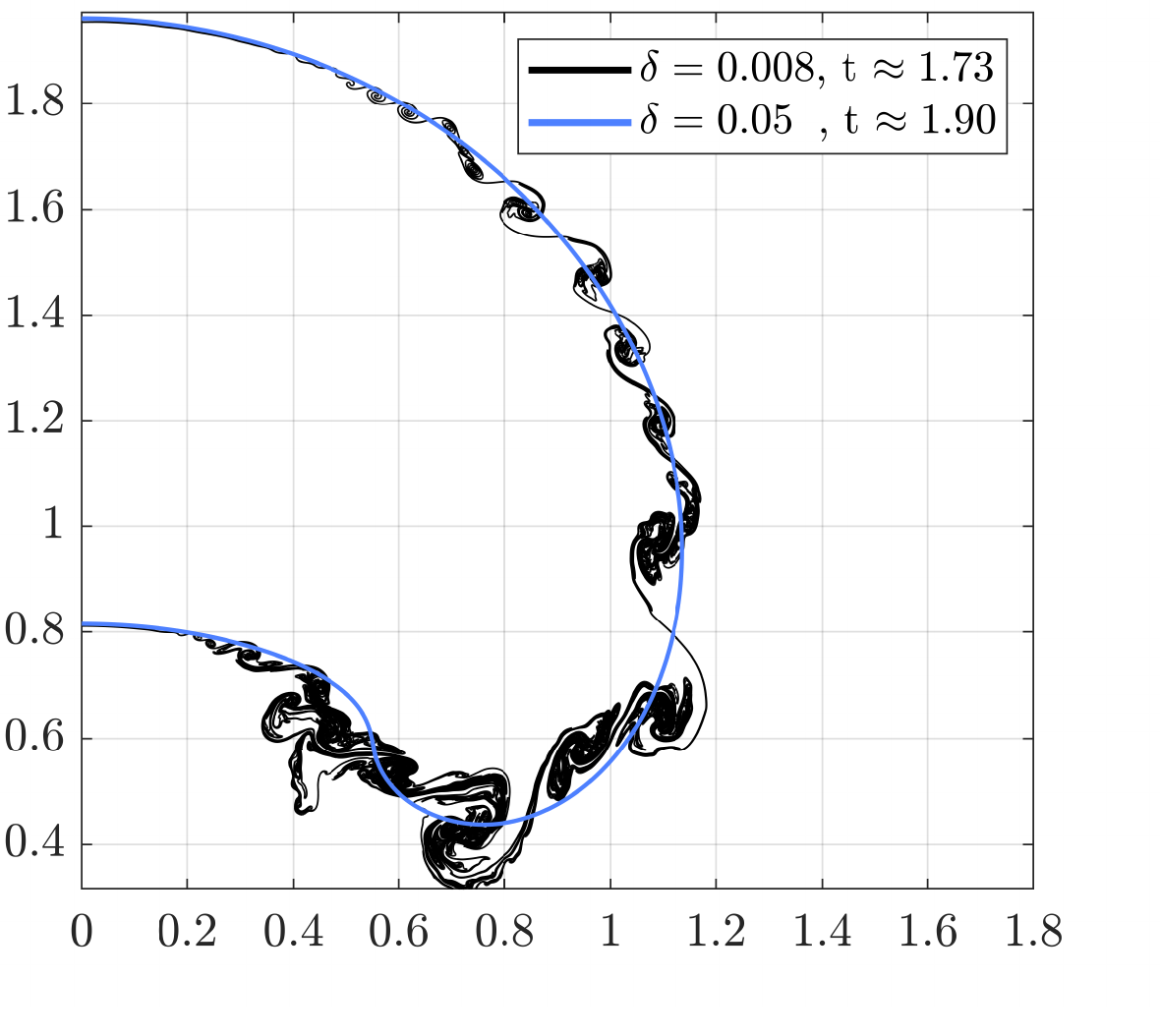}
    \caption{Sheet with $\delta=0.05$ used as a mean contour for case with $\delta=0.008$ after proper shifting in time.}
    \label{matching}
\end{figure}

\begin{figure}[H]
        \centering
        \includegraphics[trim={1cm 7.4cm 1.3cm 8.0cm},clip,width=1\textwidth]{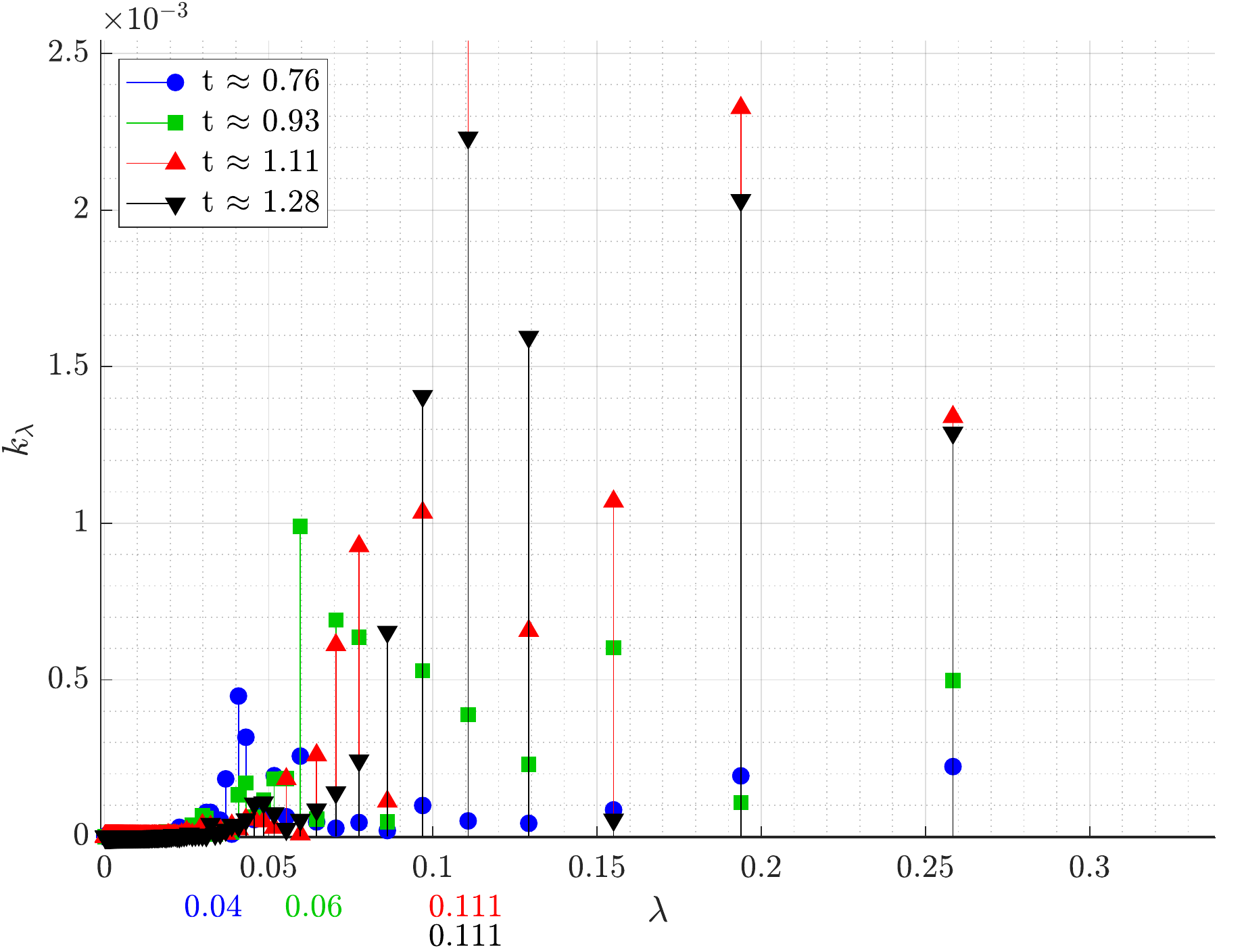}
        \put(-210,235){{\color{red}$\uparrow 5.4\times10^{-3}$}}
         \put(-210,10){{\color{white}$\blacksquare$}}
         \put(-210,4){{\color{white}$\blacksquare$}}
         \put(-210,-2){{\color{white}$\blacksquare$}}
        \caption{Kinetic energy associated with modes of particular wavelength - a selected part of the spectrum. $\delta=0.008$, $b=1$. Peaks for different times, labeled under the horizontal axis.}
        \label{k_transport}
\end{figure}
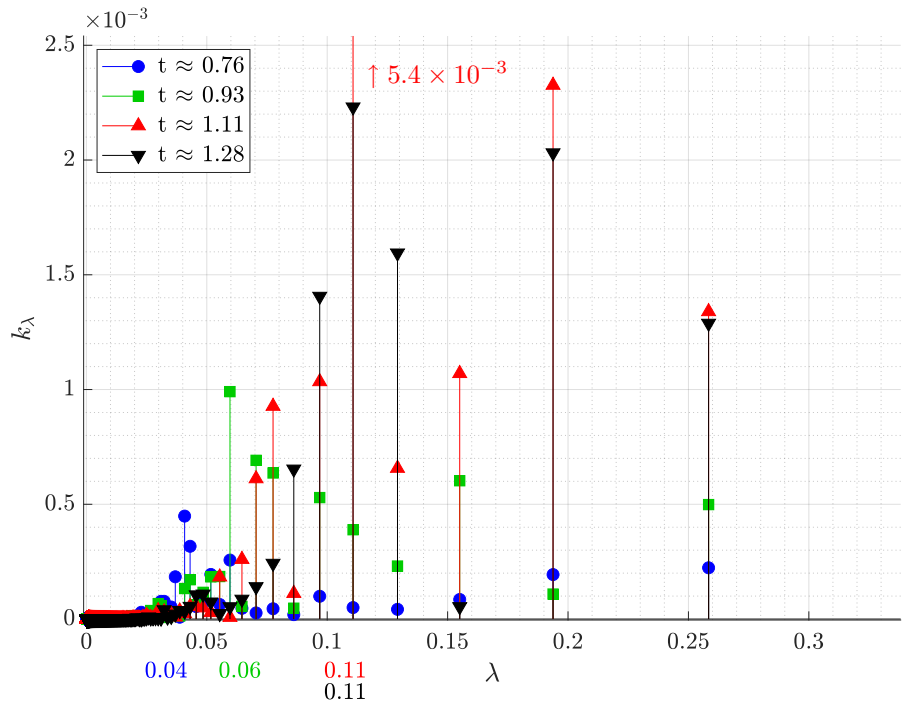

\begin{figure}[H]
        \centering
        \includegraphics[trim={1cm 6.5cm 1.3cm 8.0cm},clip,width=1\textwidth]{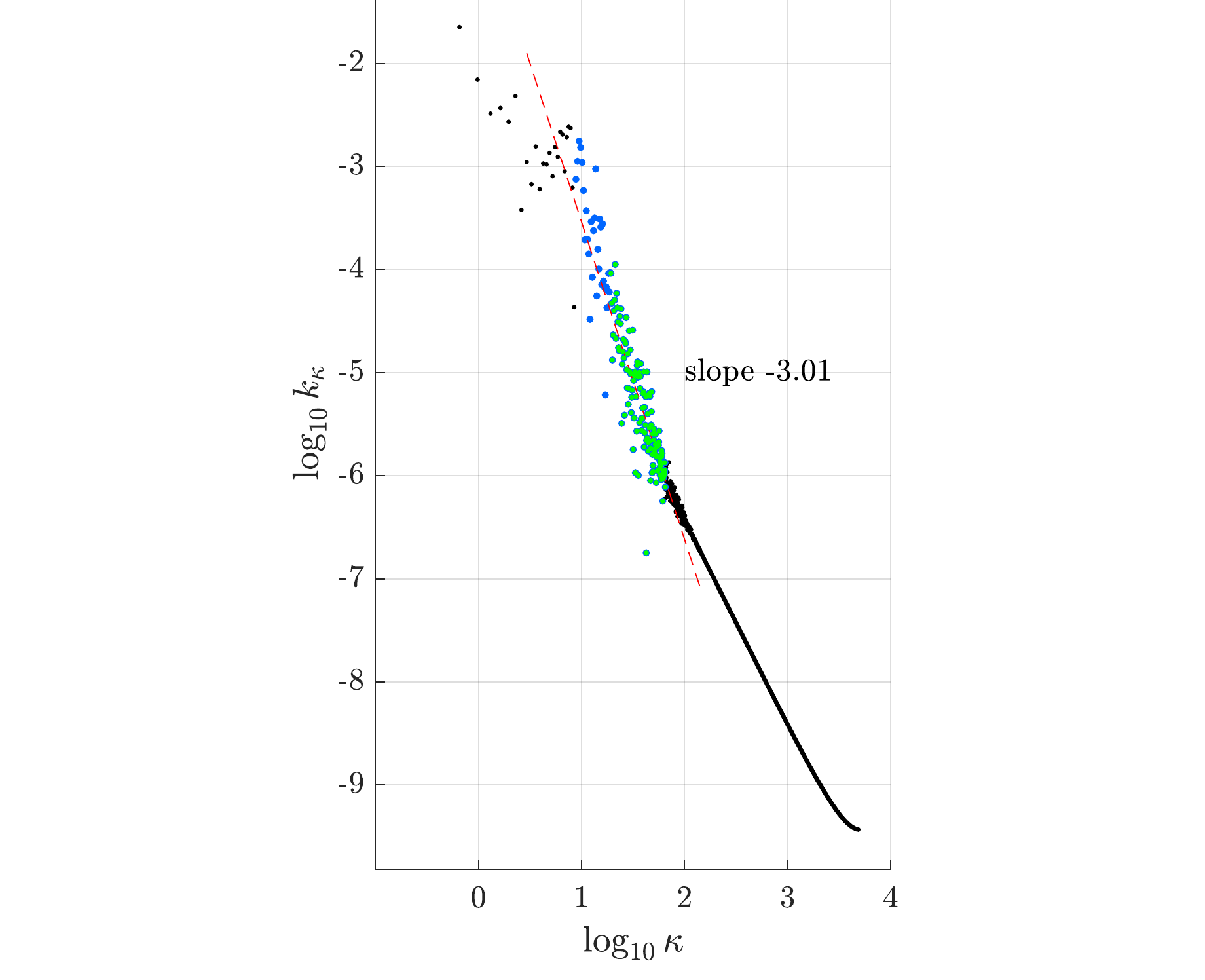}
        \caption{Energy spectrum for the whole contour in $t\approx 1.28$, $\delta=0.008$, $b=1$. The dashed line depicts least squares fit to data marked green (slope -3.01). If blue data is included, the slope decreases to -3.36. $\kappa$ with (approximately) $k_\kappa<10^{-7}$  is absent in the flow. Slope there is a numerical feature.}
        \label{fullspec}
\end{figure}

\begin{figure}
    \hspace{-1cm}
    \includegraphics[trim={3.5cm 7cm 3.5cm 7.0cm},clip,width=1.1\textwidth]{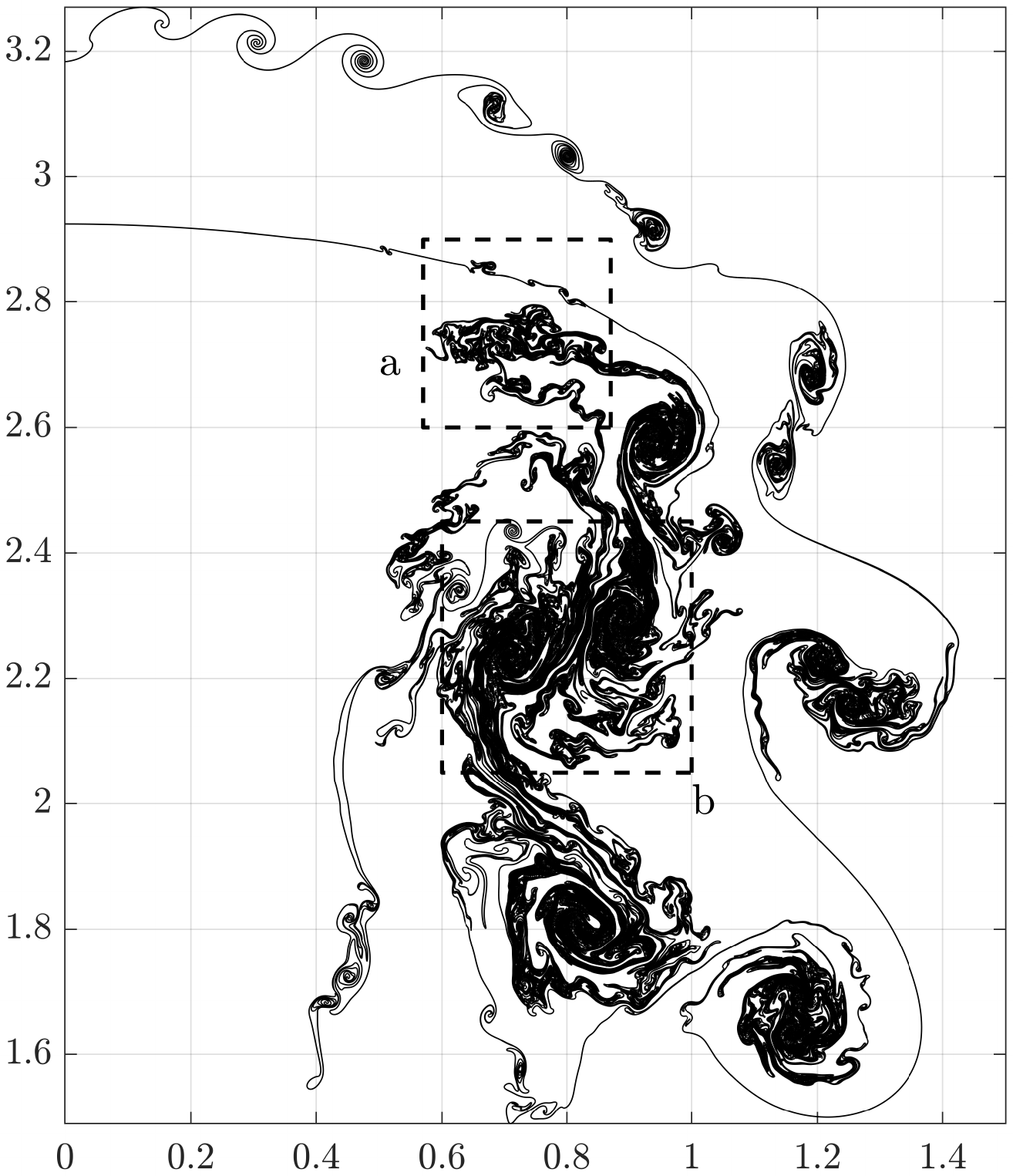}
    \caption{Vortex sheet, $\delta=0.008$, $b=1$, $t\approx 3.23$}
    \label{lastContour}
\end{figure}

\begin{figure}[H]
\vspace{-1.5cm}
        \centering
        \begin{subfigure}[b]{0.9\textwidth}
            \centering
            \includegraphics[trim={2cm 8cm 2cm 7.85cm},clip,width=1.0\textwidth]{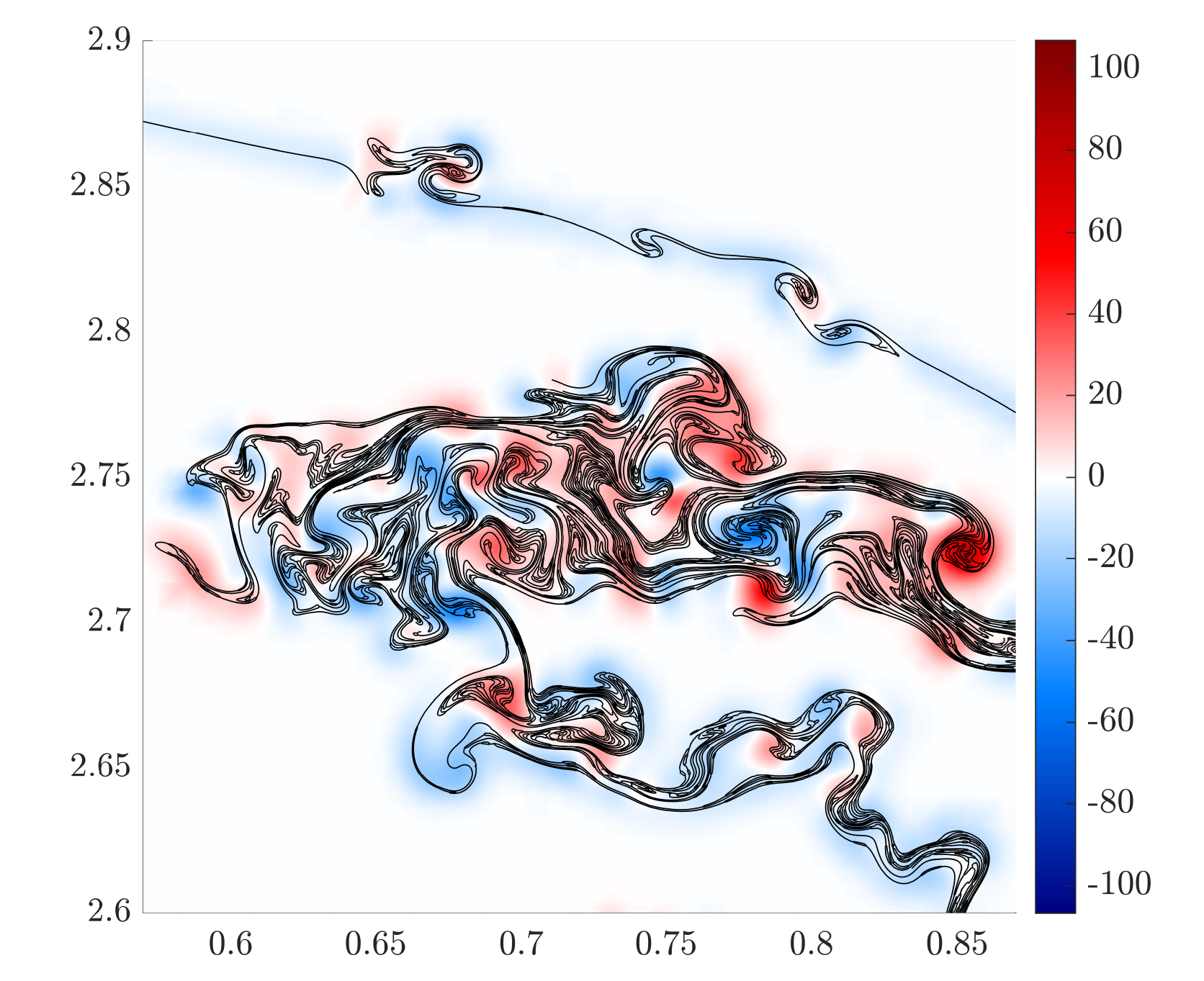}
            \caption{}
        \end{subfigure}
        \hfill 
            \begin{subfigure}[b]{0.87\textwidth}
            \centering
          \includegraphics[trim={2cm 8cm 2cm 7.85cm},clip,width=1.0\textwidth]{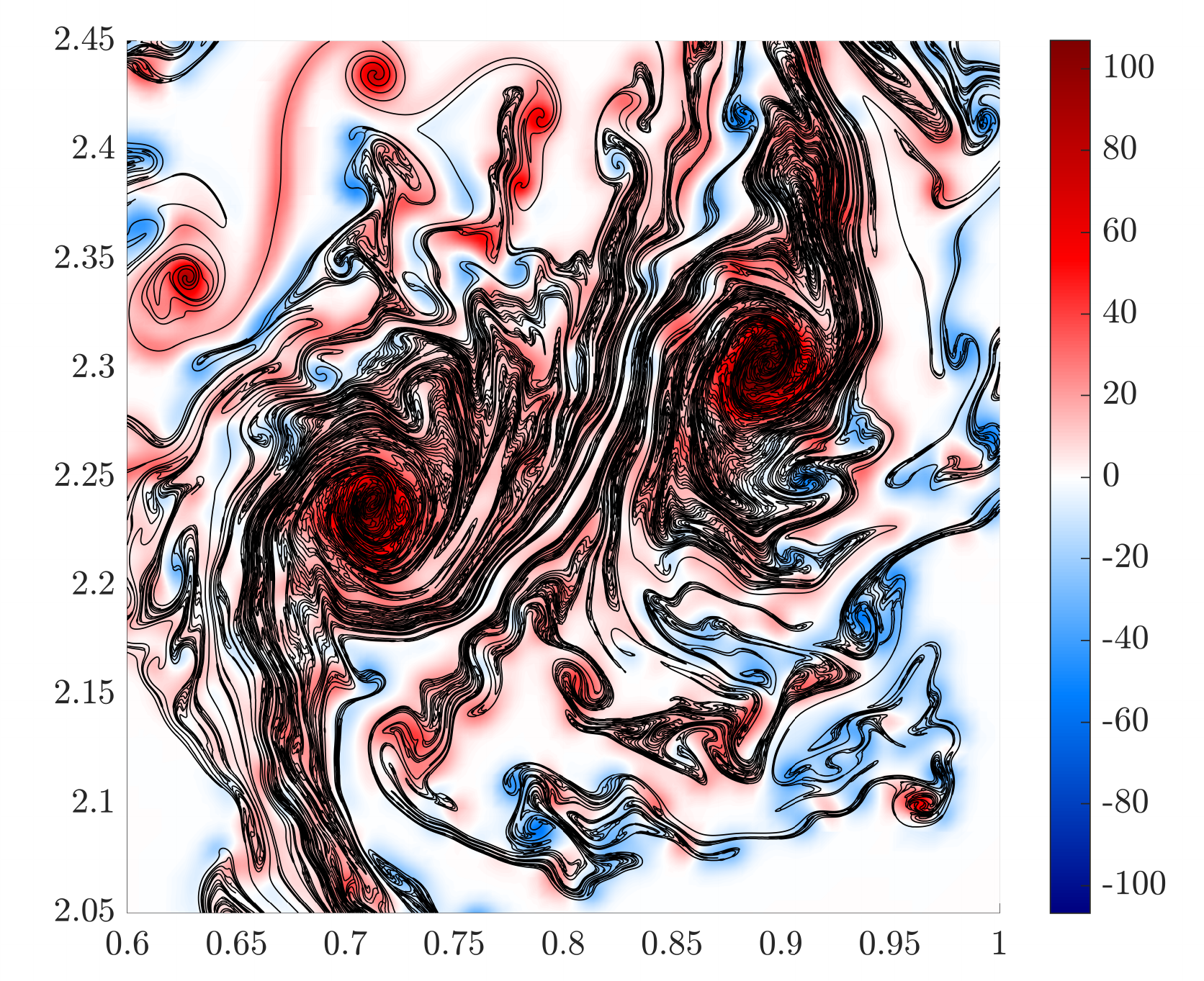}
        \caption{}
        \end{subfigure}
        \caption{Selected regions from fig. (\ref{lastContour}) zoomed with vorticity map in a background.}
        \label{lastContourZooms}
\end{figure}

\begin{figure}[H]
    \centering
    \includegraphics[trim={4cm 8cm 4cm 7.85cm},clip,width=1.0\textwidth]{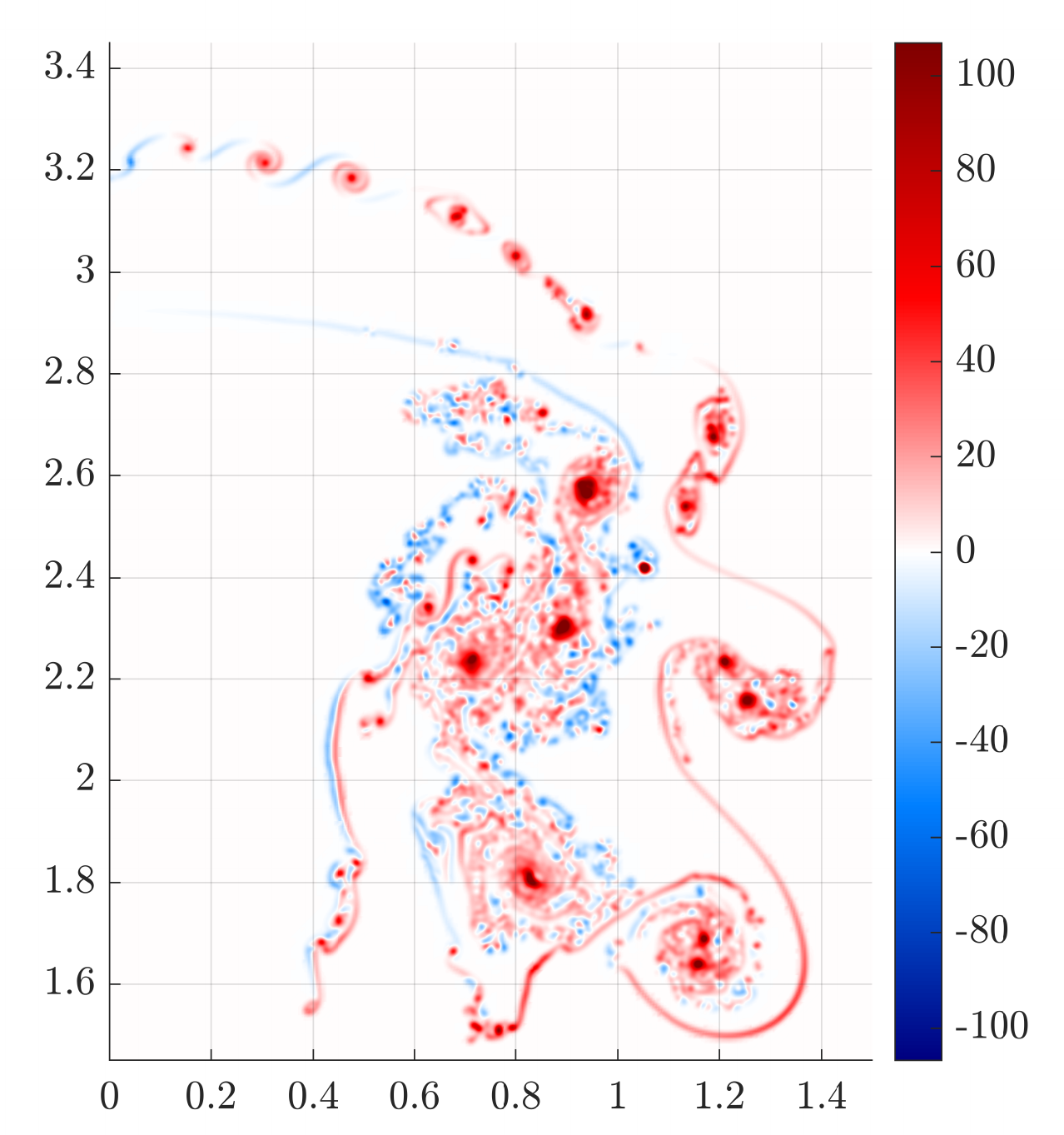}
    \caption{Vorticity field, $\delta=0.008$, $b=1$, $t\approx 3.23$}
    \label{lastVort}
\end{figure}

\begin{figure}[H]
    \centering
    \includegraphics[trim={4cm 6.0cm 3.5cm 7.85cm},clip,width=1.0\textwidth]{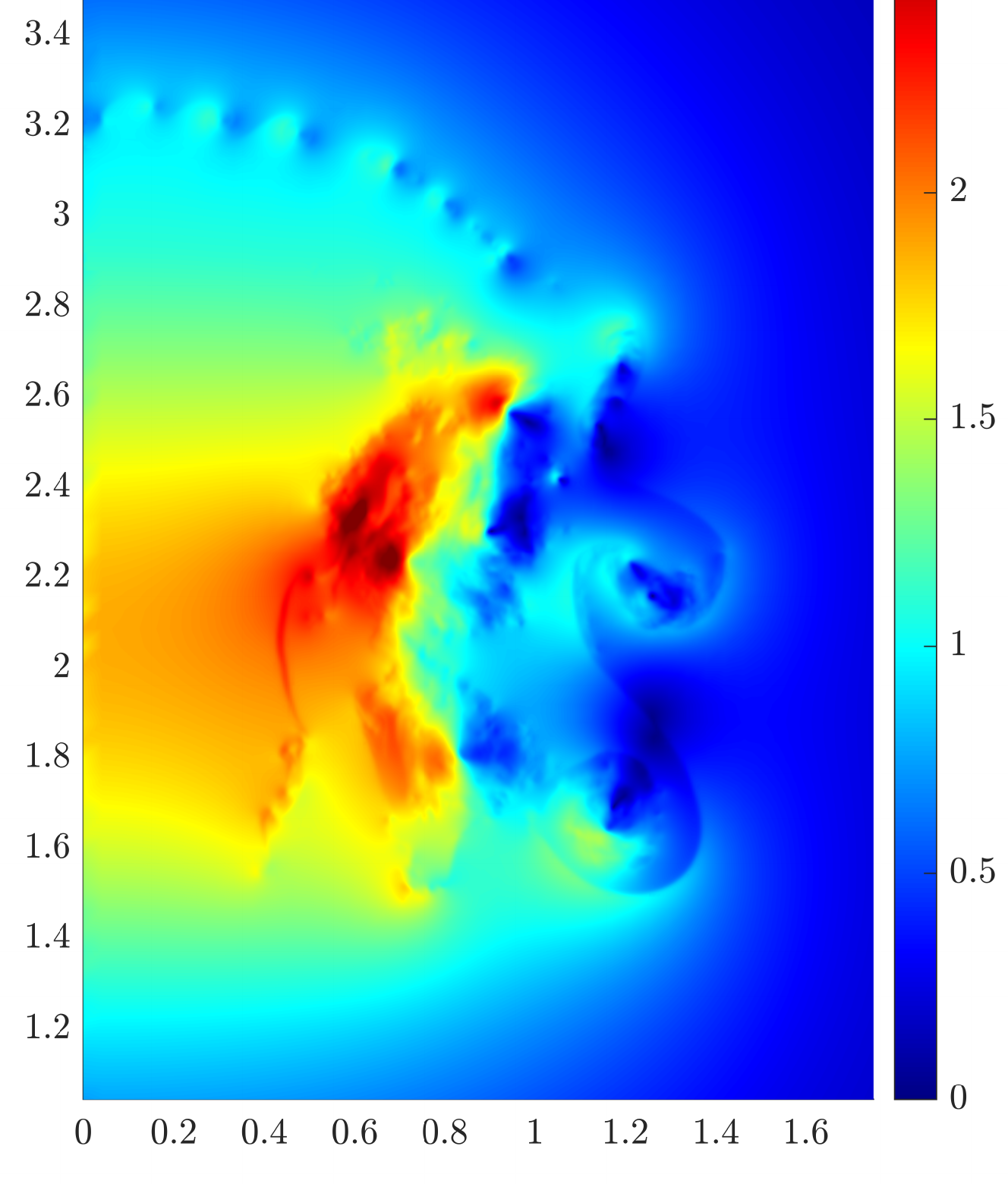}
    \caption{Velocity magnitude, $\delta=0.008$, $b=1$, $t\approx 3.23$}
    \label{lastVel}
\end{figure}

When $\delta$ is decreased to $0.008$, we observe how the sheet gets covered with tiny vortices. In the first stage, it gets wavy, then characteristic eye-reminding structures arise. They seem to effectively increase the local thickness of the sheet, stabilizing the resulting structure in high wavenumbers. The following, so-called, vortex pairing, can be seen as initial waves, but level higher, occurring in the thicker, composite sheet. We further see that they evolve into analogical, eye-reminding vortices that cover the sheet and increase its effective thickness. The resulting layer again gets unstable in even higher wave numbers and the process repeats as long until vortices get big enough to significantly affect the mean flow. Each iteration increases the scale roughly twice, what follows from vortex pairing mechanism. We clearly see the transition from low to high scales, which is probably associated with inverse energy transport.
\begin{figure}[H]
\vspace{-1.5cm}
        \centering
        \begin{subfigure}[b]{1\textwidth}
            \centering
            \includegraphics[trim={1.2cm 7.8cm 1.5cm 7.5cm},clip,width=1.0\textwidth]{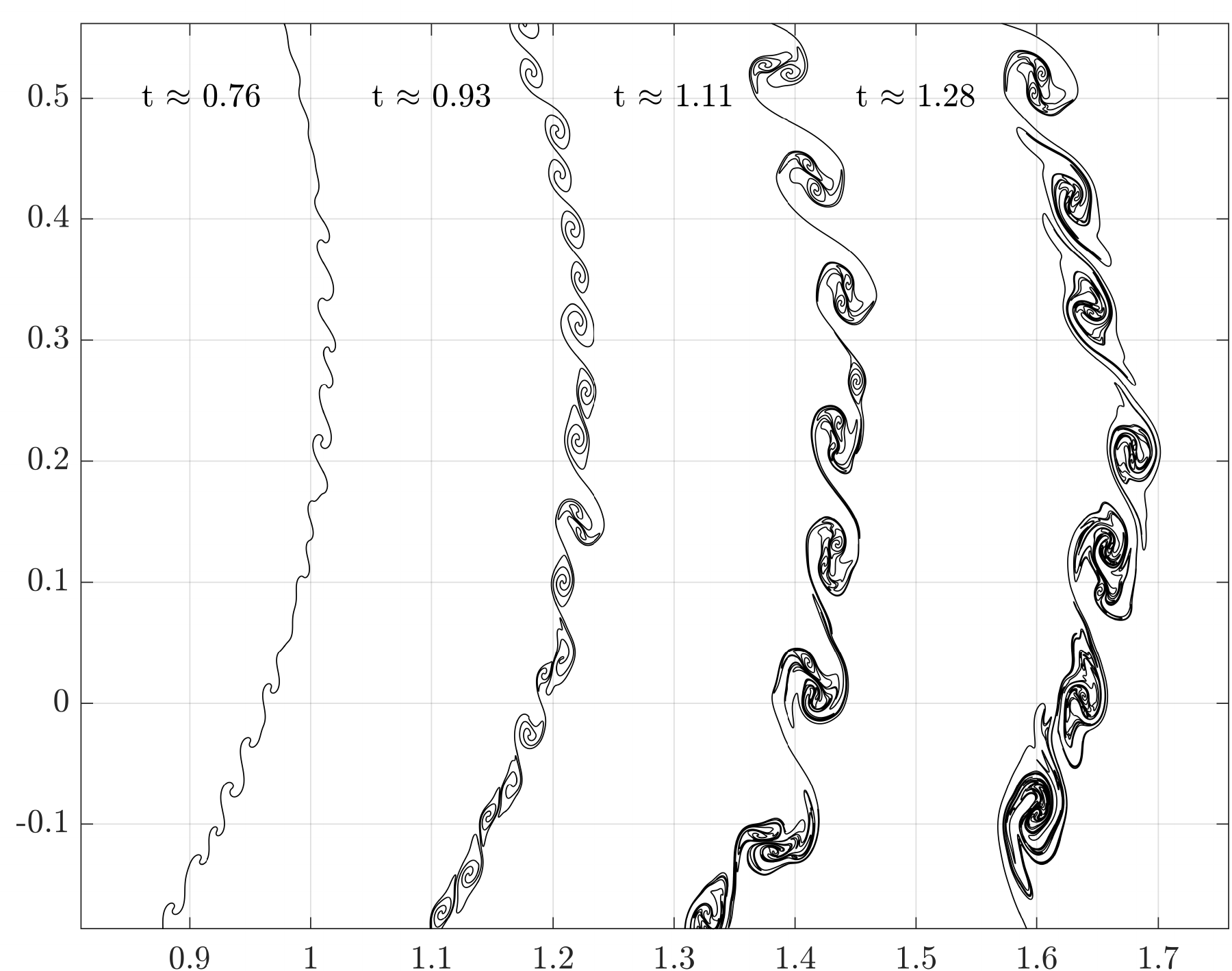}
        \end{subfigure}
        \hfill 
            \begin{subfigure}[b]{1\textwidth}
            \centering
          \includegraphics[trim={2cm 8cm 2cm 7.85cm},clip,width=1.0\textwidth]{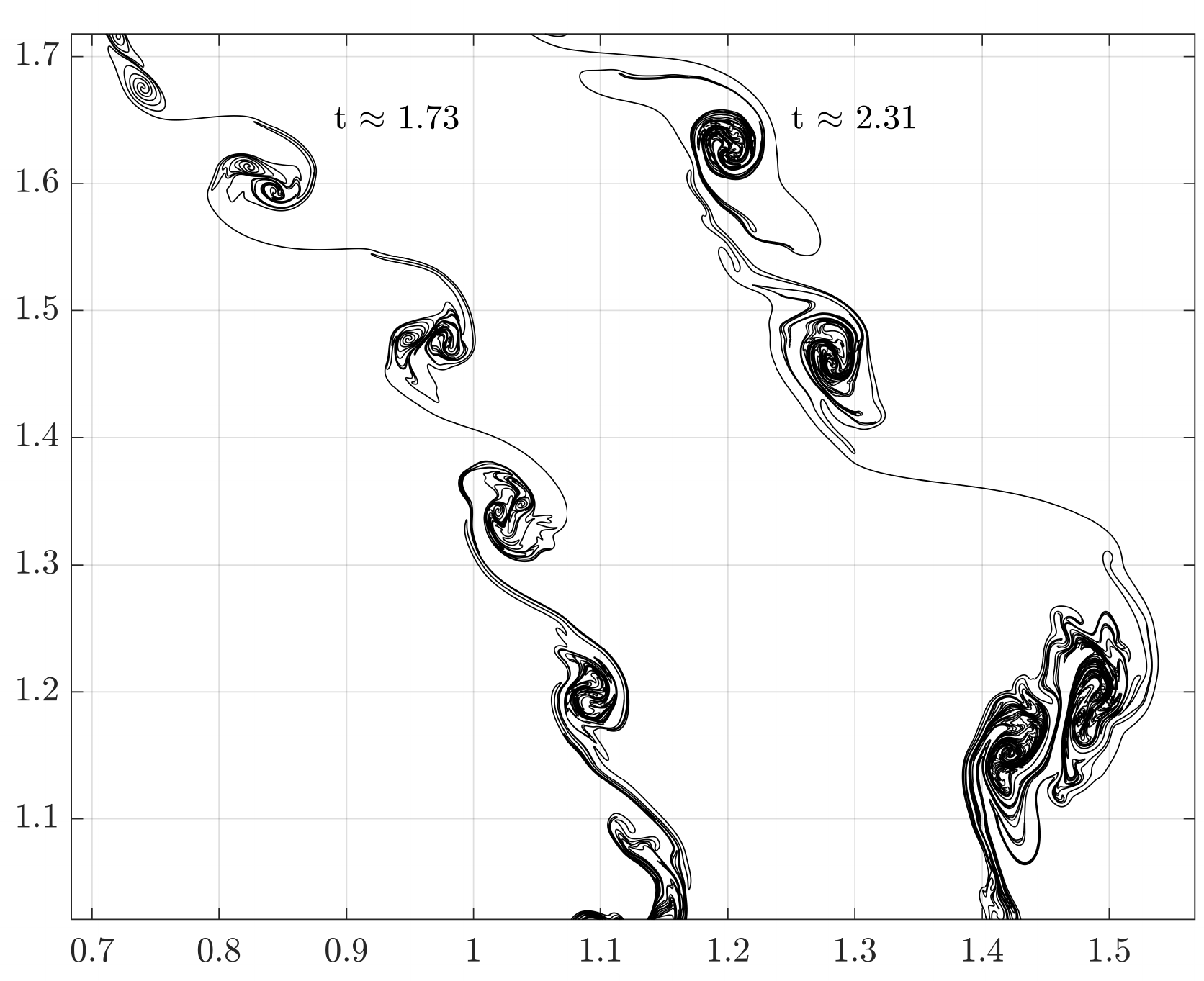}
        \end{subfigure}
        \caption{A hierarchy of Kelvin-Helmholtz instabilities ($b=1$, $\delta=0.008$). The exact position of the sheet was shifted for easier comparison. Coordinates are labeled just for size estimation.}
        \label{hierarchy}
\end{figure} 
To verify this we start by drawing the mean contour for each of the presented timesteps. Instead of trying to remove Kelvin-Helmholtz vortices by some kind of smoothing or averaging, we just use the shapes of the sheet from the $\delta=0.05$ case. Although their heights are different by few percent (what is definitely too much in comparison to the size of small vortices), they seem to be de facto shifted in time. If we choose a pair of timesteps that have the tops of the sheets aligned, the rest fits well. An example is in fig. (\ref{matching}). 
We rediscretize the mean contour, increasing the number of nodes and providing exactly equal spacing in all cases. This will allow us to use FFT and test different timesteps against exactly the same modes. Because the instability takes place only in the part of the contour, we need to extract it somehow. We take an advantage of the vertical orientation of the instability (it covers mostly the ``right side'' of the initial sphere) and look for the maxima of $\rho(s)$. For the time $t \approx 0.76$, we extract the region between the first and the last maximum and trim the last, underdeveloped 25\% of the obtained length. Let us call the resulting length $L$. In later times, the instability spreads along the sheet, being in a later stage in the initial region (right side) and in an earlier stage closer to the top. We pick this maximum of $\rho(s)$ which is right in the middle and take $L/2$ of the mean contour before and after that point. Such domains are presented in fig. (\ref{fftTrace}).
\begin{figure}[h]
    \centering
    \includegraphics[trim={2cm 8.5cm 2.0cm 8.0cm},clip,width=0.7\textwidth]{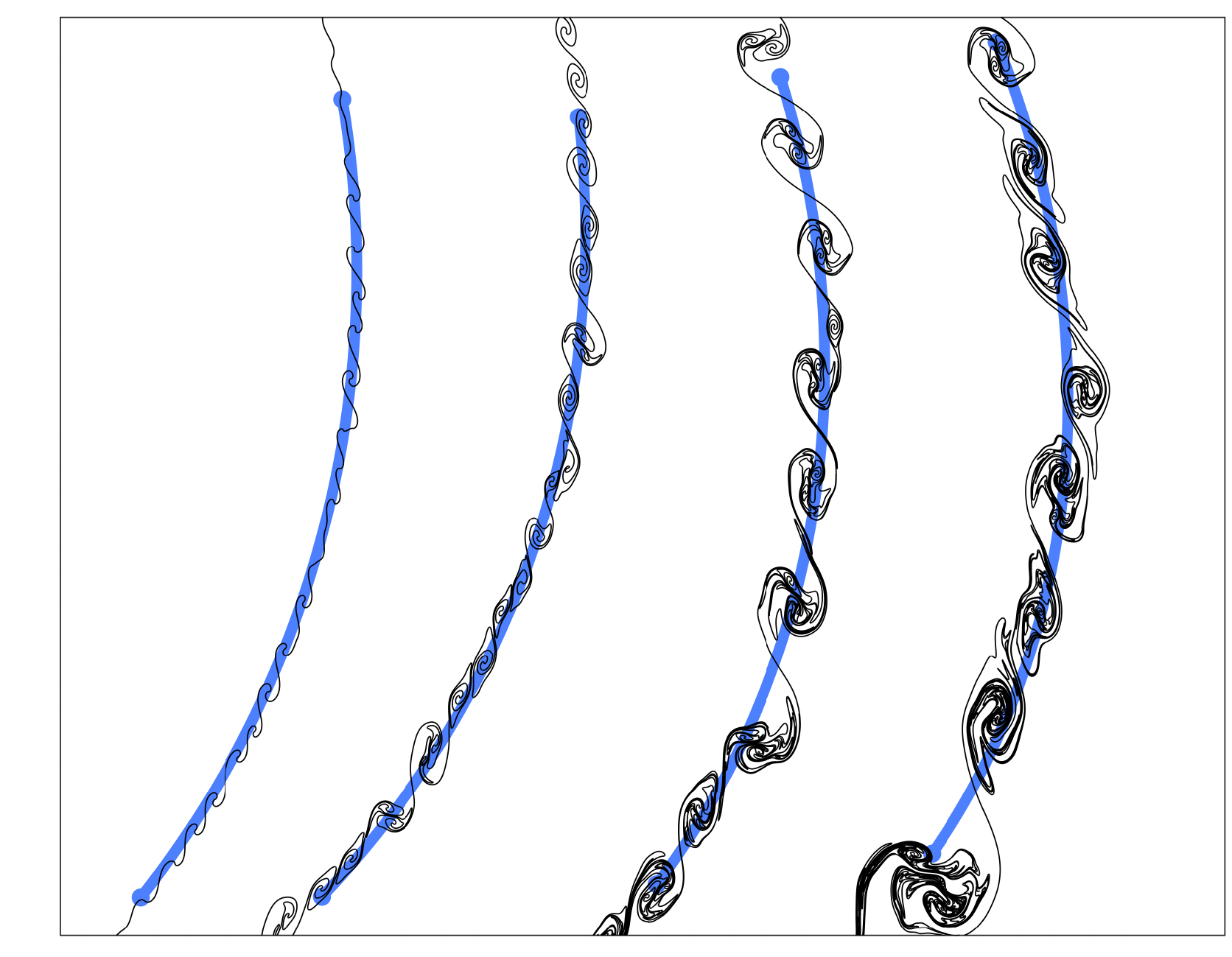}
    \caption{Paths along which Fourier transform was computed (light blue) and corresponding parts of the sheet from fig. (\ref{hierarchy})}
    \label{fftTrace}
\end{figure}
We probe the induced velocities and take a Fourier transform against its (physical) length. We compute the density of specific kinetic energy associated with particular modes, understood as follows:
\begin{equation}
    k(\kappa) = \frac{|\mathcal{F}\{u_\rho\}|^2 + |\mathcal{F}\{u_z\}|^2}{2}
\end{equation}
where $\kappa$ is wavenumber and $\mathcal{F}$ denotes Fourier transform. For convenience, we plot (\ref{k_transport}) the results against the wavelength $\lambda$ (although no longer densities, but rather whole amounts of energy associated with discrete wavelengths).
We see that the dominant mode of instability gets shifted towards higher scales, starting in $\lambda=0.04$, through $\lambda=0.06$ and reaching $0.111$ around t $= 1.28$. Values are consistent with (\ref{hierarchy}). In fig. (\ref{fullspec}) we also present a spectrum for the whole contour in $t\approx 1.28$ against wavenumber.
A further stage of the system is presented in figures (\ref{lastContour}) and (\ref{lastContourZooms}).
We can notice that, despite surgery, there are still relatively thick, coherent parts of the sheet that could possibly be simplified. Their details are finer than the features of the vorticity map, therefore it should not affect the overall evolution significantly. Below, we also present the complete map of vorticity distribution (\ref{lastVort}) as well as the induced velocity (\ref{lastVel}).
\subsection{The effectiveness of surgery}
\begin{figure}[H]
        \centering
        \begin{subfigure}[b]{0.48\textwidth}
            \centering
            \includegraphics[trim={5cm 9.7cm 5.5cm 10cm},clip,width=\textwidth]{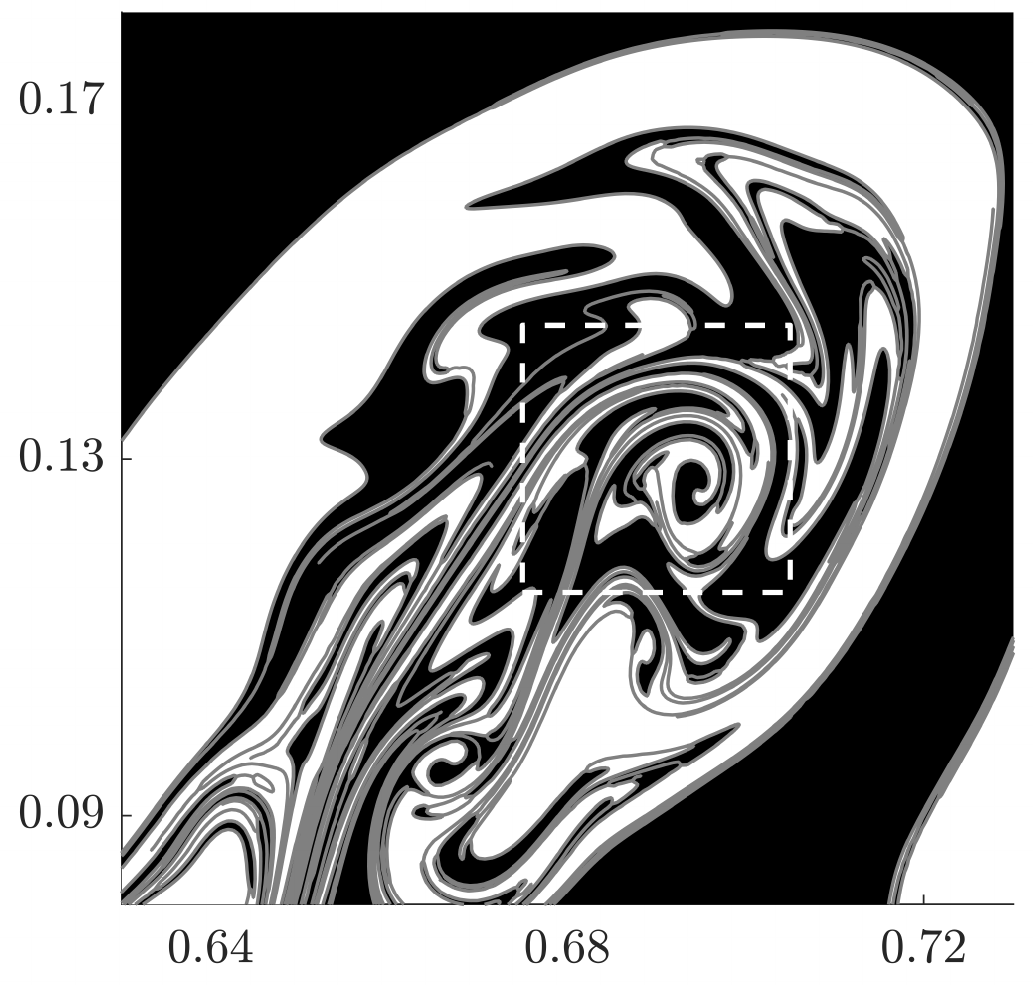}
            \caption{}
        \end{subfigure}
        \hfill 
            \begin{subfigure}[b]{0.48\textwidth}
            \centering
           \includegraphics[trim={5cm 9.7cm 5.5cm 10cm},clip,width=\textwidth]{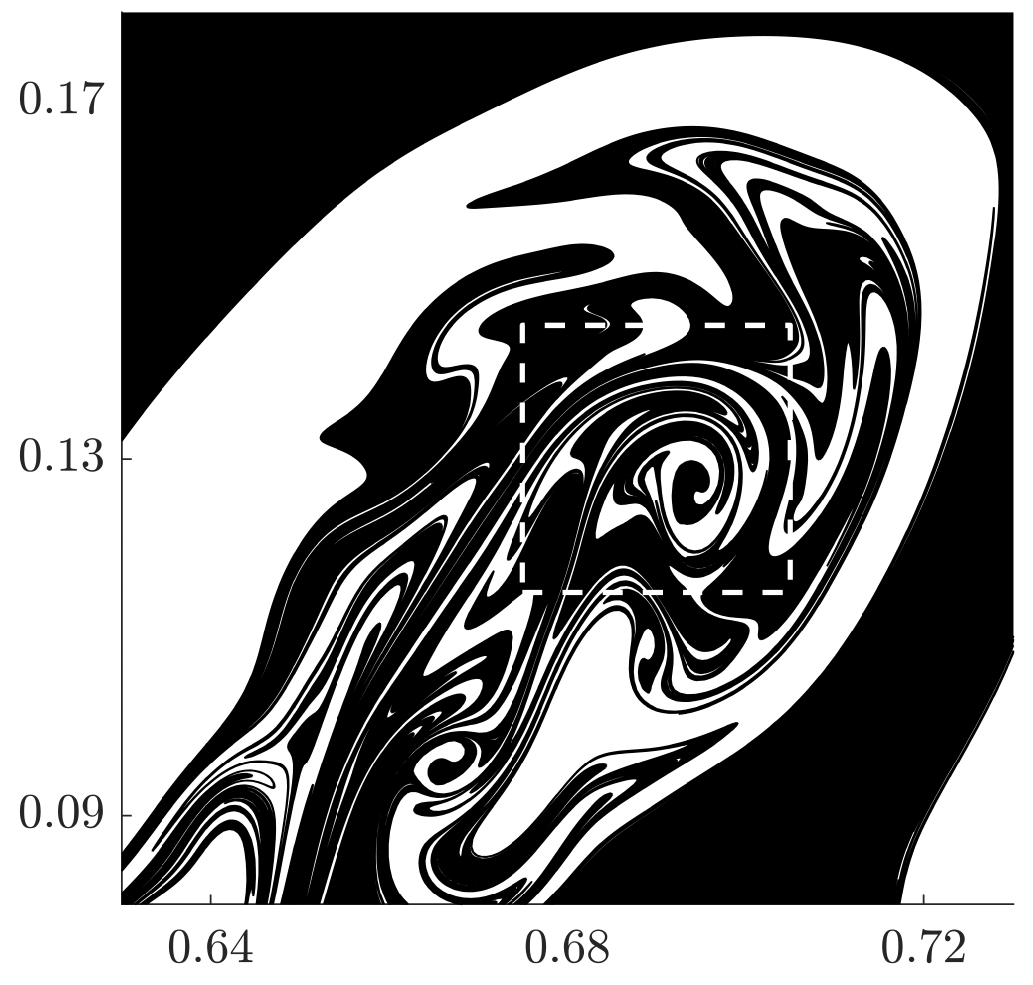}
            \caption{}
        \end{subfigure}
        \hfill 
        \begin{subfigure}[b]{0.48\textwidth}
            \centering
            \includegraphics[trim={5cm 9.7cm 5.5cm 10cm},clip,width=\textwidth]{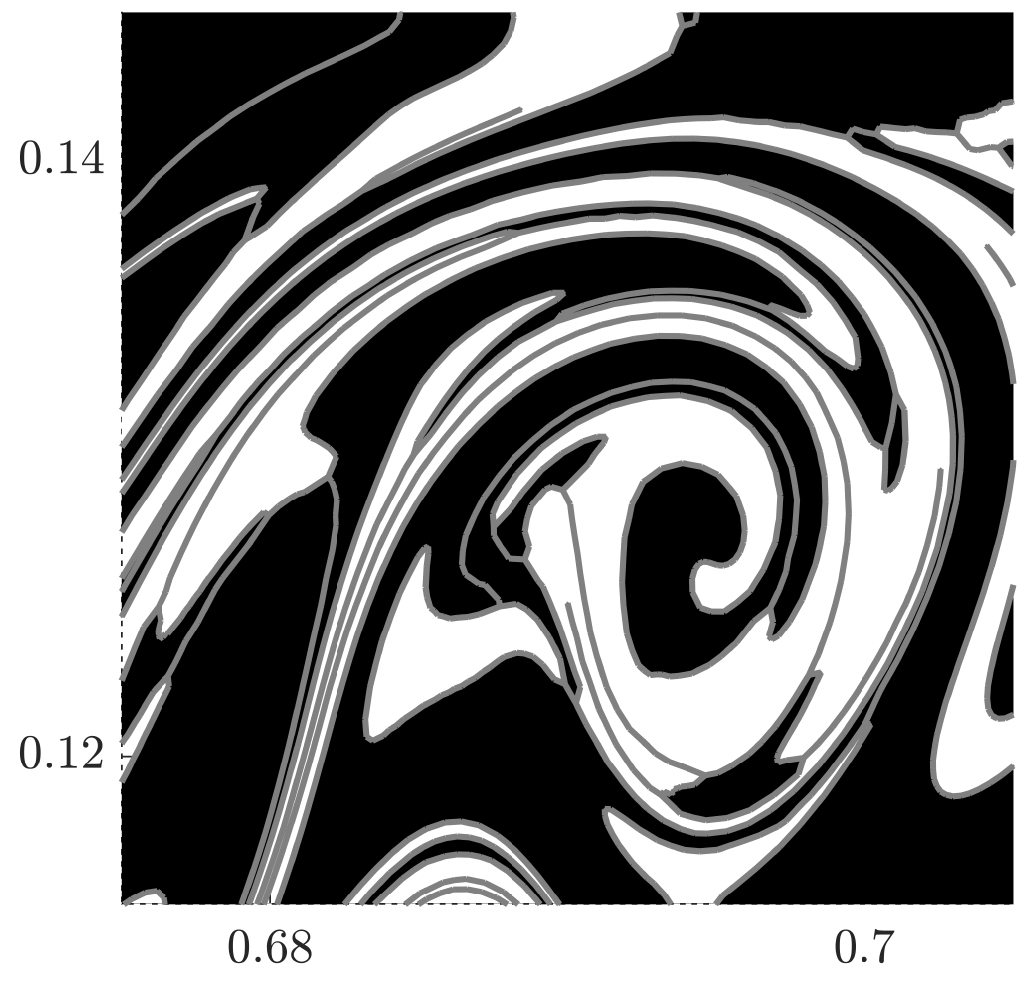}
            \caption{}
        \end{subfigure}
        \hfill 
           \begin{subfigure}[b]{0.48\textwidth}
            \centering
           \includegraphics[trim={5cm 9.7cm 5.5cm 10cm},clip,width=\textwidth]{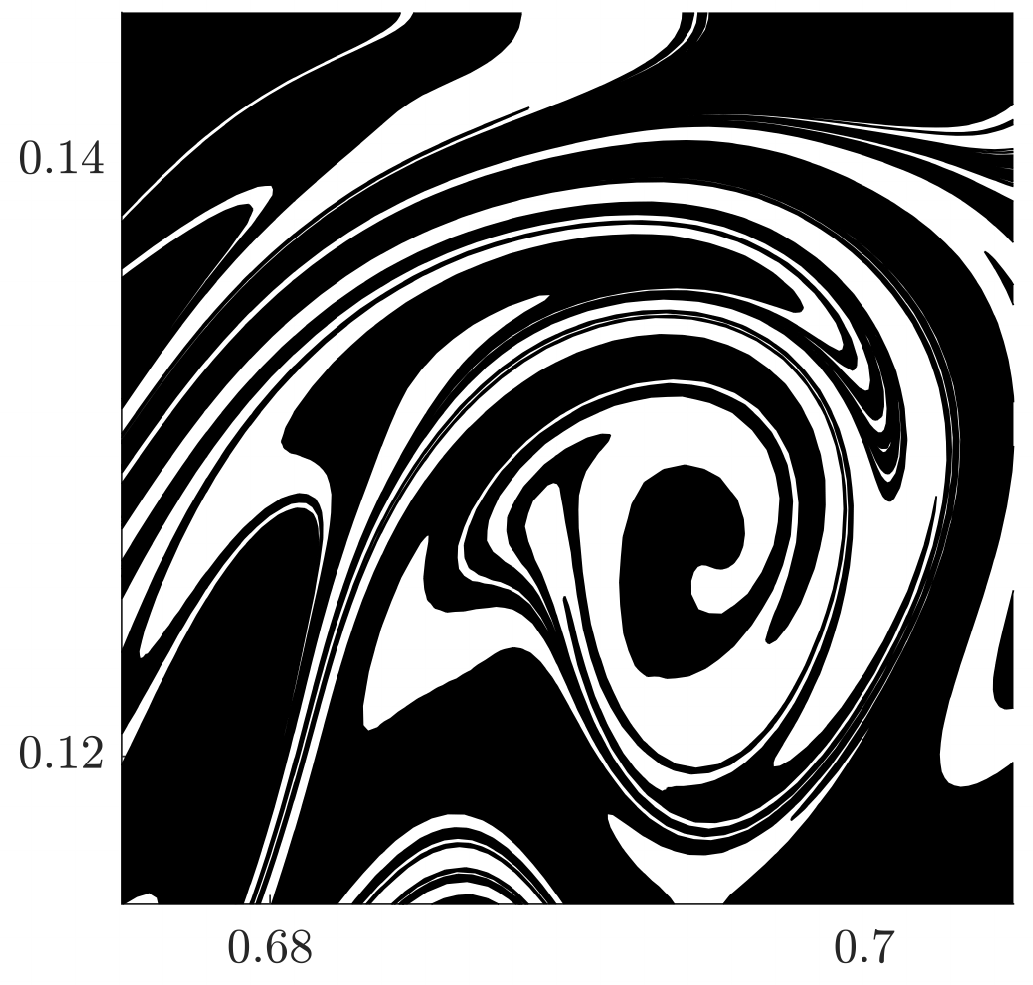}
            \caption{}
        \end{subfigure}
        \caption{Representation of a tiny structure for $\delta=0.008$ with (left) and without surgery (right from the very beginning. To get the scale, recall the characteristic length of a whole system $R=1$. Dashed square in (a) and (b) depicts region zoomed on (c) and (d). Gray lines on left figures depict vortex sheet.}
        \label{effSurg}
    \end{figure}

\begin{figure}[H]
        \centering
        \begin{subfigure}[b]{0.48\textwidth}
            \centering
            \includegraphics[trim={3cm 9.4cm 4cm 9.5cm},clip,width=\textwidth]{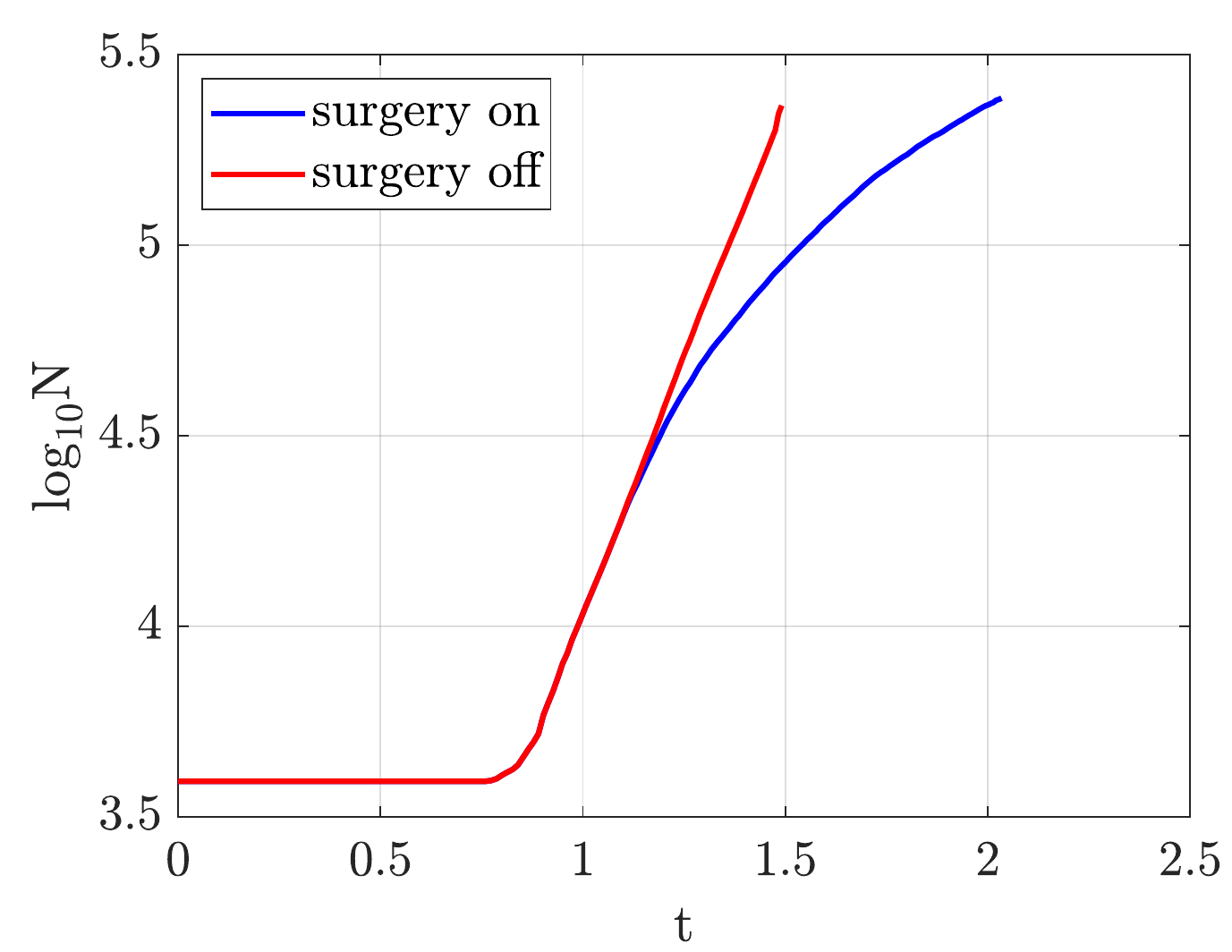}
            \caption{Number of nodes in time. Note that vertical axis is logarithmic.\vspace{\baselineskip}}
            \label{surgN}
        \end{subfigure}
        \hfill 
            \begin{subfigure}[b]{0.51\textwidth}
            \centering
            \includegraphics[trim={3cm 9.4cm 3.1cm 9.4cm},clip,width=\textwidth]{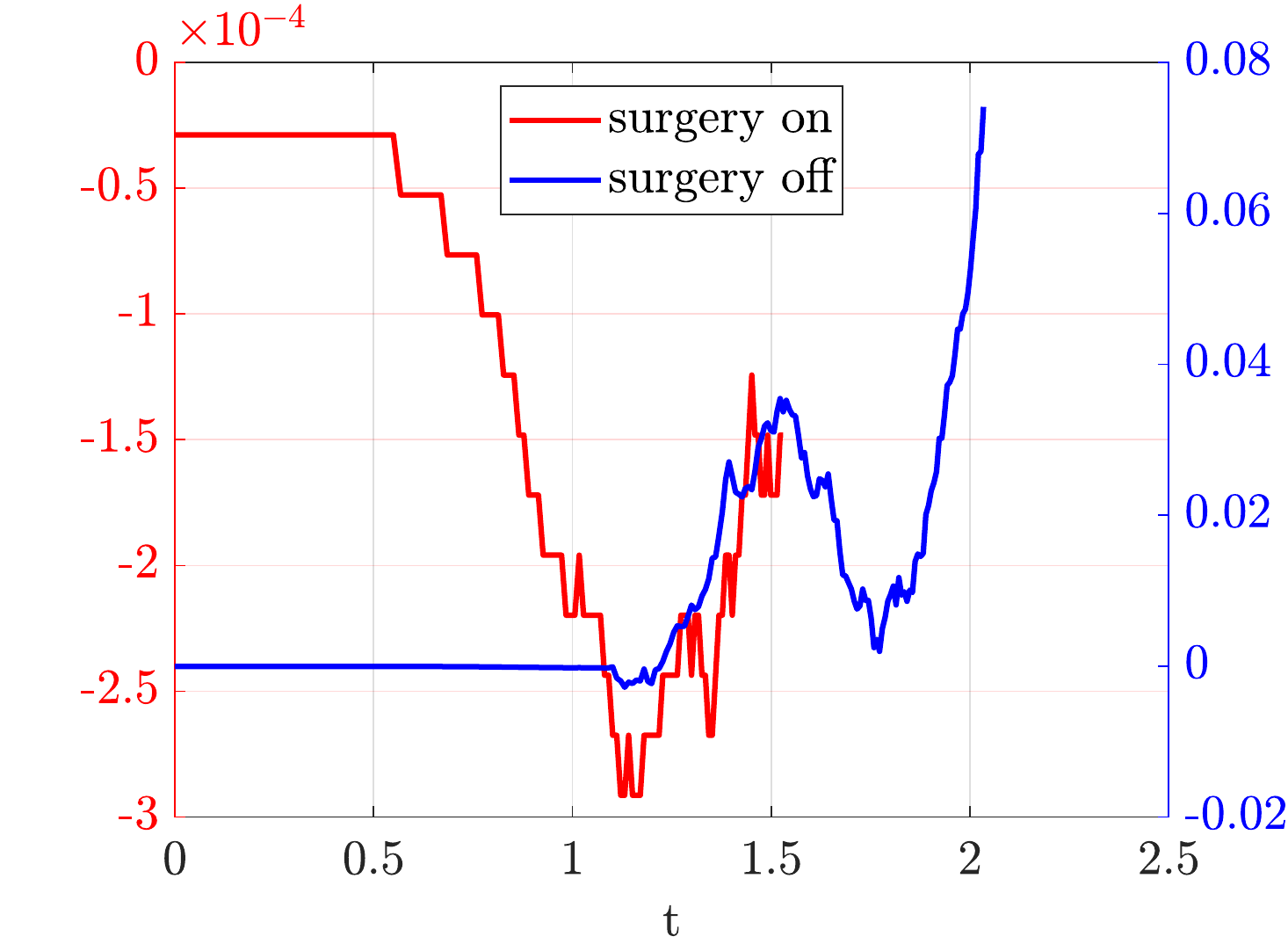}
            \caption{Relative (percentage) error of mass conservation. Notice different scaling of the axes. Single precision output.}
            \label{surgMass}
        \end{subfigure}
        \caption{The same case ($\delta=0.008$) simulated with and without surgery}
    \end{figure}

An effect of surgery is presented on the example of $\delta=0.008$ simulation. We did an additional, shorter simulation with surgery turned off from the very beginning. There is no actual difference in the position, shape, and evolution of the structures. Slight variations are presented on fig. (\ref{effSurg}). Despite a rather conservative surgery setup, the whole vortex sheet was represented with 40\% of the nodes (reduction from 210 663 to 86 615).
The operation allows to avoid an exponential increase in the number of nodes required, what is presented in fig. (\ref{surgN}). The cost is of course a slight violation of the mass conservation fig. (\ref{surgMass}).
\subsection{Accuracy and errors}
\begin{figure}[H]
        \centering
        \begin{subfigure}[b]{0.49\textwidth}
            \centering
            \includegraphics[trim={4.5cm 9.5cm 5cm 9cm},clip,width=\textwidth]{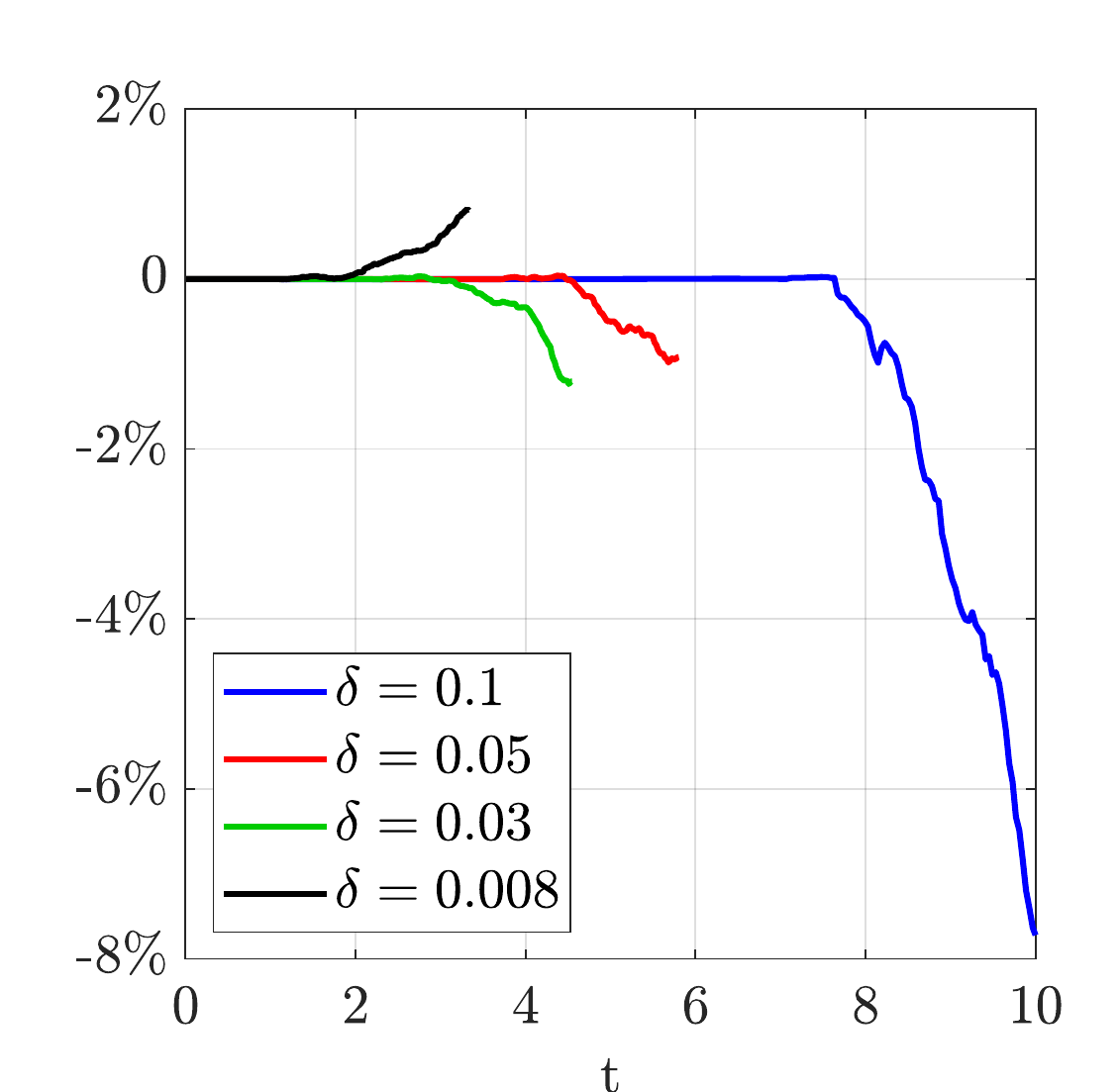}
            \caption{Mass conservation, $b=1$.}
        \end{subfigure}
        \hfill 
            \begin{subfigure}[b]{0.49\textwidth}
            \centering
            \includegraphics[trim={4.1cm 9.5cm 4.3cm 9cm},clip,width=\textwidth]{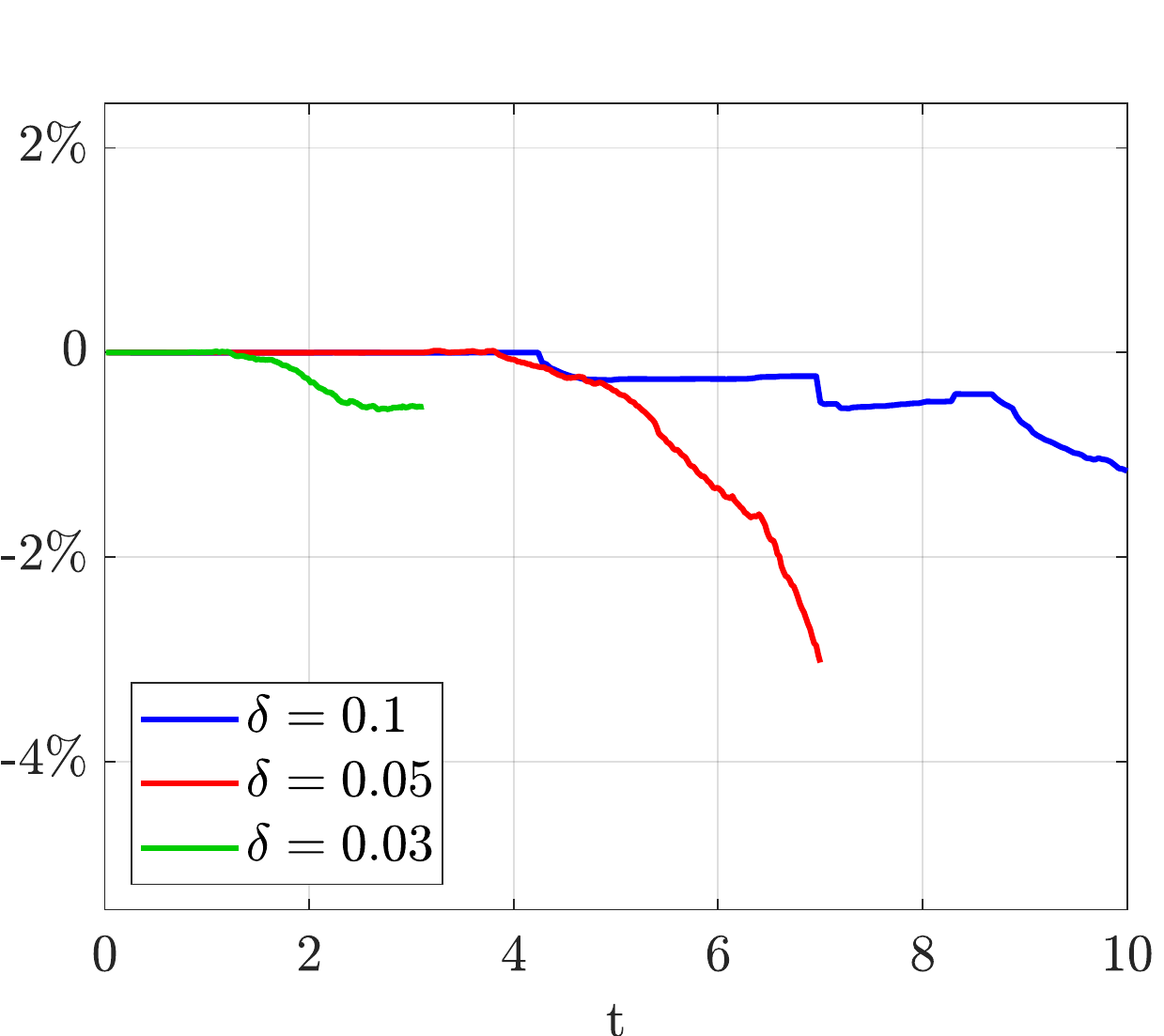}
            \caption{Mass conservation, $b=0$.}
        \end{subfigure}
        \hfill 
        \begin{subfigure}[b]{0.49\textwidth}
            \centering
            \includegraphics[trim={4.2cm 9.5cm 4cm 10.0cm},clip,width=\textwidth]{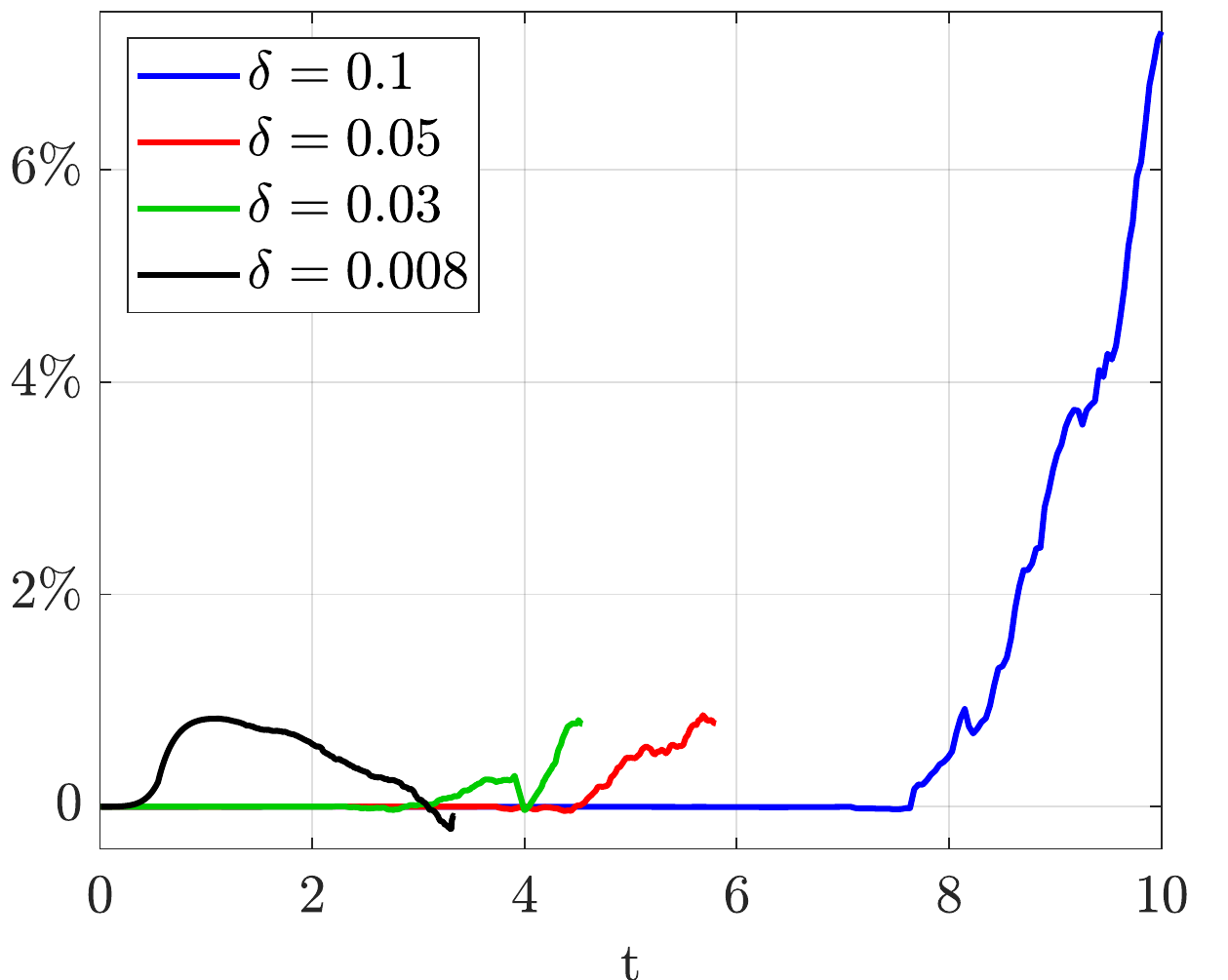}
            \caption{Moment of vorticity evolution, $b=1$.\\\hspace{\textwidth}}
        \end{subfigure}
           \begin{subfigure}[b]{0.49\textwidth}
            \centering
            \includegraphics[trim={4.1cm 9.5cm 4cm 9.2cm},clip,width=\textwidth]{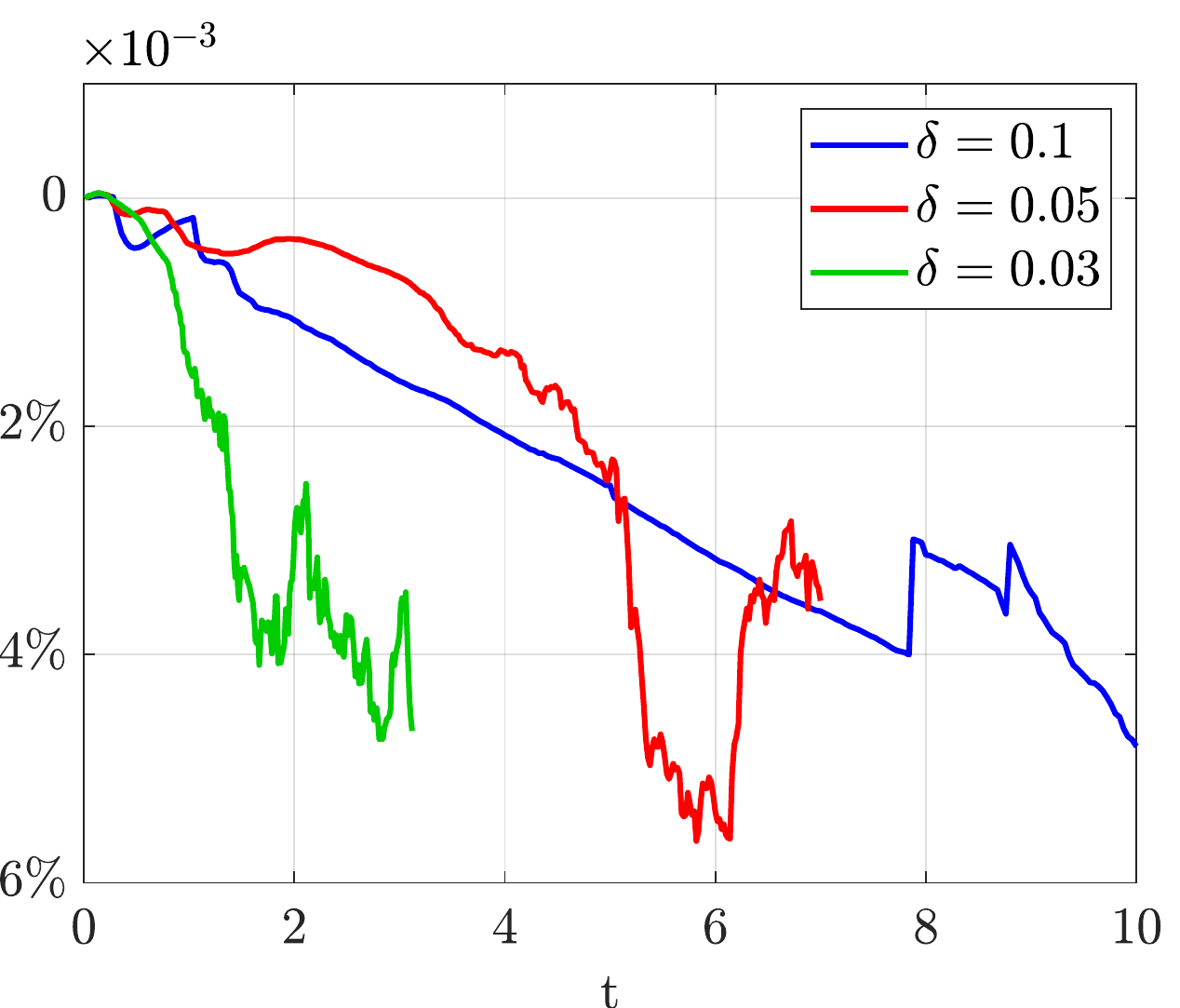}
        \caption{Moment of vorticity evolution, $b=0$. Note the order of magnitude.}
        \label{momCons0}
        \end{subfigure}
        \caption{Relative errors of predictable integrals}  
    \end{figure}

Progressive violation of mass conservation is an unavoidable result of surgery. For that reason, it should not be interpreted as a symptom of poor discretization. Figure (\ref{effSurg}) suggests, that it does not have much impact on the overall behavior of the system. While the sheet is getting more and more concentrated, it is likely that we will ``remove" a significant amount of mass, but still keep the vorticity distribution accurate enough. \\ \\
The error grows faster in cases with lower $\delta$ due to coarser discretization, which allows surgery at higher distances. The same reason causes an error in local vorticity generation. Global circulation, by the design of the numerical scheme, evolves exactly according to (\ref{totalCircEvolution}). Nevertheless, unphysical shifts of segments, introduced by surgery, affect the local distribution of production intensity. This, further, indirectly affects the moment, which, on the other hand, is conserved by surgery quite well (\ref{momCons0}). In cases where the moment of vorticity should just be conserved (no production), the error is much lower.
A probable source of error that might affect the overall behavior of the system is $\delta$. Although it is very robust in filtering out the instabilities in high wavenumbers it still affects the far field. In \cite{krasnyRing} a reported error in the vortex ring speed (comparing to the experiment) was about 15-20\%, although authors used rather stronger smoothing. \\ 
The method of fast velocity induction also introduces an error. It probably does not affect the general evolution much, but it might be far noisier than the rest, lowering the stability of the sheet. In general, we tried to keep all the errors below 1-2\%, with exception of surgery.

\section{Conclusion and further work}

The overall formation of the vortex ring appears to converge with decreasing smoothings. With lower smoothings, the system is subjected to Kelvin-Helmholtz instability, although 
volume-averaged height and radius are affected slightly.\\ \\
A significant difference is noted between the development of buoyancy-driven and mechanically generated rings. The latter, although subjected to roll-up, evolves into a coherent, stable vortex ring, with a clear wake behind. The former gets, at least partially, disintegrated by the negative vorticity that accumulated near its bottom. This qualitative difference happens because buoyancy can generate both positive and negative vorticity locally, whereas there is no vorticty source in mechanically generated case.\\ \\
The proposed methodology consists of two components: the fast velocity induction and the vortex sheet surgery.
The first component decreases computational complexity from $N^2$ to approximately $N\log N$, where $N$ is the number of nodes. Nevertheless, it introduces an error that could affect the sheet's stability. This could be a drawback in study of fundamental vortex phenomena. Increasing the order of accuracy of the method by accounting for higher-order moments of vorticity is left for future work.\\ \\
Vortex sheet surgery provides promising results in terms of efficiency. Even with a conservative setup, it allows to avoid exponential growth of nodes number. Despite the formal violation of integral mass conservation by a few percent, the results seem to keep accuracy. Implementing it for higher-order schemes could allow for more radical exploitations of flow coherences. This would not only decrease the computational complexity but also would support the study of elemental phenomena that play role in turbulence. We directly observe how vortex sheet roll-up can transport energy among the scales by increasing the effective sheet thickness. This further determines the wavenumber of the instability. The observed direction of transport is characteristic for two-dimensional flows. This might be caused by enforced axial symmetry, which is a strong constraint on vortex stretching. Loosening this restriction could be a subject of further development. \\ \\
The richness of scales present in the flow is sensitively dependent on the smoothing parameter $\delta$.
Keeping it constant, as in this study, could be an oversimplification of qualitative importance.
For example, $\delta$ could decrease associated with stretching, induced by Kelvin-Helmholtz instability. This procedure would decrease the critical wavelengths of local instability, transporting energy to the smaller scales, before vortex ironing happens. On the other hand, the thickness of the vortex sheet should also increase in time, due to diffusion in reality. Both mechanisms, on average, would lead to a local equilibrium. These possibilities are what we suggest for further investigations. \\

\printbibliography

\end{document}